\documentclass[11pt,a4paper]{article}
\usepackage{jheppub}
\usepackage{amsmath,amstext,amssymb}
\usepackage{graphicx}
\usepackage{feynmp}
\usepackage{epstopdf}
\usepackage{caption}
\usepackage{float}

\def\be{\begin{equation}}
\def\ee{\end{equation}}
\def\ba{\begin{eqnarray}}
\def\ea{\end{eqnarray}}

\usepackage{dcolumn}% Align table columns on decimal point
\usepackage{bm}% bold math
\usepackage{graphicx}
\usepackage{multirow}
\usepackage{dcolumn}% Align table columns on decimal point

\newcommand{\bq}{\textbf{q}}

\newcommand{\bk}{\textbf{k}}

\begin{document}
\today

\preprint{DESY-20-049}

\title{Analytic structure of the 8-point scattering amplitude in multi-Regge kinematics 
in $N=4$ SYM : \\conformal Regge pole and Regge cut contributions\footnote{This paper is dedicated  to the memory of L.N.Lipatov who has initiated this line of research and provided essential contributions}}

%\subheader{DESY 20-049}

\author[a]{Jochen Bartels}%,\note{Corresponding author.}}

\affiliation[a]{II. Institut f\"{u}r Theoretische Physik, Universit\"{a}t Hamburg, Luruper Chaussee 149,\\
D-22761 Hamburg, Germany}

% e-mail addresses: one for each author, in the same order as the authors
\emailAdd{jochen.bartels@desy.de}

\abstract
{Continuing our investigations of the analytic structure of the $2\to n$ scattering amplitudes in the planar limit of $N=4$ SYM in multi-Regge kinematics  we compute, in all kinematic regions, the Regge cut contributions of the $2\to 6$ process in leading order.
Compared to previous studies of the $2 \to 4$ and the $2 \to 5$ processes we 
encounter two new features: the 3-reggon cut and the product of two 2-reggeon cut contributions.}

\date{\today}

\maketitle

\section{Introduction}

It is now well established that the Bern-Dixon-Smirnow (BDS) conjecture \cite{Bern:2005iz} for the MHV n-point scattering amplitude in the planar limit of the $\mathcal{N}=4$ SYM theory is incomplete for $n \ge 6$. 
One of the first indications for this was found  in \cite{Bartels:2008ce,Bartels:2008sc}.  
Corrections to the BDS-formula have been named 'remainder functions', $R^{(n)}$, and in recent years major efforts have been made for determining these remainder functions, in particular the remainder function $R^{(6)}$ for the case $n=6$ and $n=7$ \cite{Bartels:2013jna,Bartels:2014jya}.
The function $R^{(6)}$ has been calculated for two, and three loops  \cite{Goncharov:2010jf,Lipatov:2010ad,Bartels:2010tx,
DelDuca:2009au,DelDuca:2010zg, Dixon:2011pw, Dixon:2011nj, Dixon:2012yy,Pennington:2012zj,Dixon:2013eka}. 
More recently new techniques have been developped and applied to calculate higher loops \cite{Bargheer:2015djt,DelDuca:2018hrv,DelDuca:2018raq,Caron-Huot:2019vjl,Bargheer:2019lic,DelDuca:2019tur}.

In order to fully analyse and to go beyond this loop expansion, it has turned out to be useful to consider special kinematic limits, in particular the multi-Regge limit. 
The benefit of considering this special region is the particular analytic structure of the scattering amplitudes, in particular the Regge factorization. This stucture is expected to be valid to all orders: the higher order calculations mentioned before can therefore be compared and analyzed. In subsequent papers \cite{Bartels:2013jna,Bartels:2014jya} the analytic structure of the $2\to 4$ and the $2\to 5$ scattering amplitudes has been investigated, making use of Regge theory and unitarity, and in the leading logarithmic approximation the Regge pole and Regge cut contributions have been computed in all kinematic regions. 

The present paper extends these calculations to the $2\to 6$ amplitude. This 
8-point function is of particular interest since it exhibits, for the first time,
a Reggeon cut contribution consisting of three reggeized gluons. In  \cite{Lipatov:2009nt}  it was found that
the Hamlitonian of the Regge-cuts belongs to an integrable open Heisenberg spin chain. 
Regge-cuts composed of $n$ reggeized gluons probe the spin chains consisting of 
$n$ sites. In the $2\to4$ and $2 \to 5$ scattering amplitudes only the shortest spin chain, consisting of two sites, appears; spin chains with three sites appear first in the $2\to6$ scattering amplitudes, spin chains of 4 sites in the $2\to8$ amplitude etc.  
In this paper we compute the partial waves of the $2\to6$ amplitude containing the 3-gluon cut. The energy spectrum of the Hamiltonian of the three gluon state has first been addressed in \cite{Lipatov:2009nt}. Another novel feature of the $2\to 6$ amplitude is the repetition of Regge cuts: the short cut which was found in the $2\to 4$ amplitude, now can appear twice, in the $t_2$ channel and in the $t_4$ channel.

Technically speaking the $2 \to 6$ amplitude requires new fatures. First, when using unitarity for the computation of Regge cut amplitudes, 
the new terms (Regge cuts consisting of three reggeized gluons, and the product of two short cuts)  
now require double and even higher order discontinuities, even for the leading logarithmic approximation. This provides a crucial test of the analytic structure, based on the Steinmann relations. Second, the calculation of subtraction terms now becomes more 
complicated. In \cite{Bartels:2008ce,Lipatov:2010qf} it had been pointed out that the planar approximation, when applied to the Regge limit of $\mathcal{N}=4$ SYM theory, leads to a new feature which requires special attention. As it is well known, Regge theory (Carlson theorem) requires the definition and use  of signatured amplitudes. Once signature has been introduced, in the Regge pole approximation multiparticle production amplitudes factorize.
For QCD this has been verified in the context of the BFKL equation. 
In the planar approximation (leading order large $N_c$), however, there is no space for signature: hence the study of the Regge limit of $N=4$ SYM 
theories in the planar approximation raises the question how much of the known Regge structure remains applicable.
As a first consequence of the absence of signature, it has been observed in  \cite{Bartels:2008ce,Lipatov:2010qf} that,  in the Regge pole  approximation, the  factorization of multiparticle production amplitudes is violated  in certain kinematic regions. This violation is accompanied by the appearance of unphysical singularities. It was then noticed that these singular pieces 
appear in exactly the same kinematic regions where also the Regge cut contributions contribute, and they also 
have the same phase as the Regge cut contributions. Hence they can be removed by re-defining the Regge cut contributions by introducing subtractions terms. 
In previous papers this was shown for the $2 \to 4$ and the $2 \to 5$ amplitudes. In the present paper we show that such subtractions can be found also for the  $2 \to 6$ amplitude. 
This suggests that in the planar lapproximation of $\mathcal{N}=4$ SYM theories, despite the absence of signature, the general structure of Regge amplitudes remains valid.

\section{Outline of the strategy} 

Let us first indicate how our calculations will be performed.  We will make use of the 
method developed in our previous papers on the 7-point function \cite{Bartels:2013jna,Bartels:2014jya}, and wewill  proceed in several steps:\\
(1) We describe the decomposition of the $2 \to 6$ scattering amplitude into a sum of 
several pieces, where  each term, in accordance with the Steinman relations, is characterized by a maximal set of non-overlapping energy discontinuities. In the present case we have 
a sum of 42 terms, and each term has a set of five nonoverlapping energy discontinuities.\\  
(2) In the planar approximation, each term in this decomposition contains a product of energy factors, the phases of which depend upon the kinematic region. Our list applies to the region where all energies are positive.
(In the case of signatured amplitudes one would have to form linear combinations of differenz kinematic regions).\\
(3) Whereas the Regge poles contribute to all 42 terms in this decomposition, the Regge cuts appear in specific 
terms only. Since each Regge cut can be computed from a specific set of energy discontinuities, for each term of the decomposition the content of energy discontinuities allows to decide, which of the Regge cuts might contribute.\\
(4) Regge pole and Regge cut contributions are accompanied by trigonometric factors which have their origin in the 
partial wave expansion. For the 7-point amplitude the method of finding these factors has been described in \cite{Bartels:2013jna,Bartels:2014jya}, and it is not difficult to extend these rules also to the present case (the sum of the Regge pole contributions can also be 
derived from the BDS formula  \cite{Bartels:2013jna}).
It is due to these trigonometric factors that, in certain kinemtaic regions,  unphysical singularities appear. \\  
(5) Focussing now on specific kinematic regions and inserting the corresponding phases of the energy factors, one finds extensive cancellations. It is these cancellations which
let the Regge cut terms appear only in specific kinematic regions. \\
(6) As we have said already,  the Regge pole terms come with unphysical singularities in exactly the same kinematic regions where also Regge cut terms 
appear. It is this coincidence which has suggested to introduce subtractions for the Regge cuts in such a way that the singularities are cancelled.
For the present case the determination of the subtractions  is lengthy and rather technical, and we move it into the  Appendix D. .\\
(7) Once the necessary subtractions of the Regge cut conributions have been found, the representation found in the previous steps can be modified and all singularities arising from the trigonometric factors will be removed.
We believe that the  success of finding these subtraction which consistently remove the singularities  represents an important confirmation of the correctness of our procedure.\\
(8) Once we know the form of the scattering amplitude 
we can use energy discontinuities 
(single and a few higher order energy discontinuities) to determine the individual Regge cut pieces.
It is important to note that, up to this point, the results are expected to be valid to all orders: the form of the scattering amplitude as well as the energy discintinuity relations.\\
(9) The final step is the calculation of the Regge cut terms, based upon the energy (single or double) discontinuity equations and unitarity. The evaluation of the unitarity equations will be  restricted to the leading logarithmic approximation. Most of the ingredients necessary for a NLO calculation are known and can be used,, but this will not be attempted in the present paper.

Our paper is organized as follows. First (section 3) we discuss the analytic structure (steps (1)- (3)). The trigonometric factors are listed in Appendix B,  and the sum of the Regge pole terms is given in Appendix C.
In section 4 we discuss, for the simplest case of the short Regge cut,  the problem of subtractions.
In the following two sections we then go through the different Regge cut contributions and list the kinematic regions
where they appear: section 5 for the short and long cuts, section 6 for the very long cut, the double cut and the 
3-reggeon cut. In each of these sections, we begin with the representation which directly follows from section 3 and the trigonometric factors contained in the Appendices and still contains singularities. Using then the subtractions derived in Appendix D we obtain  the regular representation.  In section 7 we compute the energy discontinuities which will allow to find the partial waves. The final step then is the computation of the corresponding unitarity integrals where we will  restrict ourselves to the leading logarithmic approximation.   Section 8 contains
our  (leading order) results for the very long cut, the 3-reggeon cut and the double cut in the different kinematic regions. Finally, in section 9 we give a brief summary and outlook.

\section{ The Analytic structure}
We begin with the analytic structure of the $2 \to 6$ amplitude in the multiregge region
(steps (1) and (2) of the previous section). Our notation is illustrated in the following figure:
\begin{figure}[H]
\centering
\epsfig{file=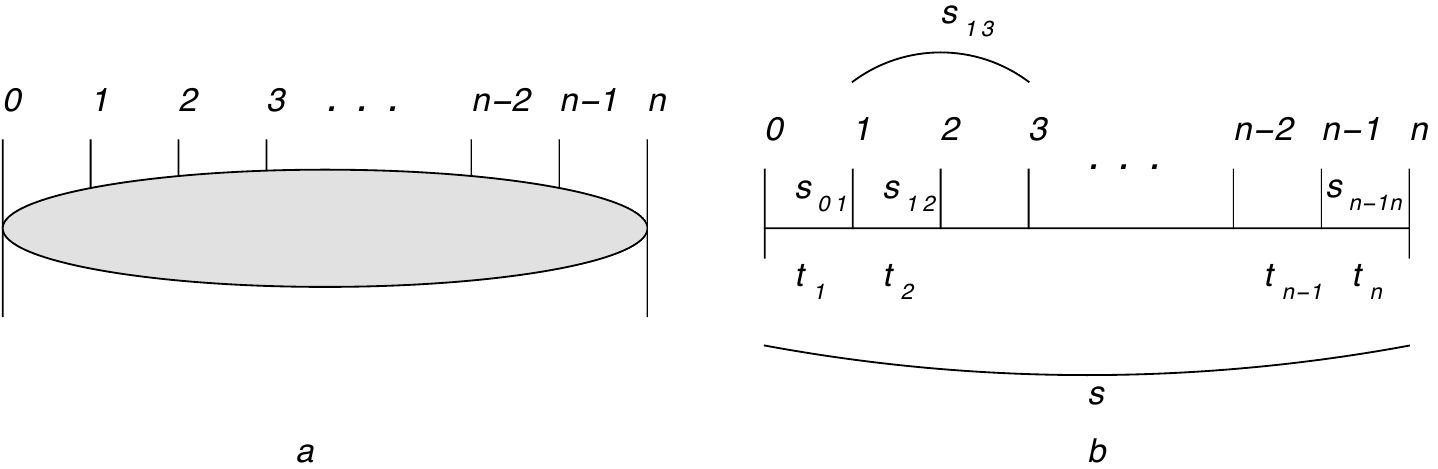,scale=1}
\caption{Notations of the $2\to n+1$ scattering amplitude}
\label{fig:notation}
\end{figure}
\noindent
In the following we will label the four produced particles also by $a$,$b$,$c$,$d$.
\subsection{Decomposition}
We write the  (unsignatured) $2\to6$ scattering amplitude in the multi-Regge kinematics as a sum of 42 terms: 
\be
T= \sum T_{ijkl}.
\label{decomp}
\ee
Each term belongs to a specific set of five simultaneous energy discontinuities in non-overlapping channels. 
We group these terms into five singlets, three doublets,  four triplets, two quartets, a singlet, and a sextet. 
Below we illustrate the discontinuity structure and list the energy factors. Here $\omega_i=j_i-1$,where  $j_i$  denotes the angular momentum in the $t_i$ channel, and $\omega_{ij}=\omega_i - \omega_j$.
First the singlets:
\begin{figure}[H]
\centering
\epsfig{file=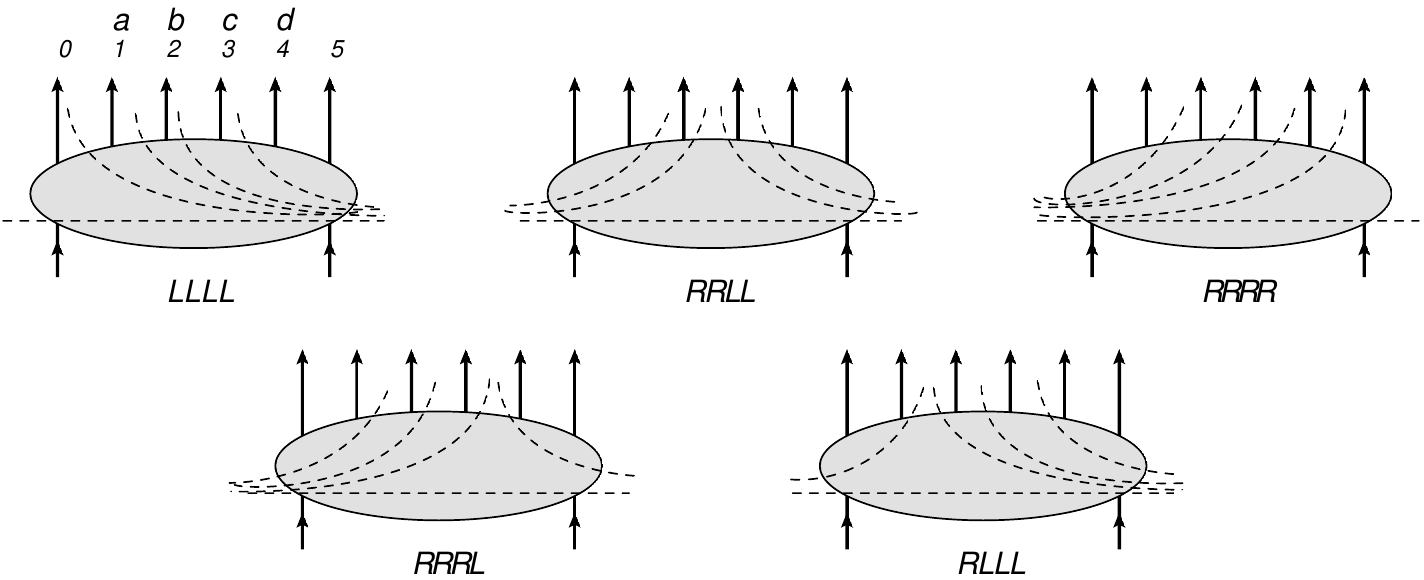,scale=1}
\caption{five singlets}
\label{fig:example_Singlets}
\end{figure}
\noindent
The corresponding energy factors are (in the kinematic region where all energies are positive): 
\begin{eqnarray}
&\;\;\;(-s_{45})^{\omega_{54}}(-s_{35})^{\omega_{43}}(-s_{25})^{\omega_{32}}(-s_{15})^{\omega_{21}}(-s)^{\omega_1}\;\;\;\;&\text{LLLL}\\
&\;\;\;(-s_{01})^{\omega_{12}}(-s_{02})^{\omega_{23}}(-s_{45})^{\omega_{54}}(-s_{35})^{\omega_{43}}(-s)^{\omega_3}\;\;\;\;&\text{RRLL}\\
&\;\;\;(-s_{01})^{\omega_{12}}(-s_{02})^{\omega_{23}}(-s_{03})^{\omega_{34}}(-s_{04})^{\omega_{45}}(-s)^{\omega_5}\;\;\;\;&\text{RRRR}\\
&\;\;\;(-s_{01})^{\omega_{12}}(-s_{02})^{\omega_{23}}(-s_{03})^{\omega_{34}}(-s_{45})^{\omega_{54}}(-s)^{\omega_4}\;\;\;\;&\text{RRRL}\\
&\;\;\;(-s_{01})^{\omega_{12}}(-s_{5})^{\omega_{54}}(-s_{35})^{\omega_{43}}(-s_{25})^{\omega_{32}}(-s)^{\omega_2}\;\;\;\;&\text{RLLL}\,.
\end{eqnarray}

Next the three doublets:
\begin{figure}[H]
\centering
\epsfig{file=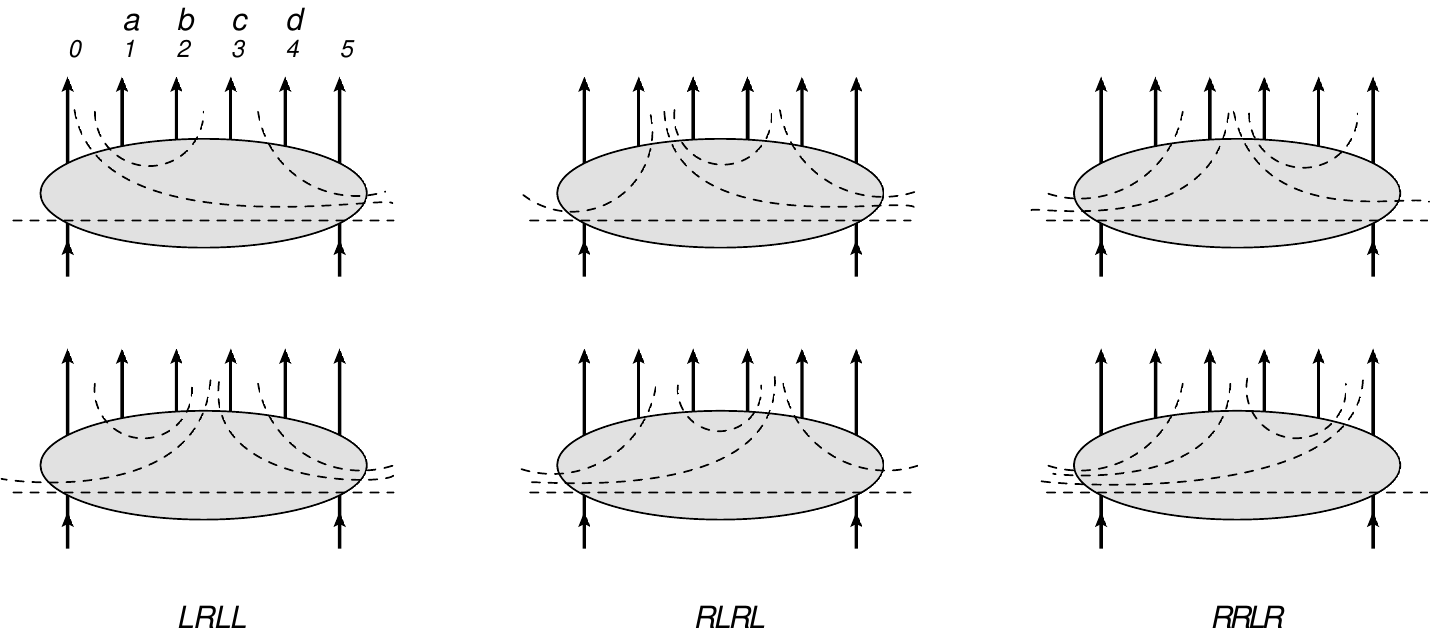,scale=1}
\caption{three doublets}
\label{fig:example5pt_dis}
\end{figure}
\begin{eqnarray}
&\;\;\;(-s_{12})^{\omega_{23}}(-s_{45})^{\omega_{54}}(-s_{35})^{\omega_{43}}(-s_{15})^{\omega_{31}}(-s)^{\omega_1}\;\;\;\;\text{LRLL(1)}\nonumber\\
&\;\;\;(-s_{12})^{\omega_{21}}(-s_{5})^{\omega_{54}}(-s_{35})^{\omega_{43}}(-s_{02})^{\omega_{13}}(-s)^{\omega_3}\;\;\;\;\text{LRLL(2)}
\end{eqnarray}

\begin{eqnarray}
&\;\;\;(-s_{01})^{\omega_{12}}(-s_{23})^{\omega_{34}}(-s_{45})^{\omega_{54}}(-s_{25})^{\omega_{42}}(-s)^{\omega_{12}}\;\;\;\;\text{RLRL(1)}\nonumber\\
&\;\;\;(-s_{01})^{\omega_{12}}(-s_{23})^{\omega_{32}}(-s_{03})^{\omega_{24}}(-s_{45})^{\omega_{54}}(-s)^{\omega_{34}}\;\;\;\;\text{RLRL(2)}
\end{eqnarray}

\begin{eqnarray}
&\;\;\;(-s_{01})^{\omega_{12}}(-s_{02})^{\omega_{23}}(-s_{34})^{\omega_{45}}(-s_{35})^{\omega_{53}}(-s)^{\omega_3}\;\;\;\;\text{RRLR(1)}\nonumber\\
&\;\;\;(-s_{01})^{\omega_{12}}(-s_{02})^{\omega_{23}}(-s_{34})^{\omega_{43}}(-s_{04})^{\omega_{35}}(-s)^{\omega_5}\;\;\;\;\text{RRLR(2)}\,.
\end{eqnarray}

The triplets are of the form:
\begin{figure}[H]
\centering
\epsfig{file=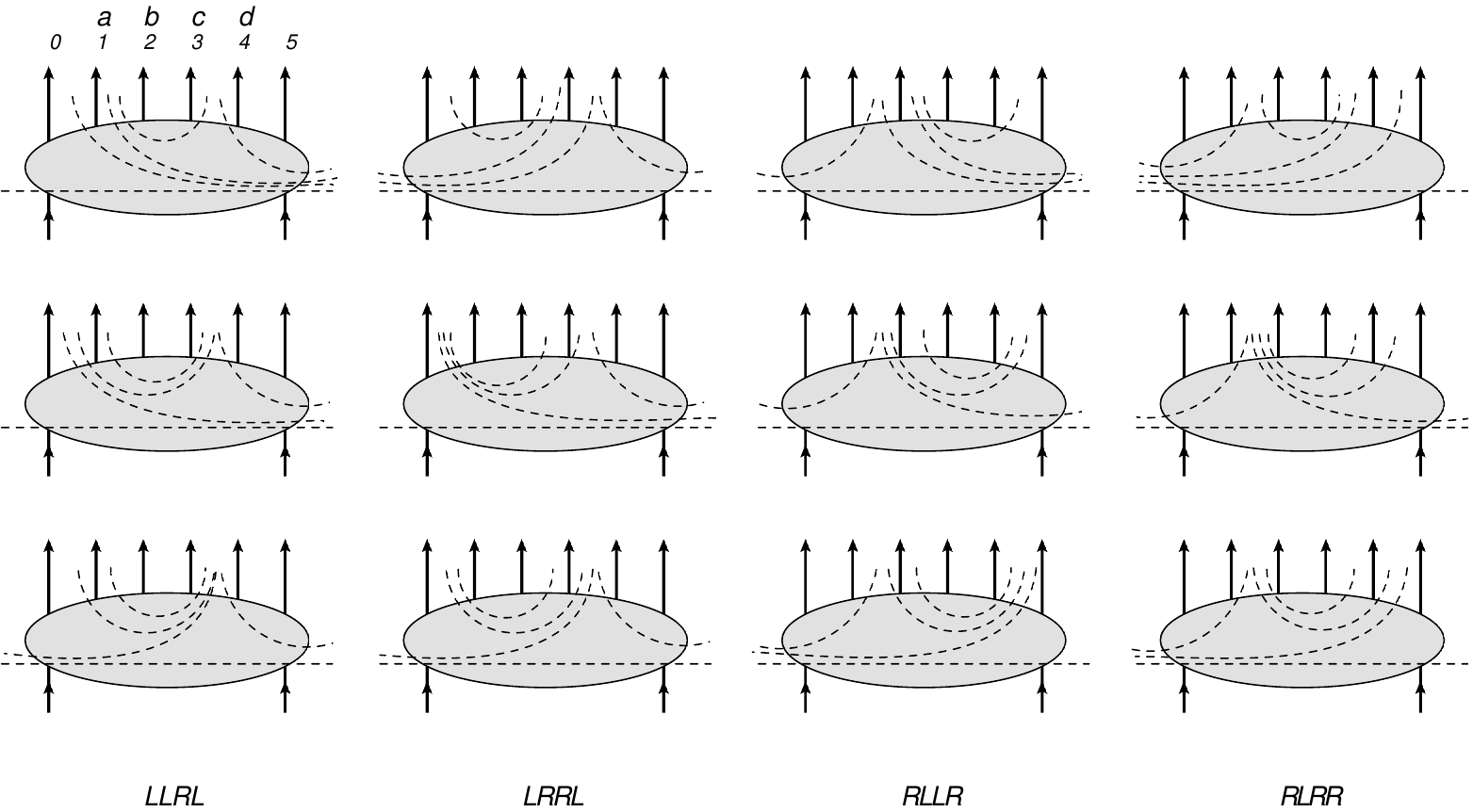,scale=0.95}
\caption{four triplets}
\label{fig:example_4triplets}
\end{figure}
\begin{eqnarray}
&\;\;\;(-s_{23})^{\omega_{34}}(-s_{45})^{\omega_{54}}(-s_{25})^{\omega_{42}}(-s_{15})^{\omega_{21}}(-s)^{\omega_1}\;\;\;\;\text{LLRL(1)\nonumber}\\
&\;\;\;(-s_{23})^{\omega_{32}}(-s_{45})^{\omega_{54}}(-s_{13})^{\omega_{24}}(-s_{15})^{\omega_{41}}(-s)^{\omega_1}\;\;\;\;\text{LLRL(2)}\nonumber\\
&\;\;\;(-s_{23})^{\omega_{32}}(-s_{45})^{\omega_{54}}(-s_{13})^{\omega_{21}}(-s_{03})^{\omega_{14}}(-s)^{\omega_4}\;\;\;\;\text{LLRL(3)}
\end{eqnarray}

\begin{eqnarray}
&\;\;\;(-s_{12})^{\omega_{21}}(-s_{45})^{\omega_{54}}(-s_{02})^{\omega_{13}}(-s_{03})^{\omega_{34}}(-s)^{\omega_4}\;\;\;\;\text{LRRL(1)}\nonumber \\
&\;\;\;(-s_{12})^{\omega_{23}}(-s_{45})^{\omega_{54}}(-s_{13})^{\omega_{34}}(-s_{15})^{\omega_{41}}(-s)^{\omega_1}\;\;\;\;\text{LRRL(2)}\nonumber\\
&\;\;\;(-s_{12})^{\omega_{23}}(-s_{45})^{\omega_{54}}(-s_{13})^{\omega_{31}}(-s_{03})^{\omega_{14}}(-s)^{\omega_4}\;\;\;\;\text{LRRL(3)}
\end{eqnarray}

\begin{eqnarray}
&\;\;\;(-s_{01})^{\omega_{12}}(-s_{34})^{\omega_{45}}(-s_{35})^{\omega_{53}}(-s_{25})^{\omega_{32}}(-s)^{\omega_2}\;\;\;\;\text{RLLR(1)} \nonumber\\
&\;\;\;(-s_{01})^{\omega_{12}}(-s_{34})^{\omega_{43}}(-s_{24})^{\omega_{32}}(-s_{04})^{\omega_{25}}(-s)^{\omega_5}\;\;\;\;\text{RLLR(2)}\nonumber\\
&\;\;\;(-s_{01})^{\omega_{12}}(-s_{34})^{\omega_{43}}(-s_{24})^{\omega_{35}}(-s_{25})^{\omega_{52}}(-s)^{\omega_2}\;\;\;\;\text{RLLR(3)}
\end{eqnarray}

\begin{eqnarray}
&\;\;\;(-s_{01})^{\omega_{12}}(-s_{23})^{\omega_{32}}(-s_{03})^{\omega_{24}}(-s_{04})^{\omega_{45}}(-s)^{\omega_5}\;\;\;\;\text{RLRR(1)}\nonumber\\
&\;\;\;(-s_{01})^{\omega_{12}}(-s_{23})^{\omega_{34}}(-s_{24})^{\omega_{42}}(-s_{04})^{\omega_{25}}(-s)^{\omega_5}\;\;\;\;\text{RLRR(2)}\nonumber\\
&\;\;\;(-s_{01})^{\omega_{12}}(-s_{23})^{\omega_{34}}(-s_{24})^{\omega_{45}}(-s_{25})^{\omega_{52}}(-s)^{\omega_2}\;\;\;\;\text{RLRR(3)}\,.
\end{eqnarray}
Next the two quartets: 
\begin{figure}[H]
\centering
\epsfig{file=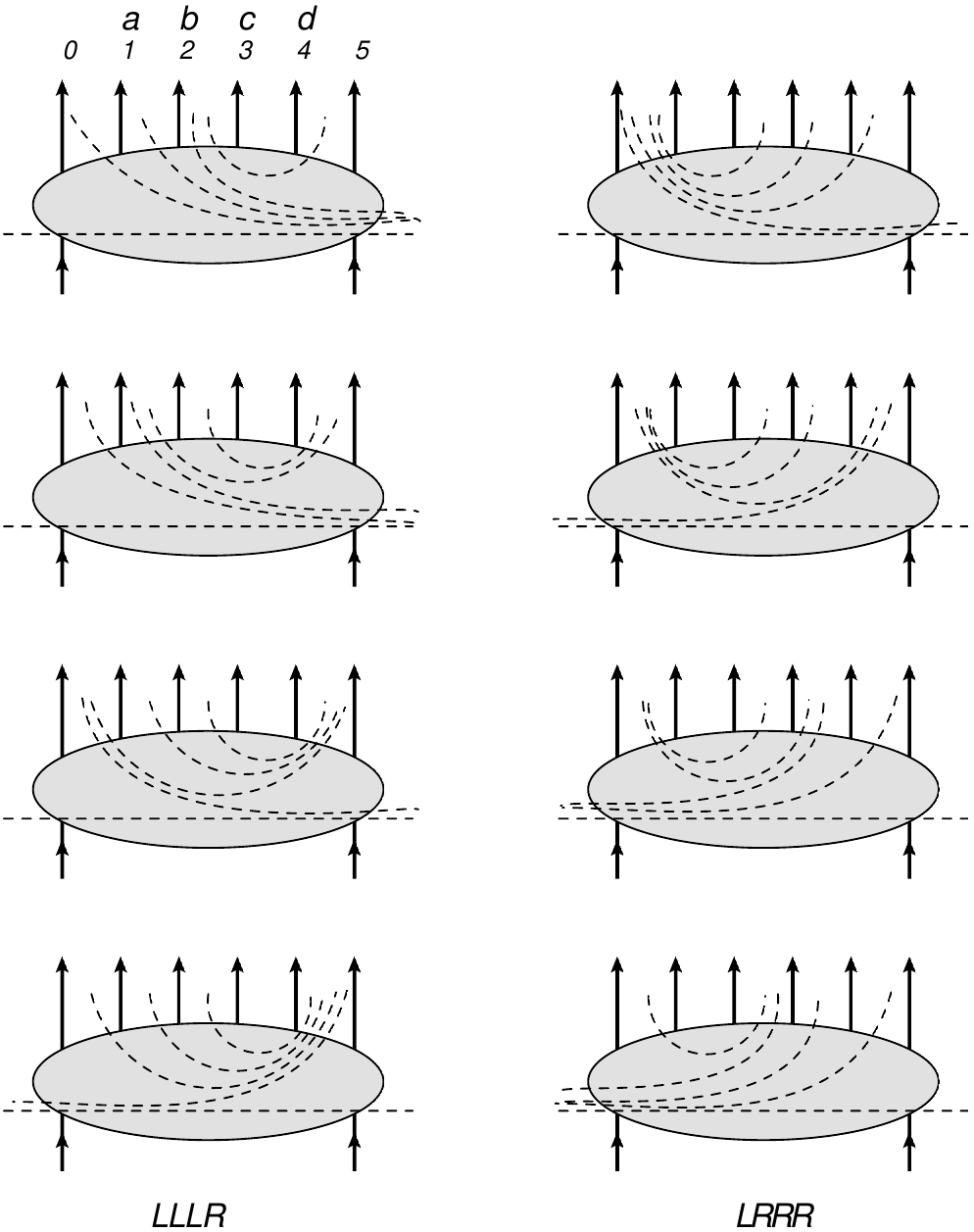,scale=1}
\caption{two quartets}
\label{fig:example_2quartets}
\end{figure}

\begin{eqnarray}
&\;\;\;(-s_{34})^{\omega_{45}}(-s_{35})^{\omega_{53}}(-s_{25})^{\omega_{32}}(-s_{15})^{\omega_{21}}(-s)^{\omega_1}\;\;\;\;\text{LLLR(1)}\nonumber\\
&\;\;\;(-s_{34})^{\omega_{43}}(-s_{24})^{\omega_{35}}(-s_{25})^{\omega_{52}}(-s_{15})^{\omega_{21}}(-s)^{\omega_1}\;\;\;\;\text{LLLR(2)}\nonumber\\
&\;\;\;(-s_{34})^{\omega_{43}}(-s_{24})^{\omega_{32}}(-s_{14})^{\omega_{25}}(-s_{15})^{\omega_{51}}(-s)^{\omega_1}\;\;\;\;\text{LLLR(3)}\nonumber\\
&\;\;\;(-s_{34})^{\omega_{43}}(-s_{24})^{\omega_{32}}(-s_{14})^{\omega_{21}}(-s_{04})^{\omega_{15}}(-s)^{\omega_5}\;\;\;\;\text{LLLR(4)}
\end{eqnarray}

\begin{eqnarray}
&\;\;\;(-s_{12})^{\omega_{21}}(-s_{02})^{\omega_{13}}(-s_{03})^{\omega_{34}}(-s_{04})^{\omega_{45}}(-s)^{\omega_5}\;\;\;\;\text{LRRR(1)}\nonumber\\
&\;\;\;(-s_{12})^{\omega_{23}}(-s_{13})^{\omega_{31}}(-s_{03})^{\omega_{14}}(-s_{04})^{\omega_{45}}(-s)^{\omega_5}\;\;\;\;\text{LRRR(2)}\nonumber\\
&\;\;\;(-s_{12})^{\omega_{23}}(-s_{13})^{\omega_{34}}(-s_{14})^{\omega_{41}}(-s_{04})^{\omega_{15}}(-s)^{\omega_5}\;\;\;\;\text{LRRR(3)}\nonumber\\
&\;\;\;(-s_{12})^{\omega_{23}}(-s_{13})^{\omega_{34}}(-s_{14})^{\omega_{45}}(-s_{15})^{\omega_{51}}(-s)^{\omega_1}\;\;\;\;\text{LRRR(4)}\,.
\end{eqnarray}

The quintet has the form (here the numbering 1..5 goes from left to right starting from the upper left diagram):
\begin{figure}[H]
\centering
\epsfig{file=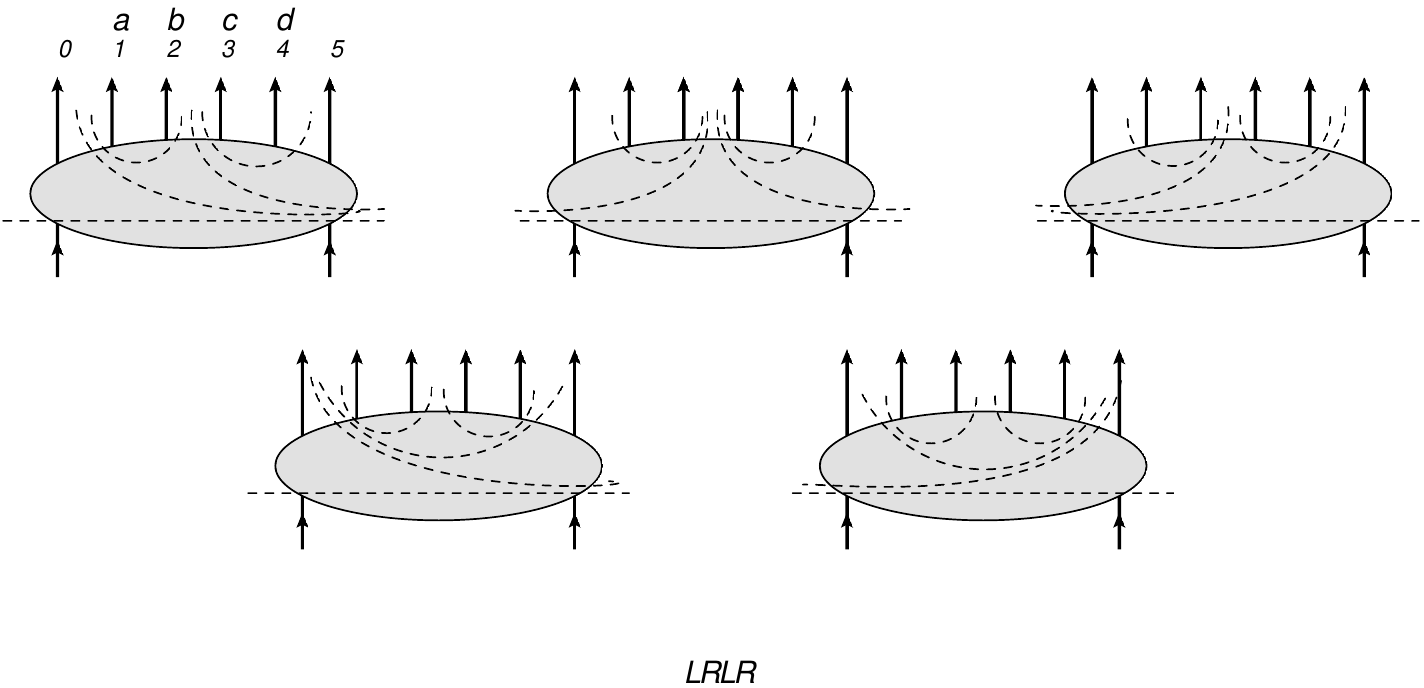,scale=1}
\caption{quintet}
\label{fig:example_Quintet}
\end{figure}

\begin{eqnarray}
&\;\;\;(-s_{12})^{\omega_{23}}(-s_{34})^{\omega_{45}}(-s_{15})^{\omega_{31}}(-s_{35})^{\omega_{53}}(-s)^{\omega_1}\;\;\;\;\text{LRLR(1)}\nonumber \\
&\;\;\;(-s_{12})^{\omega_{21}}(-s_{34})^{\omega_{45}}(-s_{02})^{\omega_{13}}(-s_{35})^{\omega_{53}}(-s)^{\omega_3}\;\;\;\;\text{LRLR(2)}\nonumber\\
&\;\;\;(-s_{12})^{\omega_{21}}(-s_{34})^{\omega_{43}}(-s_{02})^{\omega_{13}}(-s_{04})^{\omega_{35}}(-s)^{\omega_5}\;\;\;\;\text{LRLR(3)}\nonumber\\
&\;\;\;(-s_{12})^{\omega_{23}}(-s_{34})^{\omega_{43}}(-s_{14})^{\omega_{35}}(-s_{15})^{\omega_{51}}(-s)^{\omega_1}\;\;\;\;\text{LRLR(4)}\nonumber\\
&\;\;\;(-s_{12})^{\omega_{23}}(-s_{34})^{\omega_{43}}(-s_{14})^{\omega_{31}}(-s_{04})^{\omega_{15}}(-s)^{\omega_5}\;\;\;\;\text{LRLR(5)}
\end{eqnarray}

Finally the sextet (the numbering 1..6 is contained in the figure. The diagrams in the lower line are the mirror reflections of the upper line):
\begin{figure}[H]
\centering
\epsfig{file=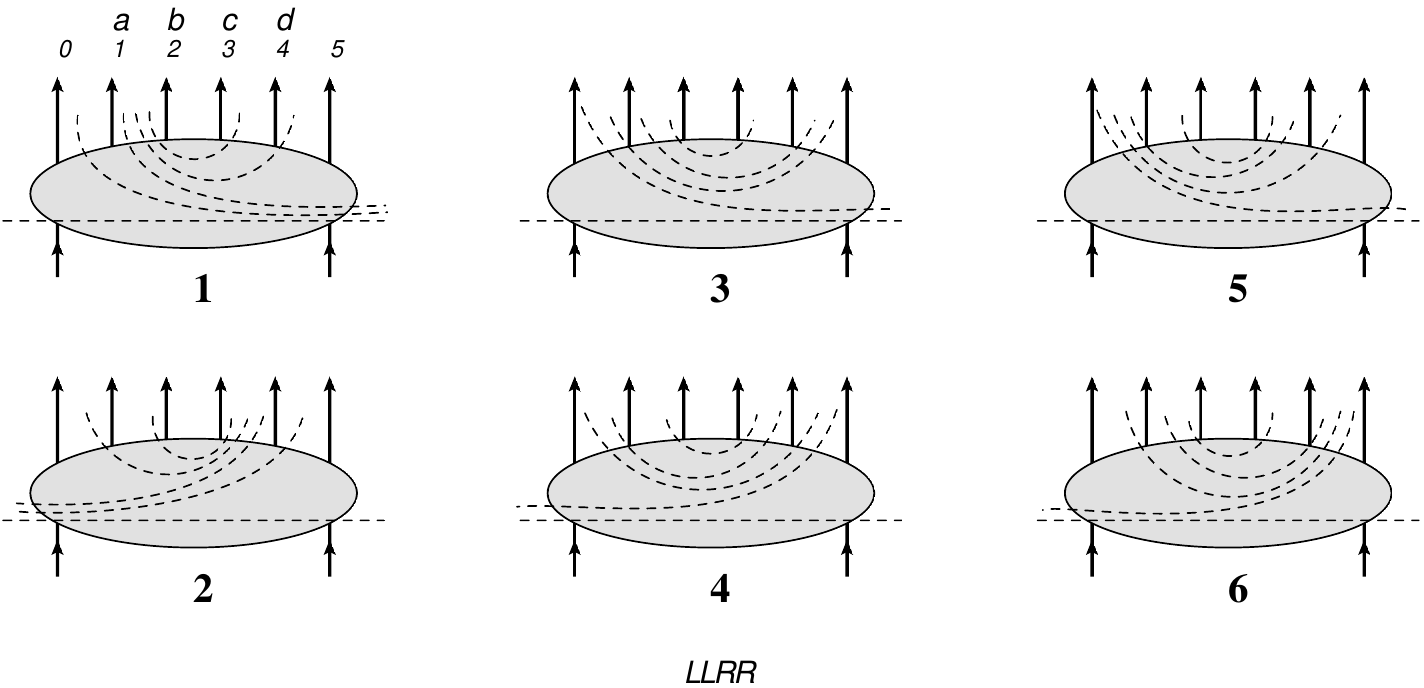,scale=1}
\caption{one sextet}
\label{fig:example_Sixtet}
\end{figure}

\begin{eqnarray}
&\;\;\;(-s_{23})^{\omega_{34}}(-s_{24})^{\omega_{45}}(-s_{25})^{\omega_{52}}(-s_{15})^{\omega_{21}}(-s)^{\omega_1}\;\;\;\;\text{LLRR(1)}\nonumber\\
&\;\;\;(-s_{23})^{\omega_{32}}(-s_{13})^{\omega_{21}}(-s_{03})^{\omega_{14}}(-s_{04})^{\omega_{45}}(-s)^{\omega_5}\;\;\;\;\text{LLRR(2)}\nonumber\\
&\;\;\;(-s_{23})^{\omega_{34}}(-s_{24})^{\omega_{42}}(-s_{14})^{\omega_{25}}(-s_{15})^{\omega_{51}}(-s)^{\omega_1}\;\;\;\;\text{LLRR(3)}\nonumber\\
&\;\;\;(-s_{23})^{\omega_{32}}(-s_{13})^{\omega_{24}}(-s_{14})^{\omega_{41}}(-s_{04})^{\omega_{15}}(-s)^{\omega_5}\;\;\;\;\text{LLRR(4)}\nonumber\\
&\;\;\;(-s_{23})^{\omega_{32}}(-s_{13})^{\omega_{24}}(-s_{14})^{\omega_{45}}(-s_{15})^{\omega_{51}}(-s)^{\omega_1}\;\;\;\;\text{LLRR(5)}\nonumber\\
&\;\;\;(-s_{23})^{\omega_{34}}(-s_{24})^{\omega_{42}}(-s_{14})^{\omega_{21}}(-s_{04})^{\omega_{15}}(-s)^{\omega_5}\;\;\;\;\text{LLRR(6)}
\end{eqnarray}

As it can be seen from Fig.2 - 7, each term can be characterized by a set of 4 subscripts 'L' or 'R'. Beginning with the left-most produced particle, the  first subscript distinguishes whether the energy discontinuity line  comes from the left or from the right. The second subscript says the same for the second produced particle, and so on.
A closer look at Fig.2 - 7 shows that these labels are not sufficient: depending on the discontinuity structure, we still have singlets, doublets etc.

For each term, the product of energy factors clearly reflects the non-overlapping multiple energy discontinuities. Often it is more convenient to rewrite these energy factors in a factorizing form. Using identities such as 
\be
s_{13} =  \frac{s_{12}s_{23}}{\kappa_{23}}, \,\,\kappa_{23}=(\bq_2-\bq_3)^2
\ee
it is then straightforward to see that in all terms the energy factors (disregarding the phases) can be written as the product
\be
s_{01}^{\omega_1}s_{12}^{\omega_2}s_{23}^{\omega_3} s_{34}^{\omega_4}s_{45}^{\omega_5} \cdot (\text{product of $\kappa$ factors}).
\ee
In  the product of $\kappa_{ii+1}$ factors the exponents depend upon the subscripts  'L' or 'R': for each produced particle labelled by   'L' or 'R':  we have $(\kappa_{ii+1})^{-\omega_{i+1}}$ or $(\kappa_{ii+1})^{-\omega_{i}}$, resp. As an example, 
terms labelled by LLLL come with the product
\be
(\kappa_{12})^{-\omega_1}(\kappa_{23})^{-\omega_2}(\kappa_{34})^{-\omega_3}(\kappa_{12})^{-\omega_4}.
\ee
In leading order, we can put these  $\kappa$ factors equal to unity.

From now on we introduce a small change of our notation. Instead of $\omega_i=j_i-1$ we now will use, as integration variable,   $\omega'_i=j_i-1$, and the unprimed  variable 
\ba 
\label{omega-i}
\omega_i&=&\omega(\bq_i^2)=\alpha(t_i)  -1\nonumber\\
&=&-\frac{\gamma_K}{4} \ln \frac{|\bq_i|^2 }{ \lambda^2}, \,\, \gamma_K= 4a,\,\, a=\frac{\alpha_s N_c}{2 \pi},\,\,t_i=-\bq_i^2
\ea
will be used  for the reggeon trajectory function in the $t_i$ channel: $\omega_i=\alpha(t_i)-1$.

\subsection{A few general remarks on the formulae of the scattering amplitude}

For each term we have a Sommerfeld-Watson integral. A more detailed discussion has been given in our previous paper \cite{Bartels:2014jya}. Each term is written in the form:
\ba
T_{LLLL} &=& s \int ...\int \frac{d \omega'_1d \omega'_2 d \omega'_3d \omega'_4 d\omega'_5} {(2\pi i)^5}   (-s_{45})^{\omega'_{54}}(-s_{35})^{\omega'_{43}}(-s_{25})^{\omega'_{32}}(-s_{15})^{\omega'_{21}}(-s)^{\omega'_1}
\nonumber\\&&\cdot F_{LLL}(t_1,t_2,t_3,t_4,t_5;\omega'_1,\omega'_2,\omega'_3,\omega'_4,\omega'_5) ,
\label{struct-LLLL}
\ea
where $\omega'_{ij}=\omega'_i-\omega'_j$  
The $F_{ijkl}$ are real-valued, and they contain, in addition to the trigonometric factors to be discussed below, the partial waves. Each $F_{ijkl}$ is written as a sum of several pieces 
which contain Regge pole or Regge cut singularities:
\be
F_{ijkl} = F_{ijkl}^{pole} + F_{ijkl}^{\text{Regge cut 1}}+ F_{ijkl}^{\text{Regge  cut 2}} + ... \,.
\label{poles-cuts}
\ee 

In order to obtain a signatured amplitude we have to form linear combinations of crossed and uncrossed amplitudes. This amounts to replacing the phase of the energy factor by a signature factor $\xi$, e.g.
\be
(-s_{35})^{\omega'_{43}} \to s_{35}^{\omega'_{43}} \xi_{43},\,\,\,\xi_{43}=e^{-i\pi \omega'_{43}} + \tau_4\tau_3.
\ee
Instead, in the planar approximation the product of signature factors is expanded in products of the $\tau_i$, and each term denotes a particular kinematic region. For example, the term without any $\tau_i$ denotes the region where all energies are positive, the term $\tau_1$ the region where the $t_1$ channel has been twisted etc. In the following we will use this notation for labelling the different kinematic regions.

We are interested in the corrections to the BDS expression for the scattering amplitude, depending on the kinematic region.. For each region
$\tau_i...\tau_j$ we write the scattering amplitude in the form
\be
A_{\tau_i...\tau_j} = A_{\tau_i...\tau_j}^{BDS} R_{\tau_i...\tau_j}.
\ee
Here the BDS part contains the Regge pole part illustrated in Fig\ref{fig:polecontr}:
\begin{figure}[H]
\centering
\epsfig{file=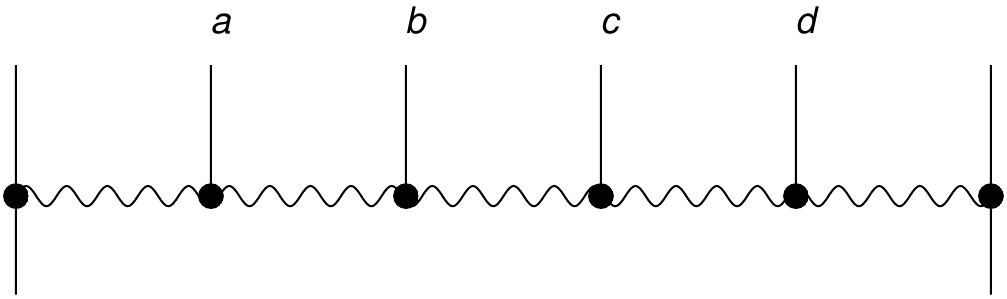,scale=0.6}
\caption{Regge pole contribution}
\label{fig:polecontr}
\end{figure}
\noindent
consisting of the energy factors (without phases) , the real valued couplings to the external incoming particles, the absolute values of the production vertices, and the two phase factors:
\be
\label{BDS-fact}
A_{\tau_i...\tau_j}^{BDS}= \pm |A_{\tau_i...\tau_j}^{BDS}| e^{i\pi \varphi_{\tau_i...\tau_j}} e^{i \delta_{\tau_i...\tau_j}}.
\ee
Here the $\delta_{\tau_i...\tau_j}$ contain the one loop Regge cuts, and in Appendix A we give a list of 
the phases $\pi \varphi_{\tau_i...\tau_j}+\delta_{\tau_i...\tau_j}$ for all kinematic regions which contain Regge cuts. In this notation, our discussion will focus on the product 
\be
\label{interest}
e^{i\pi \varphi_{\tau_i...\tau_j}} e^{i \delta_{\tau_i...\tau_j}} R_{\tau_i...\tau_j},
\ee
and the remaining factor will be called the 'BDS' part.
In our leading approximation, the exponential  $e^{i \delta_{\tau_i...\tau_j}}$ will be expanded. The same convention will be used when we discuss energy discontinuities: the BDS part will be extracted and not written explicitly.

The integral representation (\ref{struct-LLLL}) contains, for each $t_i$-channel, the integral over the corresponding angular momentum $\omega'_i=j_i-1$. If, for a given kinematic region $\tau_j...\tau_l$,  the $t_i$-channel has only a Regge pole contribution, the $\omega'_i$ integral can be done directly and  leads to a factor $s_{i-1i}^{\omega_i}$ where $\omega_i$ denotes the gluon trajetory in the $t_i$ channel   (\ref{omega-i}).
If the $t_i$-channel has a Regge cut contribution, the corresponding $\omega'_i$ can be shifted:
\be
\label{shift}
\omega'_i= \omega''_i + \omega_i
\ee
and in this way also produces a factor $s_{i-1i}i^{\omega_i}$ (a more detailed discussion of this shift will be given 
in section 5.1). As a result, all energy factors contained in the BDS part will be factored out from our scattering amplitudes,
derived in this paper, leading to the expression (\ref{interest}).
 
Finally  we address the question which Regge pieces contribute to each of the 42 terms. As we have said before, each Regge pole or cut term comes with a product of trigonometric factors, resulting from the partial wave expansions in the 
various $t$-channels. Detailed rules and their application to the $2\to5$ scattering process have been given in  the appendix of \cite{Bartels:2014jya}, and it is not dificult to generalize to the $2 \to 6$ process. A complete list of these trigonometric factors is given in Appendix B: for each of the 42 terms in our decomposition we list the trigonometric factors of the Regge pole and cut contributions. 
Regge poles appear in all terms, Regge cuts only in those terms which contain the set of energy discontinuities related to the Regge cut.  As an example, the very long two-reggeon cut in the $t_2$, $t_3$ and $t_4$  channel can appear only in those terms which have an energy discontinuity in $s_{1234}$:
\ba
&&LLLR(3),\,\, LLLR(4),\,\, LRRR(3),\,\, LRRR(4),\,\, LRLR(3),\,\, LRLR(4),\nonumber\\
&&LLRR(3),\,\,LLRR(4),\,\,LLRR(5),\,\, LLRR(6).
\ea
In order to find out in which kinematic region a given Regge cut contribution appears, we combine the trigonometric factors 
of this Regge cut and the phases coming from the energy factors of this kinematic region, and take the sum over all of the 42 terms which contain this Regge cut. As a result, one finds many cancellations, and at the end the Regge cut, if at all,  appears only in very specific kinematic regions. As one  general result we mention that, both in the region where all energies are positive and in the purely euclidean region, all Regge cut contributions cancel. 
On the other hand, the Regge poles appear in all kinematic regions, and a complete list of their contributions is given in Appendix C.

There exist important consistency checks. Beginning with the Regge pole 
contributions, the obtained results which are listed in Appendix  C can be compared with the pole expressions derived from the BDS formulae (as outlined in  \cite{Bartels:2013jna}). One finds complete agreement. As to Regge cuts, in the appendix of \cite{Lipatov:2009nt} arguments have been given that, in the planar approximation, Regge cuts can appear only if some of the produced particles are continued to the region of negative energies. As an example, in the $2 \to 4$ amplitude  the two reggeon cut in the $t_2$ channel contributes only if both the $t_1$ and $t_3$ channels are twisted, i.e. both produced particles have to be continued to have negative energy. In the $2 \to 6$ amplitude, the very long Regge cut requires twists in the $t_1$ and $t_5$ channels. 

In the following we will consider the different kinematic regions.  We find it convenient to discuss the different cut contributions  seperately. Beginning with the region where only one short Regge cut (one t-channel) contributes, we then address the regions with a long cut (two t-channels), and finally come to the regions where the very long cut (three t-channels), the double cut and the three reggeon cut appear.

\section{The problem of subtractions}

Before we begin with the different kinematic regions we say a bit more about the problem of subtractions. As we have already said before, the planar approximation leads to a problem which requires special attention. For this we return to the Regge pole contributions. 
As discussed in \cite{Lipatov:2010qf,Bartels:2013jna},
starting from the BDS formula one finds that the Regge pole contributions become singular in all those kinematic regions where Regge cut contributions appear, and these singularities 
have the same phase structure as the Regge cuts. This suggests to re-define the Regge cut by a subtraction term which removes the singularity.
For illustration we go to Appendix C where we have listed  the pole contributions of the $2 \to 6$ amplitude for all kinematic regions.
Their form can be derived from the BDS formula; alternatively we could also start from Appendix B and compute the sum of the pole contributions in the different regions.
The table begins with those regions which have no Regge cuts, and the pole contributions are just phase factors. In all subsequent regions, the pole conributions contain singularities. 

To be definite, let us consider the regions $\tau_1\tau_3$ and $\tau_1\tau_2\tau_3$, in which only the short 2-reggeon cut in the $t_2$-channel contributes.
In this region the Regge pole contribution in Appendix B is of the form:
\be 
 e^{-i\pi(\omega_2+\omega_4+\omega_5)} e^{i\pi(\omega_c+\omega_d)} \Big[
e^{i\pi(\omega_a+\omega_b)}-2i e^{i \pi \omega_2} \frac{\Omega_a\Omega_b}{\Omega_2}\Big],
\ee
which we can also write as  
\be 
\label{pole13}
 e^{-i\pi(\omega_2+\omega_4+\omega_5)} e^{i\pi(\omega_c+\omega_d)} \Big[
\cos \pi \omega_{ab} -2 i \left(- \frac{1}{2} \sin \pi(\omega_a+\omega_b) + \cos \pi \omega_2 \frac{\Omega_a\Omega_b}{\Omega_2}\right) \Big].
\ee 
Here we have used:
\ba
\omega_a&=&   -\frac{\gamma_K}{8} \ln \frac{|q_1|^2 |q_2|^2}{|k_a|^2 \lambda^2}\\
\omega_{ab}&=&\omega_a-\omega_b \nonumber\\
\Omega_a&=& \sin \pi \omega_a\,\,,\Omega_2=\sin \pi \omega_2\,,
\ea
and as discussed before, the BDS part has been removed (in particular the energy factors of fhe Regge poles).
Similarly for the region $\tau_1\tau_2\tau_3$:
\be
\label{pole123}
e^{-i\pi(\omega_4+\omega_5)} e^{i\pi(\omega_c+\omega_d)} \Big[
-\cos \pi \omega_{ab} -2 i \left(- \frac{1}{2} \sin \pi(\omega_a+\omega_b) + \cos \pi \omega_2 \frac{\Omega_a\Omega_b}{\Omega_2}\right) \Big].
\ee 
In (\ref{pole13}) and (\ref{pole123}) it is the brackets, in particular the terms proportional to $1/\Omega_2$, which are unphysical and should be removed by the Regge cut in the $t_2$ channel, $W_{\omega_2}$.

For the sum of all terms containing the short Rdegge cut $W_{\omega_2}$ we obtain after some algebra, before the 
integration over $\omega'_2$
\ba
\tau_1\tau_3:& \hspace{1cm} &2i e^{-i\pi(\omega'_2+\omega_4+\omega_5)} W_{\omega_2}e^{i\pi \omega_c} 
e^{i \pi \omega_d}\\
\tau_1\tau_2\tau_3:& \hspace{1cm}&2i e^{-i\pi(\omega_4+\omega_5)} W_{\omega_2}e^{i\pi \omega_c} 
e^{i \pi \omega_d}.
\ea 
Next we include, for the Regge cut amplitude,   the $\omega'_2$ integral with the energy $s_{12}$:
\be 
\int \frac{d \omega'_2}{2 \pi i} W_{\omega_2} (-s_{12})^{\omega'_2} = s_{12}^{\omega_2} e^{-i\pi \omega_2} \int_2 W_{\omega_2} (-s_{12})^{\omega''_2}
\ee
with 
\be 
\int_2= \int \frac{d \omega''_2}{2 \pi i}.
\ee
Here we have perfomed the shift discussed in (\ref{shift}), and extracted the factor $s_{12}^{\omega_2}$. It belongs to the 
BDS part, and in the following we will disregard it. 

Now the Regge cut part has the same phase structure as the singular 
pieces of the Regge pole term. In order to cancel these singular pieces  we put
\be
 \int_2 (-s_{12})^{\omega''_2} W_{\omega_2}= \delta W_{\omega_2} + \int_2  (-s_{12})^{\omega''_2}W_{\omega_2}^{reg}
\label{decomp-W2}
\ee
with 
\be
\label{delta-W2}
\delta W_{\omega_2}= - \frac{1}{2} \sin \pi(\omega_a+\omega_b) +  \cos \pi \omega_2 \frac{\Omega_a\Omega_b}{\Omega_2}.
\ee  
With this subtraction $\delta W_{\omega_2}$ 
we obtain for the sum of the Regge pole and the short Regge cut
\be
\tau_1\tau_3: \hspace{1cm}  e^{-i\pi(\omega_2+\omega_4+\omega_5)} e^{i\pi(\omega_c+\omega_d)} \Big[
\cos \pi \omega_{ab}+2i\int_2(-s_{12})^{\omega''_2}W_{\omega_2}^{reg}  \Big].
\ee
The first term in the square brackets defines the conformal infrared finite Regge pole contribution in this kinematic region.
In the same way we find for the region $\tau_1\tau_2\tau_3$ (using the same subtraction $\delta W_{\omega_2}$):
\be
\tau_1\tau_2\tau_3: \hspace{1cm}  e^{-i\pi(\omega_4+\omega_5)} e^{i\pi(\omega_c+\omega_d)} \Big[
-\cos \pi \omega_{ab}+2i\int_2(s_{12})^{\omega''_2}W_{\omega_2}^{reg}  \Big].
\ee
Obviously, $\delta W_{\omega_2}$ is a subtraction term of the $\omega''$-integral in the angular momentum plane.

For simplicity, we write (\ref{decomp-W2}) in the short hand notation
\be
\label{delta-W2-simp}
\int W_{\omega_2}=\delta W_{\omega_2} + \int W_{\omega_2}^{reg},
\ee
i.e. we will not explicitly write  the $\omega''_2$ integral and the energy variables multiplying $W_{\omega_2}$ or  $W_{\omega_2}^{reg}$.

It is important to keep in mind that these singular terms also appear in the energy discontinuity relations. For simplicity we consider the discontinuity in $s_{12}$  in the 
region of positive energies:
\be 
\Delta_{12} = e^{-i\pi(\omega_4+\omega_5)} e^{i \pi (\omega_c+\omega_d)}
  \Big[\int_2 s_2^{\omega''_2} W_{\omega_2} -  \frac{V_L(a) V_R(b)}{\Omega_2} \Big].
\ee
Here
\ba
&&V_L(a)= \sin \pi (\omega_2-\omega_a),\,\,V_R(a)= \sin \pi (\omega_1-\omega_a)\nonumber\\
&&V_L(b)= \sin \pi (\omega_3-\omega_b),\,\,V_R(b)= \sin \pi (\omega_2-\omega_b).
\ea
Again the singularity of the Regge pole term appears. By inserting  (\ref{decomp-W2}) and (\ref{delta-W2}), also this discontinuity becomes regular:
\be
\label{disc12}
\Delta_{12} = e^{-i\pi(\omega_4+\omega_5)} e^{i \pi (\omega_c+\omega_d)}
\Big[ \int_2 s_{12}^{\omega''_2}W_{\omega_2}^{reg}-\frac{1}{2}
 \left( e^{i\pi\omega_a} V_R(b)+e^{i\pi\omega_b} V_L(a)\right)\Big] .
\ee  

It should be stressed that the singularity in the Regge pole terms (\ref{pole13}) and (\ref{pole123}) woukd cancel if,
instead of considering the regions $\tau_1\tau_3$  and  $\tau_1\tau_2\tau_3$ separately,
we would form odd signatured amplitudes.This demonstrates that it is the planar approximation which is connected with the appearance of these singularities.

The simple case  of the short cut in the $t_2$-channel generalizes to all kinematic regions where Regge cuts appear: the Regge pole contributions have singular tems which 
have to be cancelled by subtractions of Regge cut contributions
\be
\int W_{\text{Regge cut}} = \delta W_{\text{Regge cut}}+\int W_{\text{Regge cut}}^{reg}.
\ee
Here we will use the shorthand notation discussed after  (\ref{delta-W2-simp}).

For the cases $2\to4$ and $2\to5$ the subtraction terms  have been found and discussed in previous papers. For the present case $2\to6$ it is one of the 
main challenges to find the subtraction terms for the new Regge cut conributions and to verify that, with these subtractions, both  the scattering amplitudes and the energy discontinuities become regular.

\section{The cut contributions of the short and the long Regge cuts}

We now determine the Regge cut contributions. We will go through the Regge cuts (short, long, very long, double cut, three reggeon cut) in the  different kinematic regions and combine, for each region, the phases from the energy factors listed in the prevous section with the trigonometric factors listed in Appendix B and C. These calculations are lengthy and are done using Mathematica. The results, however, become simple. In the next step we decompose the partial waves $W$:
\be
 \int W=\delta W +  \int W^{reg}  
\ee
and compute the singular pieces, $\delta W$,  from the requirement that they cancel the unphysical singularities of the Regge pole contributions. We then find the amlitudes containing only regular terms. In the next step we write down equations for the energy discontinuities, inserting the necessary subtractions. Finally, we use unitarity equations (restricting ourselves to the leading logarithmic approximation) and find explicit expressions for the Regge cut contributions.

\subsection{The short cuts}
We begin with the Regge cut contributions in the $\omega_2$ channel and conisder the regions 
$\tau_1 \tau_3$ and  $\tau_1\tau_2\tau_3$. This region has been discussed already in the last section, and we only need to compute the energy discontinuity in (\ref{disc12}). First we write the equation  in the leading approximation:
\be
\label{disc12_LL}
\Delta_{12} =
 \int_2 s_{12}^{\omega''_2}W_{\omega_2}^{reg}-\frac{\pi}{2}
 (2 \omega_2-\omega_a  -\omega_a ) .
\ee  
\noindent
For the computation of the discontinuity on the lhs we use unitarity  (Fig.\ref{fig:single-disc}a):
\begin{figure}[H]
\centering
\epsfig{file=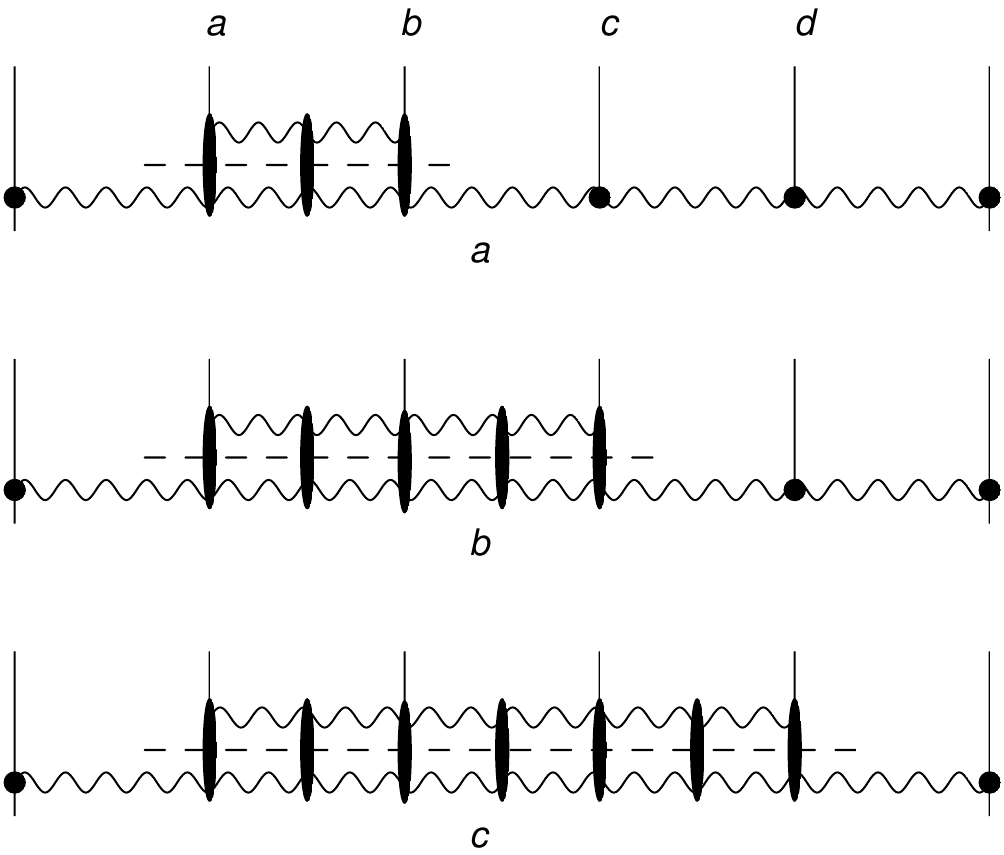,scale=0.9}
\caption{Single discontinuities in (a) $s_{12}$, (b) $s_{123}$, (c) $s_{1234}$}
\label{fig:single-disc}
\end{figure}
\noindent
and for the unitarity integrals we restrict ourselves to the leading log approximation. In leading order the production vertices in Fig.\ref {fig:single-disc} simplify:
\begin{figure}[H]
\centering
\epsfig{file=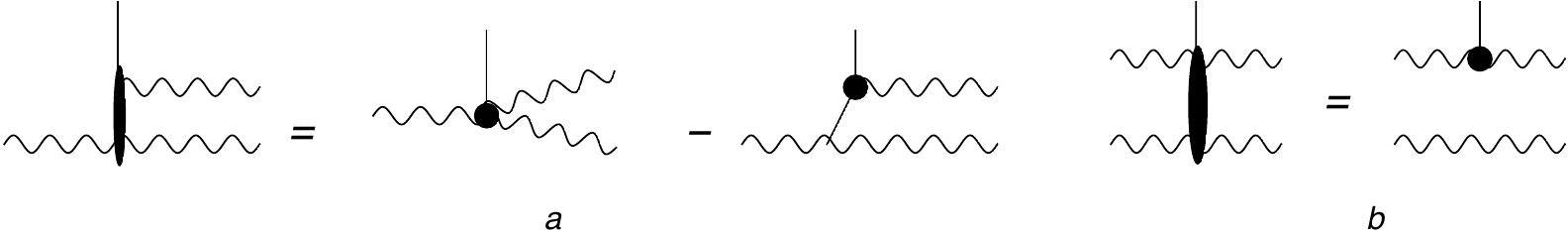,scale=0.8}
\caption{Production vertices in leading order}
\label{fig:leading-order-vertices}
\end{figure}
\noindent
In particular, in (a) the production vertices at the lhs end of the cut is a sum of a pointlike vertex and and a 
vertex involving a particle propagator. In the following this latter part will be denoted by $\Phi$. With this the result is:
\ba
\label{dd_12}
\Delta_{12}&=& \left( -\pi (\omega_2-\omega_a-\omega_b) + \frac{\pi}{2} V_{13} \right)+f_{\omega_2},
\ea
where we have used
\be
V_{ij}=   \frac{\gamma_K}{4} \ln \frac{|q_i|^2|q_j|^2}{|q_i-q_j|^2 \lambda^2}
\ee
with $V_{12}=-2 \omega_a$ etc. 
Here the first three terms denote the one loop contributions, and the leading order function $f_{\omega_2}$ which is illustrated in Fig.\ref{fig:cut-amplitudes} a 
starts from 2 loops::
\begin{figure}[H]
\centering
\epsfig{file=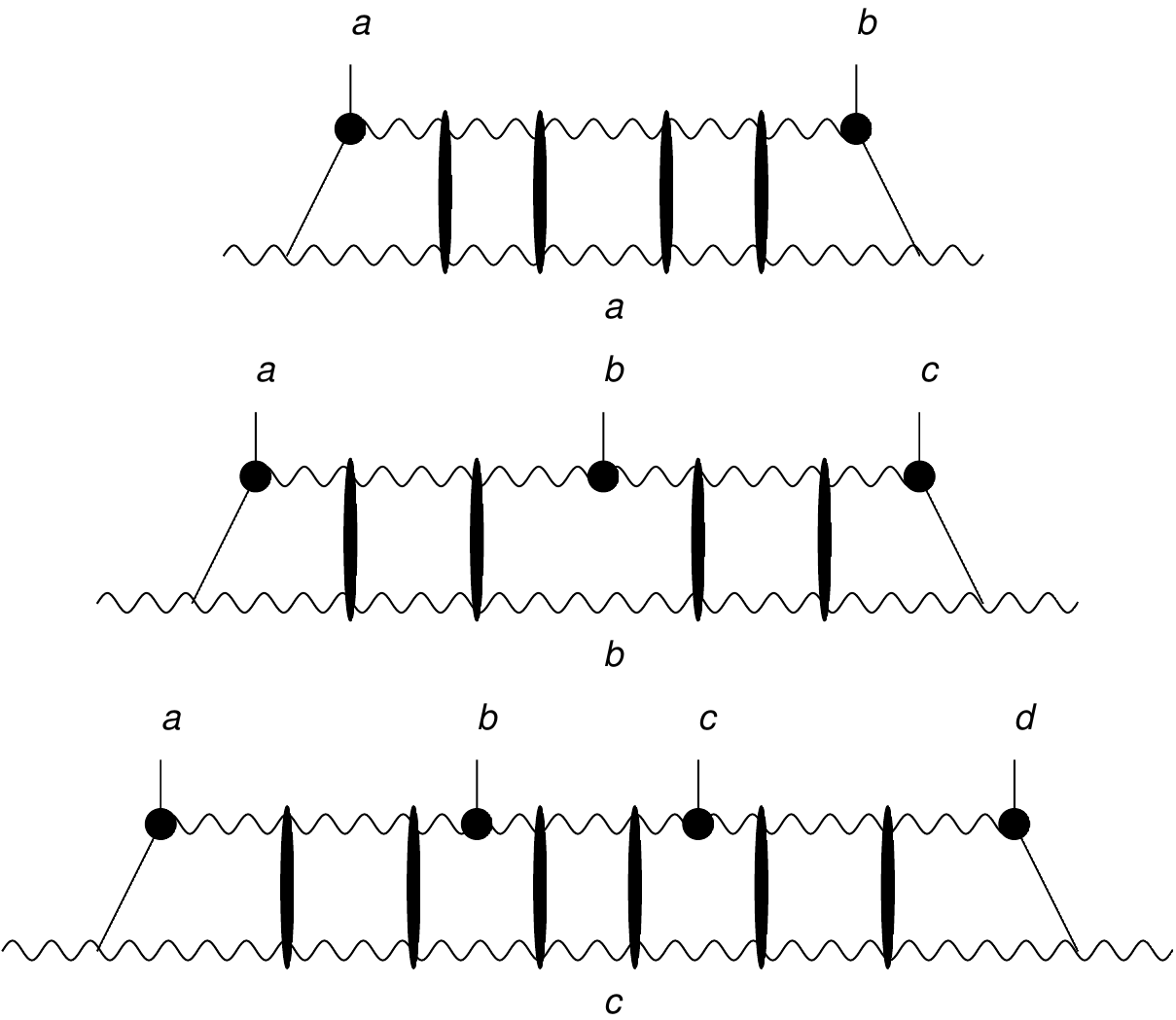,height=6cm,width=8cm}\\
\caption{Illustration of Regge cut amplitudes (a) $f_{\omega_2}$, (b) $f_{\omega_2\omega_3}$,
(c) $f_{\omega_2\omega_3\omega_4}$.}
\label{fig:cut-amplitudes}
\end{figure}

Before we continue we come back to the shift in the $\omega'_2$ which we have discussed in section 3.2.
The rungs in Fig.\ref{fig:single-disc} denote the BFKL kernel in the octet representetion which we write as \cite{Bartels:2008sc}:
\ba
\label{octet-kernel}
K^{(8_A)}(\bk,\bk',\bq)&=&\delta^{(2)}(\bk-\bk') \left(\omega(\bk^2)+\omega((\bq-\bk^2)\right)+\frac{a}{2} \frac{(q-k)k^* (q-k')^*{k'}+ c.c.}{(\bk-\bk')^2}\nonumber\\
&=&\delta^{(2)}(\bk-\bk') \Big[\omega(\bq^2)+\frac{1}{2} \left(\omega(\bk^2)+\omega((\bq-\bk)^2)-
2\omega(q^2)\right)\Big]\nonumber\\
&&+\frac{1}{2} K^{(1)}(\bq-\bk;\bq-\bk',\bk')\nonumber\\
&=&\delta^{(2)}(\bk-\bk') \Big[\omega(\bq^2)-\frac{a}{2} \ln \frac{\bk^2 (\bq-\bk)^2}{\bq^2 \bq^2} \Big]
+\frac{1}{2} K^{(1)}(\bq-\bk;\bq-\bk',\bk')\nonumber\\
&=& \delta^{(2)}(\bk-\bk') \omega(\bq^2)+ K^{2;planar} (\bq-\bk,\bk;\bq-\bk',\bk') .
\ea
The new kernel  $K^{2;planar} (\bq-\bk,\bk;\bq-\bk',\bk')$ has the important bootstrap 
property .
\be 
\int d^2\bk' K^{2;planar} (\bq-\bk,\bk;\bq-\bk',\bk') \cdot 1 =0.
\ee
In the  ladder diagrams of Fig.\ref{fig:cut-amplitudes}a the Green's function satisfies the equation
\ba
\omega' G_{\omega}^{(8_A)}(\bk,\bk',\bq_2) = \frac{\delta^{(2)}(\bk-\bk')}{\bk^2 (\bk-\bq_2)^2}
+ \frac{1}{\bk^2 (\bk-\bq_2)^2}\left(K^{(8_A})\otimes G_{\omega}^{(8_A)}\right)(\bk,\bk',\bq_2).
\ea
Writing its solution in the simplified form :
\be
 G_{\omega}^{(8_A)}
=\frac{1}{\omega'-K^{(8_A)}}=\frac{1}{\omega'-\omega(\bq_2)-K^{2;planar}}
\ee
we see that, when performing the $\omega'_2$-integral, we can shift the integration variable 
\be
\omega''_2= \omega'_2+\omega_2
\ee
and obain an extra  factor 
\be
s_{12}^{\omega_2}.
\ee
This factor belongs into the BDS part and, therefore, will be disregarded (as it has already been done in (\ref{disc12}) and (\ref{disc12_LL}) ). In the following this shift applies to all 2-reggeon cut confributions, and whenever we encounter an $\omega''_i$-integration it is understood that 
the corresponding energy factor $s_{i-1i}^{\omega_i}$ will be disregarded since it has been absorbed by the BDS part of the scattering amplitude.
 
Returning now to (\ref{dd_12}) we have for  $f_{\omega_2}$ the following analytic representation in the  $\nu,n$-representation
(Fig.\ref{fig:single-disc}a).: 
\ba
f_{\omega_2}=
\frac{g^2 N_c}{16\pi^2} \sum_n (-1)^n \int \frac{d\nu}{\nu^2 +\frac{n^2}{4}} \left( (-s_{12})^{\omega(\nu,n)} - 1\right)
\left(\frac{q_3^*k_a^*}{k_b^* q_1^*}\right)^{i\nu-\frac{n}{2}}  \left(\frac{q_3k_a}{k_b q_1}\right)^{i\nu+\frac{n}{2}},
\label{W-omega2-cut}
\ea
where we have subtracted  the infrared divergent one loop integral, $V_{ij}$.
It is convenient to introduce, in  the infrared finite and conformal invariant phases:
\be
\delta_{13}= \pi \left( V_{13} + \omega_a + \omega_b\right), 
\ee
which leads to the result for the unitarity integral:
\be 
\Delta_{12}=-\pi \omega_2 +\frac{\pi}{2}(\omega_a+\omega_b) +\frac{ \delta_{13}}{2}
+f_{\omega_2}.
\ee

With this unitarity integral  we return to (\ref{disc12_LL})
and obtain the conformal invariant result for the short Regge cut in the 
$\omega_2$ channel:
\be 
\int_2 (-s_{12})^{\omega''_2} W_{\omega_2}^{reg}= f_{\omega_2} + \frac{\delta_{13}}{2} .
\label{omega2-LL}
\ee 
Our final result for the weak coupling limit thus becomes:
\ba
\tau_1\tau_3: &\hspace{1cm}&  e^{-i\pi(\omega_2+\omega_4+\omega_5)} e^{i\pi(\omega_c+\omega_d)} \Big[
\cos \pi \omega_{ab}+2i\left( f_{\omega_2} + \frac{ \delta_{13}}{2} \right) \Big] \nonumber\\
\tau_1\tau_2\tau_3: &\hspace{1cm}&  e^{-i\pi(\omega_4+\omega_5)} e^{i\pi(\omega_c+\omega_d)} \Big[
-\cos \pi \omega_{ab}+2i \left( f_{\omega_2} + \frac{ \delta_{13}}{2} \right)  \Big].
\ea
We remind that this result represents, to leading order, the combination (\ref{interest}). In order to obtain the  complete scattering amplitude we still have to  multiply by the BDS part, i.e.the product of the Regge pole factors, $s_{01}^{\omega_1} s_{12}^{\omega_2} s_{23}^{\omega_3} s_{34}^{\omega_4} s_{45}^{\omega_5}$, and by the couplings to the external particles.

The analogous results for the short cuts in the $\omega_3$ and $\omega_4$ are easily obtained by suitable changes of variables.

\subsection{Long cuts}

Turning to the long cuts in the $\omega_2\omega_3$ and $\omega_3\omega_4$ channels
we notice that the phase structure suggests, instead of the long cut amplitudes $W_{\omega_2\omega_3;L}$ etc., to define real-valued combinations of long and short cut 
contributions. We therefore introduce the modified long cut amplitudes:
\ba
\label{tildelongcutL}
\widetilde{W}_{\omega_2\omega_3;L}&=&\frac{W_{\omega_2\omega_3;L}}{\Omega_{3'2'}} +
\frac{\Omega_a}{\Omega_{2'}} W_{\omega_3}\\
\widetilde{W}_{\omega_2\omega_3;R}&=&\frac{W_{\omega_2\omega_3;R}}{\Omega_{2'3'}} +
W_{\omega_2}\frac{\Omega_c}{\Omega_{3'}} 
\ea  
and 
\ba
\label{tildelongcutR}
\widetilde{W}_{\omega_3\omega_4;L}&=&\frac{W_{\omega_3\omega_4;L}}{\Omega_{4'3'}} +
\frac{\Omega_b}{\Omega'_3} W_{\omega_{4'}}\\
\widetilde{W}_{\omega_3\omega_4;R}&=&\frac{W_{\omega_3\omega_4;R}}{\Omega_{3'4'}} +
W_{\omega_3}\frac{\Omega_d}{\Omega_{4'}} 
\ea  
with
\be
\Omega_{i'j'}=\sin \pi (\omega'_i-\omega'_j).
\ee
Moreover, for later purposes it will be convenient to introduce the following short hand notation:
\be 
 LC(23)=  e^{i\pi \omega_3}  \widetilde{W}_{\omega_2\omega_3;R}+
e^{i\pi \omega_2}   \widetilde{W}_{\omega_2\omega_3;L} .
\ee

With these notations we find the following cut contributions:
\ba
\label{cuts-14}
\tau_1\tau_4:&& 2i e^{-i\pi (\omega_2+\omega_3+\omega_5)} e^{i \pi \omega_d}
 \Big[e^{i\pi \omega_3} \int \widetilde{W}_{\omega_2\omega_3;R}+
e^{i\pi \omega_2} \int \widetilde{W}_{\omega_2\omega_3;L}\Big]
\\
\label{cuts-124}
\tau_1\tau_2\tau_4:&& 2i e^{-i\pi (\omega_3+\omega_5)}
 e^{i \pi \omega_d}\Big[e^{i\pi \omega_3} \int \widetilde{W}_{\omega_2\omega_3;R}+
e^{i\pi \omega_2} \int \widetilde{W}_{\omega_2\omega_3;L}\nonumber\\
&&\hspace{5cm}-e^{i\pi\omega_a}\int W_{\omega_3} 
\Big]\\
\label{cuts-134}
\tau_1\tau_3\tau_4:&&2i e^{-i\pi (\omega_2+\omega_5)}
 e^{i \pi \omega_d}\Big[e^{i\pi \omega_3} \int \widetilde{W}_{\omega_2\omega_3;R}+
e^{i\pi \omega_2} \int \widetilde{W}_{\omega_2\omega_3;L}\nonumber\\  
&&\hspace{5cm}- \int W_{\omega_2}e^{i\pi\omega_c}\Big]\\
\label{cuts-1234}
\tau_1\tau_2\tau_3\tau_4:&& 2i e^{-i\pi \omega_5}
 e^{i \pi \omega_d}\Big[e^{i\pi \omega_3} \int \widetilde{W}_{\omega_2\omega_3;R}+
e^{i\pi \omega_2} \int \widetilde{W}_{\omega_2\omega_3;L}\nonumber\\
&&\hspace{2cm} -\int W_{\omega_2}e^{-i\pi\omega_c} 
-e^{-i\pi\omega_a} \int W_{\omega_3}\Big] \,.
\ea
The corresponding pole contributions are listed in Appendix C.

In (\ref{cuts-14}) - (\ref{cuts-1234}) we have used a short hand notation which we have to explain. In order 
to obtain, for example, (\ref{cuts-14}) we proceed as follows. We start from the integral of the form (\ref{struct-LLLL}) which we write as 
\ba
T_{LLLL} &=& s \int ...\int \frac{d \omega'_1d \omega'_2 d \omega'_3d \omega'_4 d\omega'_5} {(2\pi i)^5} s_{45}^{\omega'_{54}}s_{35}^{\omega'_{43}}s_{25}^{\omega'_{32}}s_{15}^{\omega'_{21}}s^{\omega'_1}
\nonumber\\
&&\cdot e^{-i\pi(\omega'_5 -\omega'_4) }
 F_{LLL}(t_1,t_2,t_3,t_4,t_5;\omega'_1,\omega'_2,\omega'_3,\omega'_4,\omega'_5) ,
\label{struct-LLLL-mod}
\ea
where $F_{LLL}$ is a sum of Regge pole and Regge cut terms
As said before, each Regge cut term comes with angular momentum integrals $\omega'_i$ for all t-channels over which the cut extends. For the long cut discussed now the $t_2$ and $t_3$ channels are involved, i.e. the 
we have the integration
\be
\int \frac{d \omega'_2 d\omega'_3}{(2 \pi i)^2} s_{12}^{\omega'_2}  s_{23}^{\omega'_3} 
\ee
All other $\omega'_i$ variables can be substituted by the corresponding Regge pole  $\omega_i$
Also the phases are written as primed, i.e. for  the first term in (\ref{cuts-14}) we have
\be
\int \frac{d \omega'_2 d\omega'_3}{(2 \pi i)^2} s_{12}^{\omega'_2}  s_{23}^{\omega'_3}  e^{-i\pi (\omega'_2+\omega_5)} e^{i \pi \omega_d}
  \widetilde{W}_{\omega_2\omega_3;R}.
\ee
After the shift $\omega'_i=\omega_i +\omega''_i$ this becomes 
\ba
\label{example}
&&s_{12}^{\omega_2}  s_{23}^{\omega_3} e^{-i\pi (\omega_2+\omega_3+\omega_5)}e^{i \pi \omega_d}  
e^{i\pi \omega_3}
\int_2 \int_3  s_{12}^{\omega''_2}   s_{23}^{\omega''_3}  e^{-i\pi \omega''_2} \widetilde{W}_{\omega_2\omega_3;R} 
\nonumber\\
&& = s_{12}^{\omega_2}  s_{23}^{\omega_3} e^{-i\pi (\omega_2+\omega_3+\omega_5)} e^{i \pi \omega_d}
 e^{i\pi \omega_3}
\int_2 \int_3  (-s_{12})^{\omega''_2-\omega''_3}  (-s_{123})^{\omega''_3}  \kappa_{2''3''}^{\omega''_3}
\widetilde{W}_{\omega_2\omega_3;R}
\ea
The energy factors in front of the integral belong to the BDS part and will be disregarded, and 
we are thus left with 
\ba
&&e^{-i\pi (\omega_2+\omega_3+\omega_5)} e^{i \pi \omega_d}
 e^{i\pi \omega_3}
\int_2 \int_3  e^{-i\pi \omega''_2} s_{12}^{\omega''_2}  s_{23}^{\omega''_3} 
\widetilde{W}_{\omega_2\omega_3;R}\nonumber\\
&&= e^{-i\pi (\omega_2+\omega_3+\omega_5)} e^{i \pi \omega_d}
 e^{i\pi \omega_3}
\int  \widetilde{W}_{\omega_2\omega_3;R}
\ea
This should explain our short hand notation '$\int$'. Throughout the main part of this paper we will make use
of this notation. 

The determination of the singular pieces has been described  in Appendix B and will not be repeated here. As the main result, we have found real-valued terms $\delta \widetilde{W}_{\omega_2 \omega_3;R}$,  $\delta \widetilde{W}_{\omega_2 \omega_3;L}$ which remove the singularities in all kinematic regions. Adding Regge pole and cut contributions and inserting the subtractions
\ba
\int \widetilde{W}_{\omega_2\omega_3;L}&=& \int\widetilde{W}_{\omega_2\omega_3;L}^{reg}+\delta\widetilde{W}_{\omega_2\omega_3;L} \\
\int \widetilde{W}_{\omega_2\omega_3;R}&=& \int \widetilde{W}_{\omega_2\omega_3;R}^{reg}+\delta\widetilde{W}_{\omega_2\omega_3;R}
\ea
we arrive, for the sum of Regge pole and Regge cuts, at the finite expressions:
\ba   
\label{allorder-longcut14} 
\tau_1\tau_4:&& e^{-i\pi (\omega_2+\omega_3+\omega_5)} e^{i \pi \omega_d}\Big[
e^{i\pi \omega_b} cos \pi \omega_{ac}
\nonumber\\
&&\hspace{1cm}+2i \left( e^{i\pi\omega_2 } \int \widetilde{W}_{\omega_2\omega_3;L}^{reg}+e^{i\pi\omega_3}    \int \widetilde{W}_{\omega_2\omega_3;R}^{reg} \right)\Big]
\\
\label{allorder-longcut124}
\tau_1\tau_2\tau_4:&&e^{-i\pi (\omega_3+\omega_5)}
 e^{i \pi \omega_d}\Big[ - e^{i \pi \omega_c}cos \pi\omega_{ba} 
\nonumber\\
&&\hspace{0.1cm}+2i  \left(e^{i\pi\omega_2 }\int  \widetilde{W}_{\omega_2\omega_3;L}^{reg}+e^{i \pi \omega_3}    \int \widetilde{W}_{\omega_2\omega_3;R}^{reg} 
-e^{i \pi \omega_a} \int W_{\omega_3}^{reg} \right)\Big] \\
\label{allorder-longcut134}
\tau_1\tau_3\tau_4:&&e^{-i\pi (\omega_2+\omega_5)}
 e^{i \pi \omega_d}\Big[  -  e^{i\pi \omega_a} \cos \pi \omega_{bc}
\nonumber\\
&&\hspace{0.1cm}+2i   \left( e^{i \pi \omega_2}\int \widetilde{W}_{\omega_2\omega_3 ;L}^{reg}+e^{i \pi\omega_3}    \int \widetilde{W}_{\omega_2\omega_3;R}^{reg} 
-  e^{i \pi \omega_c}\int W_{\omega_2}^{reg} \right) \Big]\\
 \label{allorder-longcut1234}
\tau_1\tau_2\tau_3\tau_4:&& e^{-i\pi \omega_5}
 e^{i \pi \omega_d}\Big[
e^{i\pi \omega_{ba} }
e^{i\pi\omega_{bc} } \nonumber \\
&& +2i \left(e^{i\pi\omega_2 }\int \widetilde{W}_{\omega_2\omega_3;L}^{reg}+e^{i\pi\omega_3}    \int \widetilde{W}_{\omega_2\omega_3;R}^{reg} 
-e^{-i\pi \omega_a} \int W_{\omega_3}^{reg}
- e^{-i\pi \omega_c}\int W_{\omega_2}^{reg}\right)
\Big].\nonumber\\
\ea

In the following we will use energy discontinuities and unitarity to compute the Regge cut amplitudes. Since the evaluation of the unitarity integrals will be done only in the   
leading logarithmic approximation, we will obtain only the leading order of (\ref{allorder-longcut14}) -  (\ref{allorder-longcut1234}). This means that  inside the square brackets we neglect all phase factors, and as a result only the sum  $\widetilde{W}_{\omega_2\omega_3;L}^{reg}+  \widetilde{W}_{\omega_2\omega_3;R}^{reg}$ appears. However, in view of the new Regge cuts to be discussed in the following section we have  to keep in mind that, when exanding the phase factors in powers of $(i \pi)$ , e.g. in (\ref{allorder-longcut14}):
\ba
\label{long-cut-exp}
&&2i \left( e^{-i\pi\omega_3 } \widetilde{W}_{\omega_2\omega_3;L}^{reg}+e^{-i\pi\omega_2}    \widetilde{W}_{\omega_2\omega_3;R}^{reg} \right)\nonumber\\
&&=2i \left( \widetilde{W}_{\omega_2\omega_3;L}^{reg}+  \widetilde{W}_{\omega_2\omega_3;R}^{reg}- i 
 \left(\pi \omega_3 \widetilde{W}_{\omega_2\omega_3;L}^{reg} + \pi \omega_2 \widetilde{W}_{\omega_2\omega_3;R}^{reg}\right)\right)
\ea
the second part, which in contrast to the leading part is of the order $(i\pi)^2$ , is of the same order as the new Regge cuts. Therefore, for consistency, we cannot simply ignore these 
next-to-leading order terms. We will come back to them at the end of this section. 
   
Let us now determine the sum of the Regge cut amplitudes  $\widetilde{W}_{\omega_2\omega_3;L}^{reg}+  \widetilde{W}_{\omega_2\omega_3;R}^{reg}$  by computing  the energy discontinuity of the full  amplitude (Regge pole plus cut) in the region where all energies are positive:
\ba
\Delta_{123} &=& e^{-i\pi (\omega_1+\omega_4+\omega_5)} e^{i \pi \omega_d} 
 \Big[ e^{i\pi \omega_{23}}  e^{i\pi\omega_2} \int \widetilde{W}_{\omega_2\omega_3;L}
+e^{i\pi \omega_{32}}  e^{i\pi\omega_3} \int \widetilde{W}_{\omega_2\omega_3;R}\nonumber\\
&&-e^{i\pi\omega_{23}} e^{i\pi\omega_a} \int W_{\omega_3}
-e^{i\pi\omega_{32}} \int W_{\omega_2} e^{i\pi\omega_c}\nonumber\\
&&-V_L(a) \left( e^{i\pi\omega_{23}}  \frac{V_L(b)}{\Omega_3 \Omega_{32}}+   e^{i\pi\omega_{32}} \frac{V_R(b)}{\Omega_2 \Omega_{23}} \right) V_R(c)
\Big].
\ea
Inserting, for the singular pieces, the results obtained in Appendix B all singular pieces cancel and we are left with:
\ba
\label{d_123}
\Delta_{123} &=&
e^{-i\pi (\omega_1+\omega_4+\omega_5)} e^{i \pi \omega_d} 
 \Big[ e^{i\pi \omega_{23}}  e^{i\pi\omega_2} \int \widetilde{W}_{\omega_2\omega_3;L}^{reg}
+e^{i\pi \omega_{32}}  e^{i\pi\omega_3} \int \widetilde{W}_{\omega_2\omega_3;R}^{reg}
\nonumber\\
&&-e^{i\pi\omega_{23}} e^{i\pi\omega_a} \int W_{\omega_3}^{reg}
-e^{i\pi\omega_{32}}e^{i\pi\omega_c} \int W_{\omega_2}^{reg}\nonumber\\
&&+i\left( e^{i\pi \omega_a} V_R(b) V_R(c) +V_L(a) V_L(b) e^{i \pi \omega_c} \right)\Big].
\ea

As before, the discontinuity $\Delta_{123}$ will be  computed from a unitarity integral (Fig.\ref{fig:cut-amplitudes}.b)  for which we restrict ourselves to the  weak coupling limit. Comparing the general form of the unitarity integral illustrated in Fig.\ref{fig:single-disc}b with the leading approximation in Fig.\ref{fig:cut-amplitudes}.b) we notice that the produced  particle in the center now only couples to the  upper reggeon.
We find:
\ba
\label{disc123-LL}
\Delta_{123}&=&  f_{\omega_2\omega_3}   +\frac{\delta_{14}}{2}-f_{\omega_2}  - \frac{ \delta_{13}}{2} -f_{\omega_3}  - \frac{ \delta_{24}}{2},
 \ea 
where
\be
\delta_{14}=\pi \left(V_{14}+ \omega_a+\omega_c\right)\,.
\ee
As for the short cut, we have separated the one loop terms, and $f_{\omega_2\omega_3}$ which is illustrated in Fig.\ref{fig:cut-amplitudes}b  starts from two loops. We have the integral representation:
\ba
\label{f-long}
f_{\omega_2\omega_3}&= &\frac{a}{2} \sum_{n_1,n_2} (-1)^{n_1+n_2}\int \frac{d\nu_1 d\nu_2}{(2\pi)^2}
 \frac{1}{i\nu_1+\frac{n_1}{2}} \left( \frac{k_a^*q_3^*}{q_1^* k_b^*} \right)^{i\nu_1+\frac{n_1}{2}}
\left( \frac{k_a q_3}{q_1 k_b} \right)^{i\nu_1-\frac{n_1}{2}} 
\left( \frac{s_{12}}{s_{02}} \right)^{\omega(\nu_1,n_1)}\nonumber\\
&&\hspace{-0.5cm}\cdot B(\nu_1,\nu_2,n_1.n_2) \left( \frac{s_{23}}{s_{03}} \right)^{\omega(\nu_2,n_2 )} 
\left( \frac{k_b^*q_4^*}{q_2^* k_c^*} \right)^{i\nu_2+\frac{n_2}{2}}
\left( \frac{k_b q_4}{q_2 k_c} \right)^{i\nu_2-\frac{n_2}{2}}  \frac{1}{i\nu_2- \frac{n_2}{2}}|_{\text{sub}}.
\ea
Here $B(\nu_1,\nu_2,n_1,n_2)$ denotes the production vertex of particle b. The subscript $|_{\text{sub}}$ indicates that we have subtracted the divergent one loop contribution $\frac{\pi}{2}V_{14}$. 

Returning to (\ref{d_123}) we take the weak coupling limit:
\be
\Delta_{123} =
\int \widetilde{W}_{\omega_2\omega_3;L}^{reg}+\int \widetilde{W}_{\omega_2\omega_3;R}^{reg}
-\int W_{\omega_3}^{reg}
- \int W_{\omega_2}^{reg}
\ee
and combine with (\ref{disc123-LL}). We find for the sum of $\int \tilde{W}_{\omega_2\omega_3;L}^{reg}+\int \tilde{W}_{\omega_2\omega_3;R}^{reg}$:
\ba
\label{sumWtilde}
&&\int \tilde{W}_{\omega_2\omega_3;L}^{reg}+\int \tilde{W}_{\omega_2\omega_3;R}^{reg}\nonumber\\
&&=\int_2\int_3 s_{12}^{\omega''_2} s_{23}^{\omega''_3} \left(\tilde{W}_{\omega_2\omega_3;L}^{reg}+\tilde{W}_{\omega_2\omega_3;L}^{reg}\right)\nonumber\\
&&=f_{\omega_2\omega_3}+\frac{ \delta_{14}}{2}.
\ea  

When inserting these weak coupling results into  (\ref{allorder-longcut14})-(\ref{allorder-longcut1234}) we  encounter the phases $\delta_{14}$ etc contained in the BDS formulae.
Here it is useful to note the identities: 
\ba
\delta_{14}-\delta_{24}&=&-\delta_{124}  \nonumber\\
\delta_{14}-\delta_{13}&=&-\delta_{134}  \nonumber\\
\delta_{14}-\delta_{13}-\delta_{24}&=&\delta_{1234}-\pi (2\omega_b-\omega_a-\omega_c).
\ea
For (\ref{allorder-longcut14})-(\ref{allorder-longcut1234}) we thus arrive at :
\ba
\label{LLlongcut-14}
\tau_1\tau_4:&\hspace{0.5cm}&  e^{-i\pi (\omega_2+\omega_3+\omega_5)} e^{i \pi \omega_d}\Big[  e^{i \pi \omega_b} \cos \pi \omega_{ac} +2i
\left( f_{\omega_2\omega_3}+\frac{ \delta_{14}}{2}\right)\Big] \\
\label{LLlongcut-124}
\tau_1\tau_2\tau_4:&\hspace{0.5cm}& e^{-i\pi (\omega_3+\omega_5)}
 e^{i \pi \omega_d}
\Big[- e^{i \pi \omega_c}\cos \pi \omega_{ba}  +2i \left( f_{\omega_2\omega_3}
- f_{\omega_3} - \frac{ \delta_{124}}{2}\right)\Big] \\
\label{LLlongcut-134}
\tau_1\tau_3\tau_4:&\hspace{0.5cm}& e^{-i\pi (\omega_2+\omega_5)}
 e^{i \pi \omega_d}\Big[ -e^{i \pi \omega_a}\cos \pi \omega_{bc}+2i\left(f_{\omega_2\omega_3}- f_{\omega_2} - \frac{ \delta_{134}}{2}\right)\Big]\\
\label{LLlongcut-1234}
\tau_1\tau_2\tau_3\tau_4:&\hspace{0.5cm}&  e^{-i\pi \omega_5} e^{i \pi \omega_d} 
\Big[( e^{i\pi \omega_{ba} }e^{i\pi\omega_{bc} }+ \nonumber\\
&&\hspace{1cm}+2i \left( f_{\omega_2\omega_3}-f_{\omega_2} 
- f_{\omega_3} + \frac{ \delta_{1234}}{2}-\frac{\pi}{2}(2\omega_b-\omega_a-\omega_c)\right) \Big]\,.
\ea 
We note that in the last line the term $-\frac{\pi}{2}(2\omega_b-\omega_a-\omega_c)$ 
which has its origin in the combination of phases, $\delta_{14}-\delta_{13}-\delta_{24}$,
cancels the one loop contribution of the conformal pole, $e^{i\pi \omega_{ba} }
e^{i\pi\omega_{bc} }$.

We finish this discussion of the long cut by returning to the expansion (\ref{long-cut-exp}). Starting from the single discontinuity $\Delta_{123}$ and computing the unitarity integral in the leading logarithmic approximatiion we found the sum $W_{\omega_2\omega_3;L} +W_{\omega_2\omega_3;R}$, but not separate expressions for  $W_{\omega_2\omega_3;L}$ and $W_{\omega_3\omega_4;R}$. For the leading approximation this is sufficient, however, we will see
in the next section that the new pieces - the double cut and the three reggeon cut -  are proportional to $(i\pi )^2$ and can be obtained only from double energy discontinuities.
The same is true for the second term on the rhs of (\ref{long-cut-exp}): in order to determine  $W_{\omega_2\omega_3;L}$ and $W_{\omega_3\omega_4;R}$ separately, we need 
the double discontinuities $\Delta_{23}\Delta_{123}$ and $\Delta_{12}\Delta_{123}$. Here we need to go back to our starting variables $\omega'_2$ and $\omega'_3$:
\be 
\label{d12d_123}
\Delta_{12} \Delta_{123} 
 = e^{-i\pi (\omega_1+\omega_4+\omega_5)} e^{i\pi \omega_d} 
\Big[ -\int e^{i\pi \omega'_3} W_{\omega_2\omega_3;R} +\int  \frac{\Omega_{2'3'}}{\Omega'_3} W_{\omega_2} V_R(c) + \frac{ V_L(a)V_R(b) V_R(c)}{\Omega_2}\Big]
\ee
and 
\be 
\label{d23d_123}
\Delta_{23} \Delta_{123}
 = e^{-i\pi (\omega_1+\omega_4+\omega_5)} e^{i\pi \omega_d} 
\Big[ -\int e^{i\pi \omega'_2} W_{\omega_2\omega_3;L} + \int \frac{\Omega_{3'2'}}{\Omega_2'} V_L(a)W_{\omega_3}  + \frac{ V_L(a)V_L(b) V_R(c)}{\Omega_3} \Big]
\ee
or
\ba
\label{d12_d123_ll}
\Delta_{12} \Delta_{123} 
 &=& e^{-i\pi (\omega_1+\omega_4+\omega_5)} e^{i\pi \omega_d} \Big[ \left( -\int \Omega_{2'3'}\widetilde{W}_{\omega_2\omega_3;R}^{reg} +\int \Omega_{2'3}W_{\omega_2}^{reg}e^{i\pi \omega_c} \right) \nonumber\\
&&\hspace{2cm}+\frac{1}{2} \left(e^{i\pi \omega_a} V_R(b) V_R(c) + V_L(a) V_L(b) e^{i\pi \omega_c} \right)\Big]
\ea
and 
\ba
\label{d23_d123_ll}
\Delta_{23} \Delta_{123} 
 &=& e^{-i\pi (\omega_1+\omega_4+\omega_5)} e^{i\pi \omega_d} \Big[ \left( -\int \Omega_{3'2'} \widetilde{W}_{\omega_2\omega_3;L}^{reg}+ e^{i\pi \omega_a}\int \Omega_{3'2}W_{\omega_3}^{reg}  \right) \nonumber\\ 
&&\hspace{2cm}+\frac{1}{2} \left(e^{i\pi \omega_a} V_R(b) V_R(c) + V_L(a) V_L(b) e^{i\pi \omega_c} \right)\Big]\,.
\ea
An analysis of these equation has been performed in \cite{Bartels:2019} and will not be repeated here. We only quote a few results.
First the leading order unitarity integrals for the double discontinuities $\Delta_{12}\Delta_{123}$ and $\Delta_{23}\Delta_{123}$:
\ba
\label{d12d123}
&&\Delta_{12} \Delta_{123}  = \int \frac{d \omega'_2}{2 \pi i} \int \frac{d \omega'_3}{2 \pi i} s_{12}^{\omega'_2} s_{23}^{\omega'_3} \nonumber\\
&&\cdot
V^{a} G_2 \Big[ (-\pi \omega'_2) V_0^{b} \bq_2-\bk,\bk; \bq_3-\bk',\bk') +V^{(b)} (\bq_2-\bk,\bk; \bq_3-\bk',\bk') \Big] G_2 V^c
\ea
and
\ba
\label{d23d123}
&&\Delta_{23} \Delta_{123}  =\int \frac{d \omega'_2}{2 \pi i} \int \frac{d \omega'_3}{2 \pi i} s_{12}^{\omega'_2} s_{23}^{\omega'_3} \nonumber\\
&&\cdot
V^{a} G_2 \Big[ (-\pi \omega'_3) V_0^{b} \bq_2-\bk,\bk; \bq_3-\bk',\bk') +V^{b} \bq_2-\bk,\bk; \bq_3-\bk',\bk') \Big] G_2 V^c.
\ea
Here 
\be
 V_0^{b}( \bq_2-\bk,\bk; \bq_3-\bk',\bk')= (2\pi)^2 \delta^{(2)}(\bk-\bk') \bk^2\,,
\ee
and $V^{(b)} (\bq_2-\bk,\bk; \bq_3-\bk',\bk')$ is illustrated below in Fig.\ref{fig:prod-vertex}b.
We illustrate the equation for $\Delta_{23} \Delta_{123}$ in Fig.\ref{fig:prod-vertex}:
\begin{figure}[H]
\centering
\epsfig{file=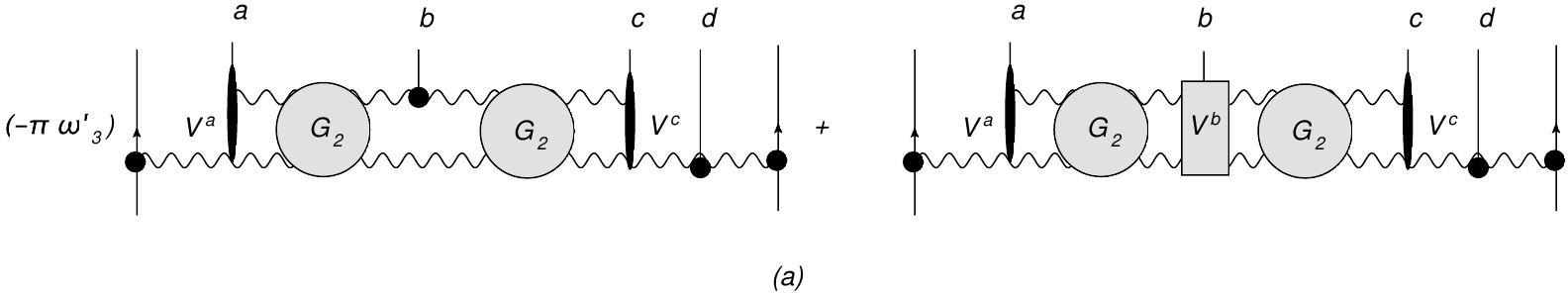,height=2cm,width=14cm}\\
\epsfig{file=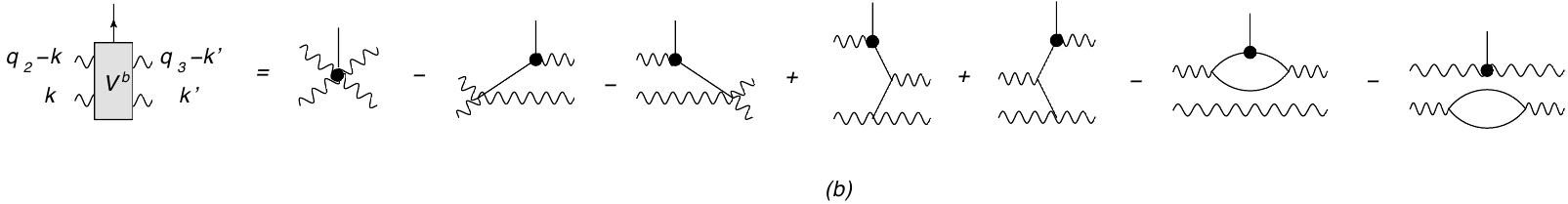,height=2cm,width=14cm}\\
\caption{(a) Illustration of the doube discontinuity  $\Delta_{23} \Delta_{123}$ and (b) of the new vertex $V^b$}
\label{fig:prod-vertex}
\end{figure}
\noindent
Making use of the bootstraop equation the first line can also be written as
\begin{figure}[H]
\centering
\epsfig{file=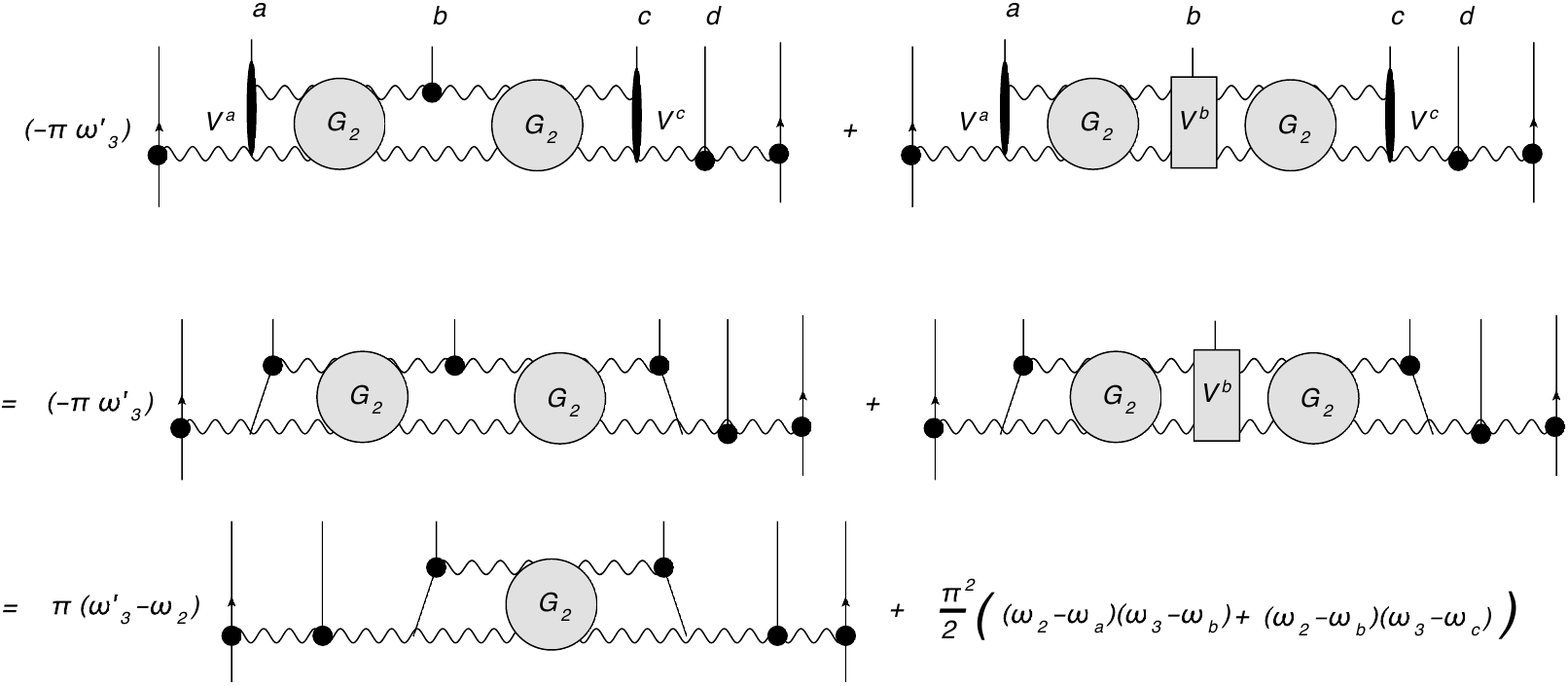,height=6cm,width=14cm}\\
\caption{Decomposition of the double discontinuity $\Delta_{23}\Delta_{123}$}
\label{fig:d23d123-ll}
\end{figure}
\noindent

\begin{figure}[H]
\centering
\epsfig{file=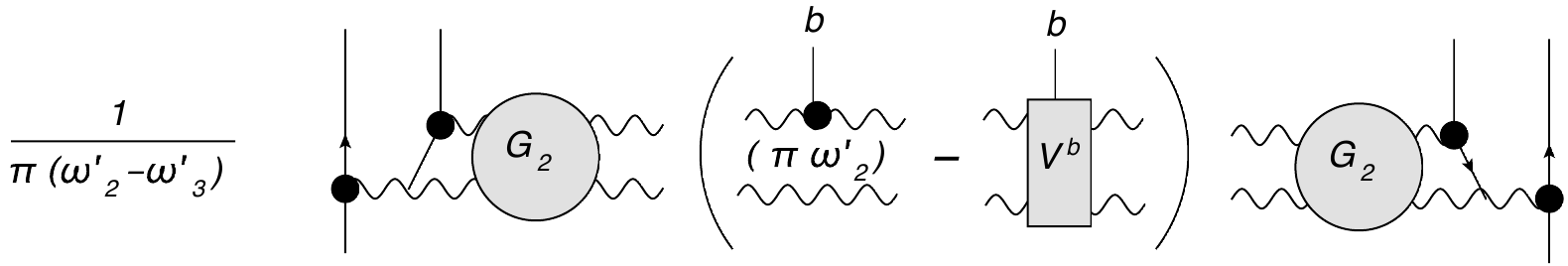,height=2cm,width=14cm}\
\caption{Graphical illustration of the two terms  of $\widetilde{W}_{23;R}$}.
\label{fig:W23R}
\end{figure}

As described in \cite{Bartels:2019}, these calculations lead to the following leading-logarithmic results for  $W_{\omega_2\omega_3;L}$ and $W_{\omega_3\omega_4;R}$:
\ba
\label{W23R-final}
\int \frac{d \omega'_2 d \omega'_3}{(2 \pi i)^2}  s_{12}^{\omega'_2}  s_{23}^{\omega'_3}     \widetilde{W}_{\omega_2\omega_3;\mathrm{R}}^{\mathrm{reg}}=
\int  \frac{d \omega'_2 d \omega'_3}{(2 \pi i)^2}s_{12}^{\omega'_2}  s_{23}^{\omega'_3}    
 \frac{\pi \omega'_2 f_{\omega_2 \omega_3}^{int} -f^{(b)}_{\omega_2\omega_3} }{\pi (\omega'_2-\omega'_3)} + {\text{two loop part}}
\ea
and
\ba
\label{W23L-final}
 \int \frac{d \omega'_2 d \omega'_3}{(2 \pi i)^2}  s_{12}^{\omega'_2}  s_{23}^{\omega'_3}     \widetilde{W}_{\omega_2\omega_3;\mathrm{L}}^{\mathrm{reg}}=
\int \frac{d \omega'_2 d \omega'_3}{(2 \pi i)^2}  s_{12}^{\omega'_2}  s_{23}^{\omega'_3}    
  \frac{\pi \omega'_3 f_{\omega_2 \omega_3}^{int} -f^{(b)}_{\omega_2\omega_3} }{\pi (\omega'_3-\omega'_2)}+ {\text{two loop part}},
\ea
where $f_{\omega_2 \omega_3}^{int}$ denotes the integrand of $f_{\omega_2\omega_3}$ in (\ref{f-long}), i.e. before the $\omega''$-integrals.
Here we have, for simplicity, disregarded the one and two loop results, and we have not yet, by shifting the $\omega'_2$ integration,   
extracted the Regge pole factors of the $t_2$ and $t_3$ channels. 
Explicit expression for the 'two loop'  terms can be found in \cite{Bartels:2019}.
In Fig.\ref{fig:W23R} we illustrate $ \widetilde{W}_{\omega_2\omega_3;\mathrm{L}}^{\mathrm{reg}}$.
\noindent
It is important to stress that in (\ref{W23R-final}) and  (\ref{W23L-final}) both terms are of the same order in $g^2$.
In the leading term of the scattering amplitude, only the sum  $W_{\omega_2\omega_3;L} +W_{\omega_2\omega_3;R}$, 
(\ref{sumWtilde}), appears which equals the sum of the two first terms only. Only if we expand the phases contained in the energy factors also the second terms show up:
\ba
&& 2i e^{-i\pi\omega_5} e^{i\pi \omega_d}    \int \frac{d \omega'_2 d \omega'_3}{(2 \pi i)^2}  s_{12}^{\omega'_2}  s_{23}^{\omega'_3}    
s_{12}^{\omega'_2}   s_{ 23}^{\omega'_3} \left( e^{-i\pi \omega'_2} 
\widetilde{W}_{\omega_2\omega_3;\mathrm{R}}^{\mathrm{reg}} +
 e^{-i\pi \omega'_3} 
\widetilde{W}_{\omega_2\omega_3;\mathrm{L}}^{\mathrm{reg}} \right) \\
  &\approx &2 ie^{-i\pi \omega_5} e^{i\pi \omega_d} \int \frac{d \omega'_2 d \omega'_3}{(2 \pi i)^2}  s_{12}^{\omega'_2}  s_{23}^{\omega'_3} 
 \left( \widetilde{W}_{\omega_2\omega_3;R}^{reg} +\widetilde{W}_{\omega_2\omega_3;L}^{reg} -i\pi \omega'_2  \widetilde{W}_{\omega_2\omega_3;R}^{reg}  -i\pi \omega'_3  \widetilde{W}_{\omega_2\omega_3;L}^{reg} \right) \,.
\nonumber
\ea
Inserting (\ref{W23L-final}) and (\ref{W23L-final}), performing the shift $\omega'_i = \omega''_i$ and dropping 
the Regge pole factors $s_{12}^{\omega'_2}  s_{23}^{\omega'_3}$ we arrive at 
\ba
&\approx&  2i e^{-i\pi (\omega_2+\omega_3+\omega_5)} e^{i\pi \omega_d} \int_2 \int_3  s_{12}^{\omega''_2}   s_{23}^{\omega''_3}  \left( f_{\omega_2 \omega_3}^{int} +i f_{\omega_2 \omega_3}^{(b)}\right)\,.
\ea
The '$\approx$' sign indicates that we have expanded the phase factors, and our equations are valid only up to the first order in $i\pi$. 
From this we see that, although the second terms on the rhs of   (\ref{W23R-final}) and  (\ref{W23L-final}) are of the same order as the first ones, i.e.they still belong to the leading approximation, in the scattering amplitude they appear proportional to $\sim i\pi$ as 'next-to-leading-order'.

Similarly, for the other kinematic regions we find: 
\ba
\tau_1\tau_4:&\hspace{0.5cm}& 2i e^{-i\pi (\omega_2+\omega_3+\omega_5)} e^{i\pi \omega_d} \int_2 \int_3  s_{12}^{\omega''_2}   s_{23}^{\omega''_3}  \left( f_{\omega_2 \omega_3}^{int} +i f_{\omega_2 \omega_3}^{(b)}\right)\\
\tau_1\tau_2\tau_4:&\hspace{0.5cm}&  2i e^{-i\pi (\omega_3+\omega_5)} e^{i\pi \omega_d} \int_2 \int_3  s_{12}^{\omega''_2}   s_{23}^{\omega''_3}  \left( f_{\omega_2 \omega_3}^{int}  +i f_{\omega_2 \omega_3}^{(b)}\right)\\
\tau_1\tau_3\tau_4:&\hspace{0.5cm}&  2i e^{-i\pi (\omega_2+\omega_5)} e^{i\pi \omega_d} \int_2 \int_3  s_{12}^{\omega''_2}   s_{23}^{\omega''_3}  \left( f_{\omega_2 \omega_3}^{int}  +i f_{\omega_2 \omega_3}^{(b)}\right)\\
\tau_1\tau_2\tau_3\tau_4:&\hspace{0.5cm}&  2i e^{-i\pi \omega_5} e^{i\pi \omega_d} \int_2 \int_3  s_{12}^{\omega''_2}   s_{23}^{\omega''_3}  \left( f_{\omega_2 \omega_3}^{int}  -i f_{\omega_2 \omega_3}^{(b)}\right)\,.
\ea
These equations show that, when energy phases are taken into account,  our sum of the two-real valued Regge cut terms can also be written in a factorized form with a complex-valued production vertex. In \cite{Bartels:2019} these results  have been used to compute the real corrections to the long cut.

\section{The very long cut, the double cut, and the 3-reggeon cut}  

We now turn to the two novel Regge cut contributions, the double cut and the 3-reggeon cut, which for the 
first time appear in the $2 \to 6$ production amplitude. As we will see, they cannot be separated from the very long cut.

\subsection{The all order amplitudes in different regions}

We now turn to the very long cut  which as we shall see cannot be separated from 
the double cut in $\omega_2\omega_4$, and to the three reggeon cut $W_{\text{3 reggeon cut}}$.  
Similar to the previous Regge cut terms,  the phase structure of our results suggests to define combinations of ghe very long cut, the long cut, and the  short cut contributions. We list them below as functions of  the $\omega'_i$:
\ba
\widetilde{W}_{LL} &=& \frac{W_{LL}}{\Omega_{3'2'} \Omega_{4'3'}} +\frac{\Omega_a}{\Omega_{2'}}  \frac{W_{\omega_3\omega_4;L}}{\Omega_{4'3'}} + \frac{\Omega_a\Omega_b}{\Omega_{2'}\Omega_{3'}}
W_{\omega_4}
\\
\widetilde{W}_{RR} &=& \frac{W_{RR}}{\Omega_{2'3'} \Omega_{3'4'}} +  \frac{W_{\omega_2\omega_3;R}}{\Omega_{2'3'}}\frac{\Omega_d}{\Omega_{4'}}+W_{\omega_2}\frac{\Omega_c\Omega_d}{\Omega_{3'}\Omega_{4'}}
\\
\widetilde{W}_{LR}&=&\frac{W_{LR}}{\Omega_{3'2'} \Omega_{3'4'}} + \frac{W_{\omega_2\omega_3;L}}{\Omega_{3'2'}} \frac{\Omega_d}{\Omega_{4'}}
+\frac{\Omega_a}{\Omega_{2'}} \frac{W_{\omega_3\omega_4;R}}{\Omega_{3'4'}} 
+\frac{\Omega_a}{\Omega_{2'}} W_{\omega_{3'}} \frac{\Omega_d}{\Omega_{4'}}
\\
\widetilde{W}_{RL}&=&\frac{W_{RL}}{\Omega_{2'3'} \Omega_{4'3'}} \,.
\ea
Looking at the energy cut structure we expect that the $W_{RL}$ mixes with the double cut, $W_{\omega_2 \omega_4}$, and 
$W_{LR}$ mixes with the triple reggeon cut, $W_{\text{3 reggeon}}$.  

Furthermore we observe that in all kinematic regions the four partial waves $\widetilde{W}_{LL}$,$\widetilde{W}_{RL}$, $\widetilde{W}_{LR}$, and $\widetilde{W}_{RR}$ come in particular 
linear combinations; the same applies to the pairs $\widetilde{W}_{\omega_2\omega_3;L}$,
$\widetilde{W}_{\omega_2\omega_3;R}$, and $\widetilde{W}_{\omega_3\omega_4;L}$,
$\widetilde{W}_{\omega_3\omega_4;R}$. We therefore define:
\ba
\label{VLC++}
&&\int VLC^{++}=\\
&& e^{i\pi(\omega_2+\omega_3)} \int\widetilde{W}_{LL} +e^{2i\pi \omega_3} \int \widetilde{W}_{RL}+e^{i\pi(\omega_2+\omega_4)}\int  \widetilde{W}_{LR}+
e^{i\pi(\omega_3+\omega_4)}\int \widetilde{W}_{RR},\nonumber
\\
\label{VLC--}
&&\int VLC^{--}=\\ 
&&e^{-i\pi(\omega_2+\omega_3)}\int \widetilde{W}_{LL} +e^{-2i\pi \omega_3}\int \widetilde{W}_{RL}+e^{-i\pi(\omega_2+\omega_4)}\int \widetilde{W}_{LR}+
e^{-i\pi(\omega_3+\omega_4)}\int \widetilde{W}_{RR}, \nonumber\\
\label{VLC+-}
&&\int VLC^{+-}=\\
&& e^{i\pi(\omega_2-\omega_3)}\int \widetilde{W}_{LL}  +\int \widetilde{W}_{RL}+e^{i\pi(\omega_2-\omega_4)} \int \widetilde{W}_{LR}+
e^{i\pi(\omega_3-\omega_4)}\int \widetilde{W}_{RR},\nonumber\\
\label{VLC-+}
&&\int VLC^{-+}=\\
&& e^{i\pi(-\omega_2+\omega_3)}\int \widetilde{W}_{LL} + \int \widetilde{W}_{RL}+e^{i\pi(-\omega_2+\omega_4)}\int \widetilde{W}_{LR}+
e^{i\pi(-\omega_3+\omega_4)}\int \widetilde{W}_{RR}\nonumber
\ea
and 
\ba 
\int LC(23)&=&  
e^{i\pi \omega_2}\int \widetilde{W}_{\omega_2\omega_3;L} +e^{i\pi \omega_3} \int\widetilde{W}_{\omega_2\omega_3;R}\\
\int LC(34)&=&  e^{i\pi \omega_3}\int \widetilde{W}_{\omega_3\omega_4;L}+
e^{i\pi \omega_4}\int \widetilde{W}_{\omega_3\omega_4;R}.  
\ea
Whereas all the partial waves $W$ and $\widetilde{W}$ are real-valued,
these  new combinations $VLC$ etc contain phases and are thus are complex-valued. 

With these notations we find the following expressions for the cut contributions in the different regions:  
\ba
\label{amp15}
\tau_1\tau_5:&& 2ie^{-i\pi(\omega_2+\omega_3+\omega_4)} 
\Big[\int VLC^{++} - e^{i\pi\omega_3}\int \frac{W_{\omega_2\omega_4}}{\Omega_{3'}}\Big]
 \\
\label{amp125}
\tau_1\tau_2\tau_5:&& 2ie^{-i\pi(\omega_3+\omega_4)} \Big[\int VLC^{++}  -e^{i\pi \omega_a}\int LC(34)
-e^{i\pi\omega_3}\int \frac{W_{\omega_2\omega_4}}{\Omega_{3'}} \Big]
\\
\label{amp145}
\tau_1\tau_4\tau_5:&& 2ie^{-i\pi(\omega_2+\omega_3)} \Big[\int VLC^{++} -\int LC(23)e^{i\pi \omega_d}
-e^{i\pi\omega_3}\int \frac{W_{\omega_2\omega_4}}{\Omega_{3'}} \Big]
\\
\label{amp135}
\tau_1\tau_3\tau_5:&& 2ie^{-i\pi(\omega_2+\omega_4)} \Big[\int VLC^{++} - \int W_{\omega_2} e^{i\pi (\omega_c+\omega_d)} - e^{i\pi (\omega_a+\omega_b}  \int W_{\omega_4}\nonumber\\
&&\hspace{2cm}-e^{i\pi\omega_3}\int \frac{W_{\omega_2\omega_4}}{\Omega_{3'}} \Big]
\ea
\ba
\label{amp1345}
\tau_1\tau_3\tau_4\tau_5:&& 2ie^{-i\pi\omega_2} \Big[\int VLC^{+-}   - \int LC(23)e^{-i\pi\omega_d}\nonumber\\
&&\hspace{2cm}+\int W_{\omega_2} e^{i\pi (\omega_c-\omega_d)} - e^{i\pi (\omega_a-\omega_b)}\int W_{\omega_4}\nonumber\\
&&\hspace{2cm}-e^{i\pi\omega_3}\int \frac{W_{\omega_2\omega_4}}{\Omega_{3'}}
- 2i  \frac{V_L(a) \Omega_b}{\Omega_2} W_{\omega_4}
\Big]
\\
\label{amp1235}
\tau_1\tau_2\tau_3\tau_5:&& 2ie^{-i\pi\omega_4} \Big[\int VLC^{-+}   - e^{-i\pi\omega_a} \int LC(34)\nonumber\\
&&\hspace{2cm}-\int W_{\omega_2} e^{i\pi (\omega_d-\omega_c)} + e^{i\pi (\omega_b-\omega_a)}\int W_{\omega_4}-\nonumber\\ 
&&\hspace{2cm}-e^{i\pi\omega_3}\int  \frac{W_{\omega_2\omega_4}}{\Omega_{3'}}
- 2i  W_{\omega_2} \frac{\Omega_c V_R(d)}{\Omega_4}
\Big]
\\
\label{amp1245}
\tau_1\tau_2\tau_4\tau_5:&&2i e^{-i\pi \omega_3} \Big[
e^{2i\pi \omega_3} \int VLC^{--} -e^{-i\pi \omega_a}\int LC(34) -\int LC(23) e^{-i\pi \omega_d}  \nonumber\\
&&\hspace{3cm}  +\int W_{\omega_3} e^{i \pi (\omega_a +\omega_d)}-e^{i\pi\omega_3} \int \frac{W_{\omega_2\omega_4}}{\Omega_{3'}} \nonumber\\
&&+2i \left( \int W_{\text{3 reggeon cut}}- e^{i\pi \omega_3}\int \frac{\Omega_{2'3'}\Omega_{4'3'} \widetilde{W}_{LR}}{\Omega_{3'}}\right) -2i e^{i\pi \omega_3}\int  \frac{\Omega_a \Omega_d}{\Omega_{3'}} W_{\omega_3}\nonumber\\
&&
\hspace{3cm}-2i\left( \int  \widetilde{W}_{\omega_2\omega_3;L}\frac{ \Omega_{3'2'}\Omega_d}{\Omega_{3'}}
+\int  \frac{\Omega_a \Omega_{3'4'}}{\Omega_{3'}} \widetilde{W}_{\omega_3\omega_4;R} \right) 
\Big]
 \ea
\ba
\label{amp12345}
\tau_1\tau_2\tau_3\tau_4\tau_5:&& 2i\Big[ e^{-2i\pi \omega_3} \int VLC^{++}  -e^{i\pi \omega_a} \int  LC(34)^* - \int  LC(23)^* e^{i\pi \omega_d}\nonumber\\
&&\hspace{-0.5cm}+
 \int  W_{\omega_2} e^{i\pi(\omega_c-\omega_d)} + 
 \int  W_{\omega_3} e^{-i\pi(\omega_a+\omega_d)}+e^{i\pi(\omega_b-\omega_a)}  \int W_{\omega_4}  -e^{+i\pi\omega_3} \int  \frac{W_{\omega_2\omega_4}}{\Omega_{3'}}\nonumber\\
&&
\hspace{-0.5cm}-2i \left( \int  W_{\text{3 reggeon cut}}-e^{-i\pi \omega_3} \int  \frac{\Omega_{2'3'}\Omega_{4'3'}\widetilde{W}_{LR}}{\Omega_{3'}}\right) +2i  e^{-i\pi \omega_3} \int  W_{\omega_3} \frac{\Omega_a\Omega_d}{\Omega_{3'}} \nonumber\\
&&
\hspace{3cm}+2i\left(  \int  \widetilde{W}_{\omega_2\omega_3;L} \frac{\Omega_{3'2'}\Omega_d}{\Omega_{3'}}
+  \int  \frac{\Omega_{3'4'}\Omega_a}{\Omega_{3'}} \widetilde{W}_{\omega_3\omega_4;R} \right)\nonumber\\
&&\hspace{-0.5cm}- 2i \left(  \int  W_{\omega_2}\frac{\Omega_c V_R(d)}{\Omega_4}
+  \int W_{\omega_4}\frac{V_L(a) \Omega_b}{\Omega_2} \right)\Big]\,.
\nonumber\\
\ea

\subsection{The regular amplitudes}
As the next step we have to remove the singular pieces of the partial waves. As before we write, e.g. 
\ba
\int \widetilde{W}_{LL}&=& \int \widetilde{W}_{LL}^{reg}+\int \delta \widetilde{W}_{LL}  \nonumber\\
\int VLC^{++}&=&\int VLC^{++;reg}+\int \delta VLC^{++}\,,
\ea
and insert the singular pieces $\delta \widetilde{W}$ etc. In contrast to the previous long cut, now also the subtraction involves  $\omega'_i$ integrations.
For the long Regge cuts, the double cut, and the three reggeon cut the derivation of the singular pieces is somewhat lengthy and will be described in Appendix D.
Using the results and combining them with the Regge pole terms listed in Appendix A we arrive at the simpler expressions:
\ba
\label{amp15reg}
\tau_1\tau_5:&& e^{-i\pi(\omega_2+\omega_3+\omega_4)} 
 \Big[ e^{i\pi (\omega_b+\omega_c)} \cos \pi(\omega_a-\omega_d)\nonumber\\
&&+2i \Big[ \int VLC^{++;reg}   \nonumber\\ 
&& - i \left(  \int W_{\omega_2}^{reg} -\frac{1}{2} \sin \pi (\omega_a-\omega_b)\right)
\left(  \int  W_{\omega_4}^{reg} -\frac{1}{2} \sin \pi (\omega_d-\omega_c)  \right) \big] \Big]
 \\
\label{amp125reg}
\tau_1\tau_2\tau_5:&& e^{-i\pi(\omega_3+\omega_4)} \Big[ -e^{i\pi (\omega_c+\omega_d)} \cos \pi(\omega_a-\omega_b) \nonumber\\
&&+2i\Big[  \int VLC^{++;reg}   -e^{i\pi \omega_a} \int  LC(34)^{reg} \nonumber\\
&&-i \left( \int  W_{\omega_2}^{reg} -\frac{1}{2} \sin \pi (\omega_a-\omega_b)\right) 
\left(  \int W_{\omega_4}^{reg} -\frac{1}{2} \sin \pi (\omega_d-\omega_c) \right)  \big]\Big]
\\
\label{amp145reg}
\tau_1\tau_4\tau_5:&& e^{-i\pi(\omega_2+\omega_3)} \Big[-e^{i\pi (\omega_a+\omega_b)} \cos \pi(\omega_d-\omega_c)\nonumber\\
&&
+2i\Big[  \int  VLC^{++;reg} - \int  LC(23)^{reg}e^{i\pi \omega_d}\nonumber\\
&&-i\left( \int W_{\omega_2}^{reg} -\frac{1}{2} \sin \pi (\omega_a-\omega_b)\right) 
\left( \int  W_{\omega_4}^{reg} -\frac{1}{2} \sin \pi (\omega_d-\omega_c) \right) \big]\Big]
\\
\label{amp135reg}
\tau_1\tau_3\tau_5:&& e^{-i\pi(\omega_2+\omega_4)} \Big[-e^{i\pi (\omega_a+\omega_d)} \cos \pi(\omega_b-\omega_c)\nonumber\\
&&+2i\Big[  \int VLC^{++;reg} -  \int W_{\omega_2}^{reg} e^{i\pi (\omega_c+\omega_d)} - e^{i\pi (\omega_a+\omega_b}  \int W_{\omega_4}^{reg}\nonumber\\
&&
- i\left(  \int W_{\omega_2}^{reg} -\frac{1}{2} \sin \pi (\omega_a-\omega_b)\right) 
\left( \int  W_{\omega_4}^{reg} -\frac{1}{2} \sin \pi (\omega_d-\omega_c) \right)\big] \Big]
\ea
\ba
\label{amp1345reg}
\tau_1\tau_3\tau_4\tau_5:&& e^{-i\pi\omega_2} \Big[e^{i\pi (\omega_c-\omega_d)} \cos \pi(\omega_b-\omega_a)
\nonumber\\
&&
+2i \Big[ \int  VLC^{+-;reg}   -  \int  LC(23)^{reg}e^{-i\pi\omega_d}\nonumber\\
&&+ \int W_{\omega_2}^{reg} e^{i\pi (\omega_c-\omega_d)} - e^{i\pi (\omega_a-\omega_b)} \int  W_{\omega_4}^{reg}\nonumber\\
&&- i\left( \int  W_{\omega_2}^{reg} -\frac{1}{2} \sin \pi (\omega_a-\omega_b)\right) 
\left( \int  W_{\omega_4}^{reg} -\frac{1}{2} \sin \pi (\omega_d-\omega_c) \right)
\big] \Big]
\\
\label{amp1235reg}
\tau_1\tau_2\tau_3\tau_5:&& e^{-i\pi\omega_4} \Big[e^{i\pi (\omega_b-\omega_a)} \cos \pi(\omega_c-\omega_d) \nonumber\\
&&+2i\Big( \int  VLC^{-+;reg}   - e^{-i\pi\omega_a} \int  LC(34)^{reg}\nonumber\\
&&- \int  W_{\omega_2}^{reg} e^{i\pi (\omega_d-\omega_c)} + e^{i\pi (\omega_b-\omega_a)}  \int   W_{\omega_4}^{reg}\nonumber\\
&&- i \left( \int  W_{\omega_2}^{reg} -\frac{1}{2} \sin \pi (\omega_a-\omega_b)\right) 
\left( \int  W_{\omega_4}^{reg} -\frac{1}{2} \sin \pi (\omega_d-\omega_c) \right)
\Big]\Big]\\
\label{amp1245reg}
\tau_1\tau_2\tau_4\tau_5:&& e^{-i\pi \omega_3} \Big[ e^{i\pi(\omega_b+\omega_c-\omega_a-\omega_d)}
\nonumber\\
&&
+2i\Big[ e^{i\pi( \omega_3-\omega_2)}  \int \widetilde{W}_{LL}^{reg}+ \int  \widetilde{W}_{RL}^{reg}+e^{i\pi(2 \omega_3-\omega_2-\omega_4)}  \int  \widetilde{W}_{LR}^{reg}+e^{i\pi( \omega_3-\omega_4)}  \int \widetilde{W}_{RR}^{reg}\nonumber\\
 &&-e^{-i\pi \omega_a} \int  LC(34)^{reg} - \int LC(23)^{reg} e^{-i\pi \omega_d} + \int W_{\omega_3}^{reg} e^{i \pi (\omega_a +\omega_d)} \nonumber\\
&&
-i \left(  \int W_{\omega_2}^{reg} -\frac{1}{2} \sin \pi (\omega_a-\omega_b)\right) 
\left( \int  W_{\omega_4}^{reg} -\frac{1}{2} \sin \pi (\omega_d-\omega_c) \right)\nonumber\\ 
&&+2i \Big[  \int W_{\text{3 reggeon cut}}^{reg}\nonumber\\
&&\hspace{0.5cm}-i \Omega_3 \cos \pi \omega_3 \left(  \int W_{\omega_2}^{reg} -\frac{1}{2} \sin \pi (\omega_a-\omega_b)\right) 
\left( \int  W_{\omega_4}^{reg} -\frac{1}{2} \sin \pi (\omega_d-\omega_c) \right) \nonumber\\
&&- 2i \Omega_3^2 \cos \pi \omega_3 \cos \pi (\omega_a + \omega_b) \delta W{\omega_3}
\Big]\Big] \Big]
\ea
\ba
\label{amp12345reg}
\tau_1\tau_2\tau_3\tau_4\tau_5:&& -\cos \pi (\omega_b+\omega_d-\omega_a-\omega_c)\nonumber\\
&&\hspace{-0.9cm}+2i\Big[ e^{i\pi( \omega_2-\omega_3)} \int  \widetilde{W}_{LL}^{reg}+ \int \widetilde{W}_{RL}^{reg}+e^{-i\pi(2 \omega_3-\omega_2-\omega_4)} \int  \widetilde{W}_{LR}^{reg}+e^{i\pi( \omega_4-\omega_3)} \int  \widetilde{W}_{RR}^{reg}\nonumber\\
 &&\hspace{-0.9cm}-e^{-i\pi \omega_a} { \int LC(34)^{reg}}^* - \int  {LC(23)^{reg}}^* e^{-i\pi \omega_d} + \int W_{\omega_3}^{reg} e^{i \pi (\omega_a +\omega_d)} \nonumber\\
&&
\hspace{-0.9cm}-i \left(  \int W_{\omega_2}^{reg} -\frac{1}{2} \sin \pi (\omega_a-\omega_b)\right) 
\left(  \int W_{\omega_4}^{reg} -\frac{1}{2} \sin \pi (\omega_d-\omega_c) \right)\nonumber\\
&&\hspace{-0.5cm}-2i \Big[ \int  W_{\text{3 reggeon cut}}^{reg}
\nonumber\\
&&\hspace{-0.9cm}+i \Omega_3 \cos \pi \omega_3 \left( \int  W_{\omega_2}^{reg} -\frac{1}{2} \sin \pi (\omega_a-\omega_b)\right) 
\left( \int  W_{\omega_4}^{reg} -\frac{1}{2} \sin \pi (\omega_d-\omega_c) \right) \nonumber\\
&&\hspace{-0.9cm}+ 2i \Omega_3^2 \cos \pi \omega_3 \cos \pi (\omega_a + \omega_b) \delta W{\omega_3}
\Big]\Big] \Big] \,.
\ea

Before we address the energy discontinuities and unitarity equations we remind of our discussion at end of section 5,  the expansion in powers of  $(i \pi)$. As long as we consider only the leading order, we  disregard all phase factors and restrict ourselves to terms proportional to $\sim i\pi$, i.e. we retain only the very long cut terms $\widetilde{W}_{LL}^{reg}$ etc,  the long cut terms $\widetilde{W}_{\omega_2 \omega_3;L}^{reg}$ etc , and the short cut terms $W_{\omega_2}$ etc. However, now we are interested in the double cut and the 3-reggeon cut  
which are of the order $(i\pi)^2$. In all kinematic regions where the very long cut appears
we have  the double cut term
\be
 \left( \int  W_{\omega_2}^{reg} -\frac{1}{2} \sin \pi (\omega_a-\omega_b)\right) 
\left( \int  W_{\omega_4}^{reg} -\frac{1}{2} \sin \pi (\omega_d-\omega_c) \right)\nonumber
\ee
and, in the last two kinematic regions, also the three reggeon cut:
\be
  \int  W_{\text{3 reggeon cut}}^{reg}\,.
\ee
These terms are of the order $(i\pi)^2$, and 
in order to be consistent we can no longer disregard the phase factors coming from the 
energy factors. For example, for terms of the order  $(i\pi)^2$ we have  contributions from the phase factors multiplying the Regge cuts. For the long cut pieces, $LC(23)$ and $LC(34)$, some consequences have already been discussed in sedction 5. For the very long cut we will come back in section 7.3.

\section{Energy discontinuities for the very long cut, the double cut and the 3-reggeon cut}

Having determined the scattering amplitudes in the different kinematic region we now need to calculate the Regge cut contributions. For this we now turn to energy discontinuities. 

\subsection{The very long cut: the discontinuity in  $s_{1234}$ and the corresponding unitarity integral}

Let us start with the discontinuity in $s_{1234}$ (Fig.\ref{fig:single-disc}c) which determines the sum of the partial waves of the very long cut. The discontinuity is found to be:
\ba
\label{disc-1234}
&&\Delta_{1234} = e^{-i\pi(\omega_1+\omega_5)}\cdot\nonumber\\
&& 
\cdot \Big[ e^{i\pi(2\omega_2-\omega_4)}\int  \widetilde{W}_{LL} + e^{i\pi(3\omega_3-\omega_2-\omega_4)}  \int  \widetilde{W}_{RL}+
e^{-i\pi \omega_3}\int  \widetilde{W}_{LR} + e^{i\pi(2\omega_4-\omega_2)} \int  \widetilde{W}_{RR}  \nonumber\\
&&\hspace{1cm}-e^{i\pi (\omega_4-\omega_2-\omega_3)} \int  LC(23)  e^{i\pi \omega_d} -e^{i\pi (\omega_2-\omega_3-\omega_4)}  e^{i\pi\omega_a}\int  LC(34)
\nonumber\\
&&
\hspace{1cm}+e^{i\pi (\omega_2+\omega_4 - \omega_3)}  e^{-i\pi \omega_a}\int  W_{\omega_3} e^{-i\pi \omega_d}-e^{i\pi (2\omega_3-\omega_2-\omega_4)}\int  \frac{W_{\omega_2\omega_4}}{\Omega_{3'}}
\nonumber\\
&&
\hspace{1cm}-\frac{1}{\Omega_3} 
\left(e^{i\pi (\omega_3-\omega_2)}  \frac{V_L(a)  \Omega_b}{\Omega_2}
 -2i  V_L(a) V_L(b)\right)  
  \left(e^{i\pi (\omega_3-\omega_4)}\frac{ \Omega_cV_R(d)}{\Omega_4}  -2iV_R(c)V_R(d)\right)
\nonumber\\ 
&&\hspace{1cm}+2i e^{i\pi (\omega_2+\omega_4-\omega_3)}\left( -\int  W_{\text{3 reggeon}} 
+ e^{-i\pi \omega_3} \int  \frac{\Omega_{2'}\Omega_{4'}}{\Omega_{3'}}  \widetilde{W}_{LR}\right.\nonumber\\
&&\hspace{1cm}\left. +\int  \widetilde{W}_{\omega_2\omega_3;L} 
\frac{ \Omega_d \Omega_{3'2'}}{\Omega_{3'}} +  \int \frac{ \Omega_a \Omega_{3'4'}}{\Omega_{3‘}}  \widetilde{W}_{\omega_3\omega_4;R}
 +e^{-i\pi \omega_3}
\int W_{\omega_3} \frac{\Omega_a\Omega_d}{\Omega_3}
\right) \nonumber\\
&&\hspace{1cm}+2i e^{i \pi (\omega_3-\omega_2-\omega_4)} \left( -\int  W_{\omega_2} \frac{\Omega_c V_R(d)}{\Omega_4}- 
\frac{V_L(a)\Omega_b}{\Omega_2}\int  W_{\omega_4} \right.\nonumber\\
&&\hspace{4cm}\left. +\int W_{\omega_2} e^{i\pi(\omega_4-\omega_c)}V_R(d) +e^{i\pi(\omega_2-\omega_b)}V_L(a) \int W_{\omega_4}\right) \Big] \,.\nonumber\\
\ea
This expression still contains many singularities which have to be removed by inserting the subtractions for the cut contributions. Making use of the subractions derived in the appendix D, we arrive at the finite expression:
\ba
\label{disc-1234-finite}
&&\Delta_{1234} =\nonumber\\
&& e^{-i\pi(\omega_1+\omega_5)} 
\cdot \Big[ e^{i\pi(2\omega_2-\omega_4)}\int  \widetilde{W}_{LL}^{reg} + e^{i\pi(3\omega_3-\omega_2-\omega_4)}  \int  \widetilde{W}_{RL}^{reg}+e^{i\pi(2\omega_4-\omega_2)} \int  \widetilde{W}_{RR}^{reg}\nonumber\\
&&+ e^{-i\pi \omega_3} \left(1+ 2i e^{i\pi ( \omega_2+\omega_4)} \cos \pi \omega_3 \sin \pi (\omega_2+\omega_4 - \omega_3) -(2i)^2  e^{i\pi ( \omega_2+\omega_4)} \Omega_2 \Omega_4  \right)\int  \widetilde{W}_{LR}^{reg}   \nonumber\\
&&-e^{i\pi (\omega_4-\omega_2-\omega_3)} \int  LC(23)^{reg}  e^{i\pi \omega_d} -e^{i\pi (\omega_2-\omega_3-\omega_4)}  e^{i\pi\omega_a}\int  LC(34)^{reg}
\nonumber \\
&&+e^{i\pi (\omega_2+\omega_4 - \omega_3)} \left( e^{-i\pi\omega_a}e^{-i\pi \omega_d}- \frac{1}{2} (2i)^2\Omega_a \Omega_d\right)
\int  W_{\omega_3}^{reg} \nonumber\\
&&-2i e^{i\pi (\omega_2+\omega_4-\omega_3)}\int W_{\text{3 reggeon}}^{reg} \nonumber\\
&&-i e^{i\pi (\omega_3-\omega_2-\omega_4}) \left( 1+2 \cos \pi \omega_3 (e^{-i\pi (\omega_2+\omega_4)} + 4\Omega_2 \Omega_4)  \right. \nonumber\\
&&\left. \hspace{1cm} \cdot \left( \int W_{\omega_2}^{reg} - \frac{1}{2}\sin \pi (\omega_a-\omega_b)\right) \left(\int W_{\omega_4}^{reg} - \frac{1}{2}\sin \pi (\omega_d-\omega_c)\right)\right)\nonumber\\
&&+ \text{Regge pole terms}\, ,
\ea  
where
\ba
&&\text{Regge pole terms}=\frac{1}{2} (2i)^2 \Big[ \sin \pi (\omega_a-\omega_b) V_R(c) V_R(d) +V_L(a)) V_L(b)  \sin \pi (\omega_d-\omega_c) \nonumber\\
&& -V_L(a) \left( 2 \cos \pi \omega_b \cos \pi \omega_c \Omega_3 -\sin \pi (\omega_b+ \omega_c) \cos \pi \omega_3   \right)V_R(d)\Big] \nonumber\\
&&+ (2i)^3 \Big[ V_L(a) \left( \Omega_b V_R(c) + V_L(b) \Omega_c \right) V_R(d)\nonumber\\
&&  + \Omega_2 \delta W_{\omega_2} V_R(c) V_R(d) + V_L(a) V_L(b) \Omega_4 \delta W_{\omega_4} -
\frac{1}{2} V_L(a) \Omega_3 \delta W_{\omega_3} V_R(d) \Big]\,.
\ea

In the weak coupling limit we have the much simpler expression:
\ba 
\label{d1234-leading}
\Delta_{1234}& = &\int_2 \int_3 \int_4 s_{12}^{\omega''_2} s_{23}^{\omega''_3} s_{34}^{\omega''_4} \left( \widetilde{W}_{LL}^{reg}+\widetilde{W}_{LR}^{reg}+\widetilde{W}_{RL}^{reg}+\widetilde{W}_{RR}^{reg}\right)\nonumber\\
&&- \int_3 \int_4 s_{23}^{\omega''_3} s_{34}^{\omega''_4} LC(34)^{reg}
 - \int_2 \int_3   s_{12}^{\omega''_2} s_{23}^{\omega''_3}  LC(23)^{reg}+ \int_3  s_{23}^{\omega''_3}W_{\omega_3}^{reg}\,.
\ea

Finally we have to compute, via unitarity,  the weak coupling limit of the discontinuity on the lhs. The result is:
\ba
\label{d_1234_LL}
\Delta_{1234}&=& \left( f_{\omega_2\omega_3\omega_4} -  f_{\omega_2\omega_3} - f_{\omega_3\omega_4} +f_{\omega_3}\right)
+\frac{\pi}{2} \left( V_{24} +V_{15} - V_{14} -V_{25} \right) \nonumber\\
&&= f_{\omega_2\omega_3\omega_4}+\frac{\delta_{15}}{2}  - f_{\omega_2\omega_3}-\frac{\delta_{14}}{2} - f_{\omega_3\omega_4}-\frac{\delta_{25}}{2} +f_{\omega_3}+\frac{\delta_{24}}{2}\, ,
\ea
where we have used
\ba
\delta_{14}&=&\pi \left(V_{14}+\omega_a+\omega_c\right)\nonumber\\
\delta_{25}&=&\pi \left(V_{25}+\omega_b+\omega_d\right)\nonumber\\
\delta_{24}&=&\pi \left(V_{24}+\omega_b+\omega_c\right)\nonumber\\
\delta_{15}&=&\pi \left(V_{15}+\omega_a+\omega_d\right)\,.
\ea
The amplitiude of the very long cut (Fig.\ref{fig:cut-amplitudes}c ), $f_{\omega_2\omega_3\omega_4}$, has the form
\ba
\label{f-verylong}
f_{\omega_2\omega_3\omega_4}= &&\frac{a}{2} \sum_{n_2,n_3,n_4} (-1)^{n_2+n_3-n_4}\int \frac{d\nu_2 d\nu_3d\nu_4}{(2\pi)^3}\Big[
 \frac{1}{i\nu_2+\frac{n_2}{2}} \left( \frac{k_a^*q_3^*}{q_1^* k_b^*} \right)^{i\nu_2+\frac{n_2}{2}}
\left( \frac{k_a q_3}{q_1 k_b} \right)^{i\nu_2-\frac{n_2}{2}} \nonumber\\
&&\left( \frac{s_{12}}{s_{02}} \right)^{\omega(\nu_2,n_2)}
\cdot B(\nu_2,\nu_2,n_3.n_3) 
\left( \frac{k_b^*q_4^*}{q_2^* k_c^*} \right)^{i\nu_3+\frac{n_3}{2}}
\left( \frac{k_b q_4}{q_2 k_c} \right)^{i\nu_3-\frac{n_3}{2}}  \left( \frac{s_{23}}{s_{03}} \right)^{\omega(\nu_3,n_3)} 
\\
&&\cdot B(\nu_3,\nu_4,n_3,n_4) 
\left( \frac{k_c^*q_5^*}{q_3^* k_d^*} \right)^{i\nu_4+\frac{n_4}{2}}
\left( \frac{k_c q_5}{q_3 k_d} \right)^{i\nu_4-\frac{n_4}{2}}  \left( \frac{s_{34}}{s_{04}} \right)^{\omega(\nu_4,n_4 )}\frac{1}{i\nu_4- 
\frac{n_4}{2}}\Big] |_{\text{sub}}.\nonumber
\ea
As we have done for the other cut amplitudes, $f_{\omega_2}$ and $f_{\omega_2\omega_3}$ in $f_{\omega_2\omega_3\omega_4}$  we have separated the  one loop contribution. 

Combining this with (\ref{d1234-leading}) and inserting our weak coupling results for $LC(23)$, $LC(34)$  we arrive at the weak coupling result for the sum of the four partial waves:
\be
\int_2 \int_3 \int_4 s_{12}^{\omega''_2}  s_{23}^{\omega''_3}  s_{34}^{\omega''_4} \left(  \widetilde{W}_{LL}^{reg}+\widetilde{W}_{LR}^{reg}+
\widetilde{W}_{RL}^{reg}+\widetilde{W}_{RR}^{reg}\right) =f_{\omega_2\omega_3\omega_4} +\frac{ \delta_{15}}{2}\,.
 \ee 
 
\subsection{Multiple discontinuities and unitarity integrals}

To proceed further we have to move to double and even to triple energy discontinuities. We begin with the double cut which is obtained from the double discontinuity $\Delta_{12} \Delta_{34}$.

\subsubsection{The double cut: $\Delta_{12}\Delta_{34}$}

In order to find  $W_{\omega_2 \omega_4}$ we compute the double discontinuity in $s_{12}$ and $s_{34}$  (Fig.\ref{fig:double-disc-1}a):
\begin{figure}[H]
\centering
\epsfig{file=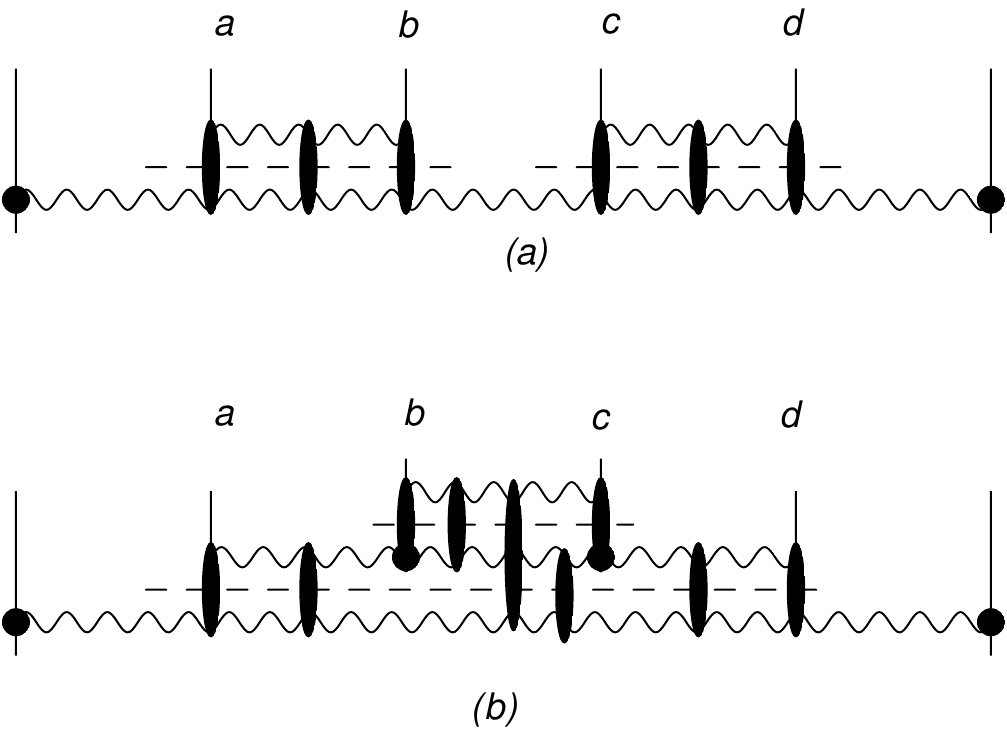,scale=1.0}
\caption{Double discontinuities (a) in  $s_{12}$ and $s_{34}$, (b) in $s_{23}$ and $s_{1234}$}.
\label{fig:double-disc-1}
\end{figure}
\ba
\label{d_2d_4}
&&\Delta_{12} \Delta_{34}
 = e^{-i\pi(\omega_1 + \omega_3 + \omega_5)} 
\Big[ e^{i\pi(\omega_2 + \omega_4)} \int W_{\omega_2\omega_4} \\
&&- \frac{V_L(a) V_R(b)}{\Omega_2} e^{i\pi \omega_4} \int W_{\omega_4}
- e^{i\pi \omega_2} \int W_{\omega_2}  \frac{V_L(c) V_R(d)}{\Omega_4}
+  \frac{V_L(a) V_R(b) V_L(c)V_R(d)}{\Omega_2\Omega_4} \Big]\, .\nonumber
\ea
With the following  ansatz for $W_{\omega_2 \omega_4}$:
\be
W_{\omega_2 \omega_4}= W_{\omega_2} W_{\omega_4} 
\ee
we arrive at the factorizing result:
\be
\label{d_12_34}
\Delta_{12} \Delta_{34}
 =e^{-i\pi(\omega_1 + \omega_3 + \omega_5)} 
\Big[e^{i\pi \omega_2} \int W_{\omega_2} - \frac{V_L(a) V_R(b)}{\Omega_2} \Big]
\Big[e^{i\pi \omega_4} \int W_{\omega_4} - \frac{V_L(c) V_R(d)}{\Omega_4} \Big]\,.
\ee
Using our results for the subtractions of $W_{\omega_2}$ and $W_{\omega_4}$ from (\ref{decomp-W2})  and (\ref{delta-W2})
we find
\ba
\label{d12-d34-rhs}
\Delta_{12} \Delta_{34}
 =e^{-i\pi(\omega_1 + \omega_3 + \omega_5)} \Big[ e^{i\pi\omega_2} W_{\omega_2}^{reg}-\frac{1}{2}\left( e^{i\pi \omega_a} V_R(b) + e^{i\pi \omega_b} V_L(a)\right)\Big]\cdot\nonumber\\
\cdot  \Big[ e^{i\pi\omega_4} W_{\omega_4}^{reg}-\frac{1}{2}\left( e^{i\pi \omega_d} V_L(c) + e^{i\pi \omega_c} V_R(d)\right)\Big]\,.
\ea
As expected, all singular terms cancel. We therefore conclude that the factorizing form of 
$W_{\omega_2 \omega_4}$ is fully consistent with the double discontinuity.  

As a result, in (\ref{amp15}) - (\ref{amp12345}) we find the following subtraction for $W_{\omega_2 \omega_4}$:
\be
\int W_{\omega_2 \omega_4} = \int W_{\omega_2}^{reg} \int W_{\omega_4}^{reg} + \int \delta  W_{\omega_2 \omega_4}.
\ee
with 
\be
\int \delta  W_{\omega_2 \omega_4}=\delta W_{\omega_2} \int W_{\omega_4} + \int W_{\omega_2} \delta W_{\omega_4}
+\delta W_{\omega_2} \delta W_{\omega_4}\, .
\ee

We finally compute,  from unitarity, the double discontinuity in the weak coupling limit.   
We find
\ba
\label{d_12-d_34_LL}
\Delta_{12}\Delta_{34}&=&\Big[f_{\omega_2} +\frac{\pi}{2} V_{13} -\pi(\omega_2-\omega_a-\omega_b) \Big]
\Big[f_{\omega_4} +\frac{\pi}{2} V_{35} -\pi(\omega_4-\omega_c-\omega_d) \Big]\, .
\ea
Comparison  with  the rhs of (\ref{d12-d34-rhs}) 
\be
rhs=
\Big[\int W_{\omega_2}^{reg}-\pi \omega_2+\frac{\pi}{2} (\omega_a+\omega_b)\Big] \cdot 
\Big[\int W_{\omega_4}^{reg}-\pi \omega_4+\frac{\pi}{2} (\omega_c+\omega_d)\Big]
\ee
leads to the result:
\be
\int_2 \int_4 s_{12}^{\omega''_2} s_{34}^{\omega''_4} W_{\omega_2}^{reg} W_{\omega_4}^{reg}
=\left( f_{\omega_2}+\frac{ \delta_{13}}{2}\right) \left(f_{\omega_4}+\frac{ \delta_{35}}{2}\right)\,,
\ee 
in agreement with (\ref{omega2-LL}).  

 \subsubsection{The 3-reggeon cut: $\Delta_{{23}} \Delta_{1234}$ }

Next we address   the double discontinuity in $s_{23}$ and  $s_{1234}$ which determines the three reggeon cut (Fig.\ref{fig:double-disc-1}b). This double discontinuity has the form:
\ba
\label{d-23-1234}
&&\Delta_{23} \Delta_{1234}=e^{-i\pi(\omega_1+\omega_5)} \nonumber\\
&&\cdot \Big[\left(- e^{i\pi(\omega_2+\omega_4-\omega_3)} + 2  e^{i\pi \omega_3}\cos \pi (\omega_2+\omega_4)\right)
\int W_{\text{3 reggeon}} \nonumber\\
&&- e^{-i\pi (\omega_2+\omega_4)} \left(1-(2i)^2 
 e^{i\pi (\omega_2+\omega_4)}\Omega_2 \Omega_4\right) \int \frac{\Omega_{3'2'} \Omega_{3'4'}}{\Omega_{3'}} 
\widetilde{W}_{LR} \nonumber\\
&&+\left( e^{i\pi (\omega_4-\omega_2)}\int  \widetilde{W}_{\omega_2\omega_3;L} \frac{\Omega_{3'2'}\Omega_{3'd}}{\Omega_{3'}} 
+e^{i\pi (\omega_2-\omega_4)} \int \frac{\Omega_{3'4'} \Omega_{3'a}}{\Omega_{3'}}  \ \widetilde{W}_{\omega_3\omega_4;R}\right) \nonumber\\
&&-2i 
\left( e^{i\pi \omega_2} \Omega_d 
\int\frac{\Omega_{3'2'} \Omega_{3'4'}}{\Omega_{3'}} \widetilde{W}_{\omega_2\omega_3;L}+ e^{i\pi \omega_4} \Omega_a \int \frac{\Omega_{3'2'} \Omega_{3'4'}}{\Omega_{3'}}\widetilde{W}_{\omega_3\omega_4;R}\right)\nonumber\\
&&
-\left( e^{i\pi(2\omega_3-\omega_2-\omega_4)} \frac{\Omega_{a3} \Omega_{d3}}{ \Omega_3}- \frac{\Omega_{23}\Omega_{43}}{\Omega_3} \left(2i e^{i\pi(\omega_3-\omega_a-\omega_d)} \Omega_3 +(2i)^2 \Omega_a \Omega_d \right)\right)
\int W_{\omega_3}  \nonumber\\
&&-2i \frac{V_L(a) V_L(b)V_R(c)V_R(d)}{\Omega_3} 
 \Big] \,.
\ea 
Inserting all subtractions (cf. Appendix D) we arrive at the regular expression:
\ba
\label{d-23-1234-final}
&&\Delta_{23} \Delta_{1234}=e^{-i\pi(\omega_1+\omega_5)} \cdot \nonumber\\
&&\Big[\left(- e^{i\pi(\omega_2+\omega_4-\omega_3)} + 2 e^{i\pi \omega_3}\cos \pi (\omega_2+\omega_4)\right) \int W_{\text{3 reggeon}}^{reg}\nonumber\\
&&+i \Omega_{23}  \Omega_{43} \left( e^{i\pi (\omega_2+\omega_4)}+2e^{2 i\pi  \omega_3} \cos \pi(\omega_2-\omega_4) \right) \int \widetilde{W}_{LR}^{reg}\nonumber\\
&&-e^{i\pi \omega_d} e^{i\pi (\omega_4-\omega_2)} \Omega_{23} \int \widetilde{W}_{23;L}^{reg}  - e^{i\pi \omega_a} e^{i\pi (\omega_2-\omega_4)} \Omega_{43} \int \widetilde{W}_{34;R}^{reg}\nonumber\\
&&+ \left(e^{-i\pi(\omega_2+\omega_4)}\sin \pi (\omega_a+\omega_d-\omega_3)+2i (e^{i\pi(\omega_3-\omega_a-\omega_d)} \Omega_{23}\Omega_{43} +\frac{1}{2} \Omega_a \Omega_d) \right.\nonumber\\
&&\hspace{2cm}\left. +4i e^{i\pi \omega_3} \Omega_{23} \Omega_{43} \Omega_a\Omega_d
\right)  \int W_{\omega_3}^{reg}\nonumber\\
&&+i \left( e^{i\pi (\omega_2+\omega_4)}+2e^{2 i\pi  \omega_3 \cos \pi(\omega_2-\omega_4)} \right) \Omega_3 \cos \pi \omega_3\nonumber\\
&&\hspace{1cm} \cdot \left( \int W_{\omega_2}^{reg} -\frac{1}{2} \sin \pi (\omega_a-\omega_b)\right) 
\left( \int W_{\omega_4}^{reg} -\frac{1}{2} \sin \pi (\omega_d-\omega_c)\right)\nonumber\\
&& -iV_L(a) \left( e^{i\pi \omega_b} V_R(c) +V_L(b)  e^{i\pi \omega_c} \right) V_R(d)\\
&&+i \left( e^{i\pi \omega_2} \sin \pi (\omega_b-\omega_a) V_R(c) V_R(d)+  V_L(a) V_L(b) e^{i\pi \omega_4} \sin \pi (\omega_c-\omega_d) \right)\, .
\ea
In leading order we find the much simpler expression:
\ba
\Delta_{23} \Delta_{1234}&=&e^{-i\pi(\omega_1+\omega_5)} \Big[ \int W_{\text{3 reggeon}}^{reg} -  \int \Omega_{2'3'} \widetilde{W}_{23;L}^{reg}-  \int \Omega_{4'3'} \widetilde{W}_{34;R}^{reg}\nonumber\\
&&+ \int \pi (\omega_a+\omega_d-\omega_3') W_{\omega_3}^{reg}\Big] \,.
\ea
It will be convenient to consider the following combination (in leading order, cf. (\ref{d23_d123_ll})):
\ba
\label{dec-3reggeon}
&&\Delta_{23} \Delta_{1234}+\Delta_{23} \Delta_{123}+\Delta_{23} \Delta_{234}=\nonumber\\
&&e^{-i\pi(\omega_1+\omega_5)} \Big[\int W_{\text{3 reggeon}}^{reg} + 
 \int \pi (\omega_a+\omega_d-\omega_3') W_{\omega_3}^{reg} \nonumber\\
&&\hspace{1cm}+ \frac{\pi^2}{2} \left( (\omega_2-\omega_a) (\omega_3-\omega_b)+(\omega_2-\omega_b) (\omega_3-\omega_c)
\right. \nonumber\\
&&\hspace{1cm}\left.
+(\omega_3-\omega_b) (\omega_4-\omega_c)+(\omega_3-\omega_c) (\omega_4-\omega_d)\right) \Big]\,.
\ea

\begin{figure}[H]
\centering
\epsfig{file=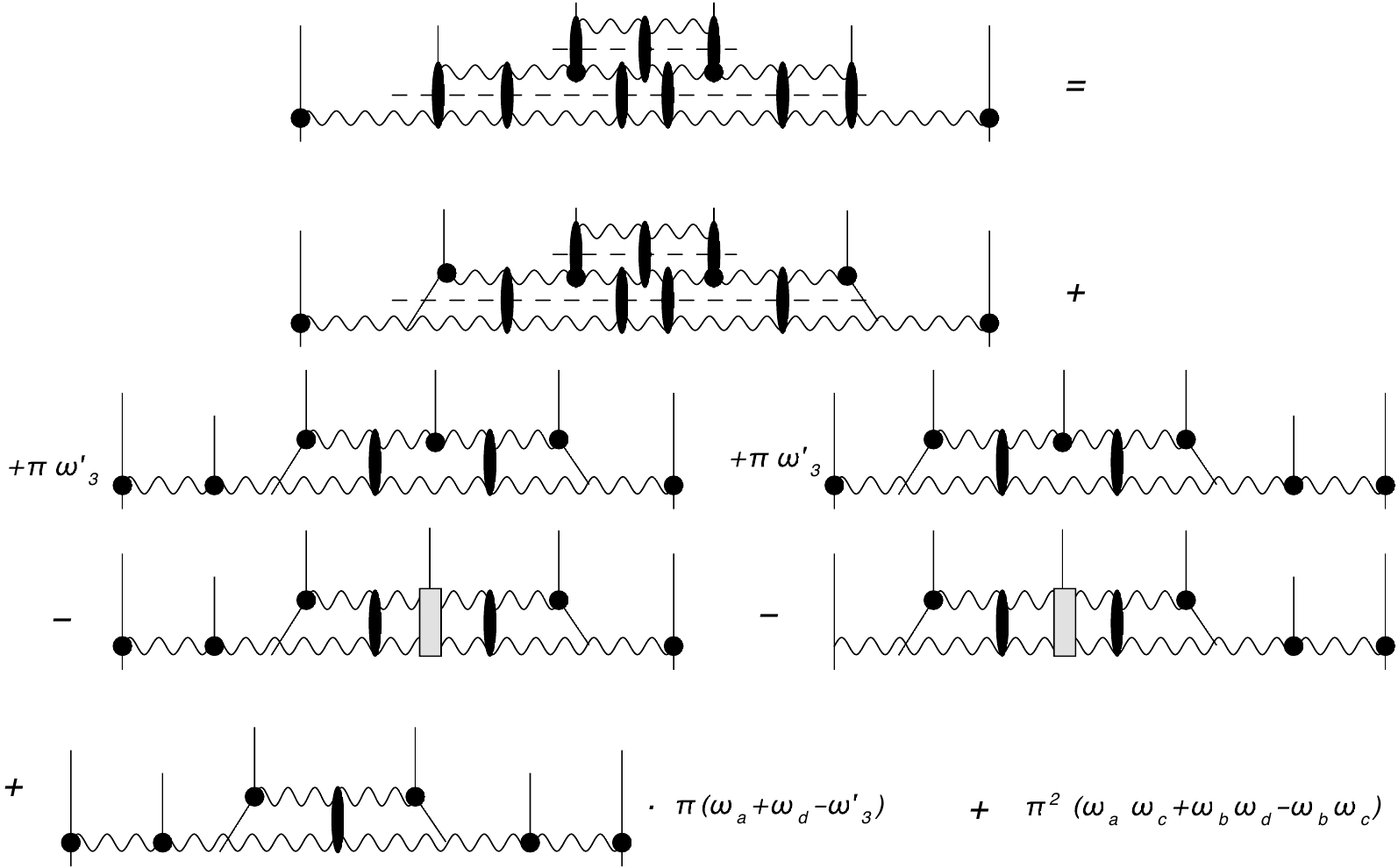,scale=0.9}
\caption{Leading order unitarity integral for the double discontinuity in the subenergies $s_{23}$ and $s_{1234}$}
\label{fig:double-disc-3}
\end{figure}

Now  let us make use of unitarity and compute the lhs, first the double discontinuity $\Delta_{23} \Delta_{1234}$. Diagrammatically, this double discontinuity is illustrated in Fig.\ref{fig:double-disc-1}b. Making repeated use of the bootstrap equations we find, after some algebra, 
the result illustrated in Fig.\ref{fig:double-disc-3}. 
For the double discontinuities $\Delta_{23}\Delta_{123}$ and $\Delta_{23}\Delta_{234}$ we can use Fig.\ref{fig:d23d123-ll}. Combining all these contributions with (\ref{dec-3reggeon}) we end up with 
\be
\int W_{\text{3 reggeon}}^{reg} = \int  f_{\text{3 reggeon}}+ {\text{2 loop terms}}\,,
\ee
where $ f_{\text{3 reggeon}}$ denotes the second line of  Fig.\ref{fig:double-disc-3}.
The 2-loop terms are
\ba
\pi^2 \Big[\omega_a \omega_b + \omega_c \omega_d +\frac{1}{2} (\omega_a \omega_c+\omega_b \omega_d)
-2\omega_b \omega_c
 +\omega_3 (\frac{\omega_a+\omega_d}{2} -2(\omega_b+\omega_c))+\omega_2 \omega_b +\omega_4 \omega_c \Big]\nonumber\\
\ea
and $ f_{\text{3 reggeon}}$ has the form:
\ba
\label{f-3-reggeon}
\int f_{\text{3 reggeon}}= \Phi^a \frac{1}{\omega''_2-K^{2;planar}} V^b \frac{1}{\omega''_3-K^{3;planar}} V^c \frac{1}{\omega''_4-K^{2;planar}} \Phi^d|_{\text{more than two loops}}\,.\nonumber\\
\ea
Here $\Phi^a$, $\Phi^d$ are  the nonlocal parts of the production vertices of particles 'a' and 'd' (cf. Fig.\ref{fig:leading-order-vertices}), and $V^b$ and $V^c$  the  full production  vertices  (Fig.\ref{fig:leading-order-vertices}) of particles 'b' and 'c'. 
For the kernel $K^{3;planar}$
of the  Green's function in the 3 gluon state in the $t_3$ channel we have (Fig.\ref{fig:3-gluon kerrnel}):
\begin{figure}[H]
\centering
\epsfig{file=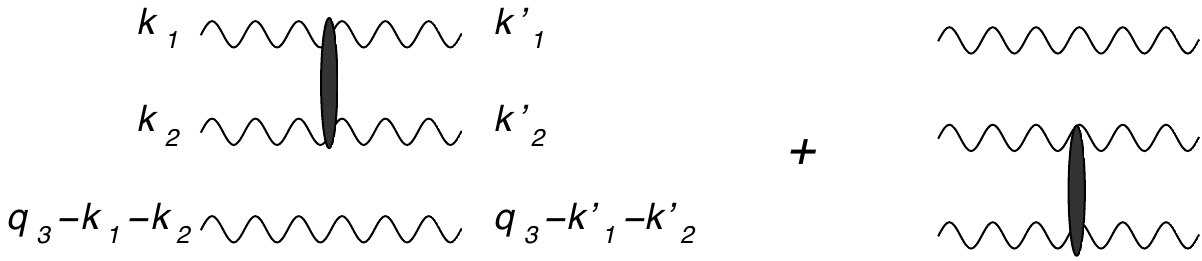,scale=0.7}
\caption{Kernel of the 3-gluon octet state }.
\label{fig:3-gluon kerrnel}
\end{figure}
\ba
\label{3-kernel}
&&\delta^{(2)}(\bk_1-\bk'_1) \delta^{(2)}(\bk_2-\bk'_2) \Big[\omega(\bk_1^2)+\omega(\bk_2^2)+
\omega((\bq_3-\bk_1-\bk_2)^2)\Big]\nonumber\\
&&+\frac{a}{2} \Big[K_{BFKL}(\bk_1,\bk_2;\bk'_1,\bk'_2)+K_{BFKL}(\bk_2,\bq_3-\bk_1-\bk_2;\bk'_2,\bq_3-\bk'_1-\bk'_2)\Big],
\ea 
where the BFKL kernel $K_{BFKL}$ has the well known form (in complex notation):
\be
K_{BFKL}(\bk_1,\bk_2; \bk'_1,\bk'_2)=\frac{k_1{k_2}^* {k'_1}^* k'_2+c.c.}{(\bk-\bk')^2}\,.
\ee
We rewrite the first line:
\ba
&&\delta^{(2)}(\bk_1-\bk'_1) \delta^{(2)}(\bk_2-\bk'_2) \Big[\omega(\bk_1^2)+\omega(\bk_2^2)+\omega((\bq_3-\bk_1-\bk_2)^2)\Big] \nonumber\\
&&=\delta^{(2)}(\bk_1-\bk'_1) \delta^{(2)}(\bk_2-\bk'_2) \Big[ \omega(\bq_3^2) +\frac{1}{2} \left( [\omega(\bk_1^2)+\omega((\bq_3-\bk_1-\bk_2)^2)-2\omega(\bq_3^2) \right)\nonumber\\
&& + \frac{1}{2} \left([\omega(\bk_1^2)+2 \omega(\bk_2^2)+\omega((\bq_3-\bk_1-\bk_2)^2)\right)\Big]\,.
\ea
Here the bracket in the second line is combined to the infrared finite expression
\be
\frac{1}{2}\left(  \omega(\bk_1^2)+\omega((\bq_3-\bk_1-\bk_2)^2)-2\omega(\bq_3^2) \right)= -\frac{1}{2} \ln \frac{\bk_1^2,
 (\bq_3-\bk_1-\bk_2)^2}{\bq_3^2 \bq_3^2},
\ee
and the last line is combined with the BFKL kernels. Alltogether (\ref{3-kernel})
takes the form:
\be
\text{(\ref{3-kernel})}=\delta^{(2)}(\bk_1-\bk'_1) \delta^{(2)}(\bk_2-\bk'_2)\omega(\bq_3^2) +K^{3;planar}
\ee
with the infrared finite kernel:
\ba
&&K^{3;planar}(\bk_1,\bk_2,\bq_3-\bk_1-\bk_2;\bk'_1,\bk'_2,\bq_3-\bk'_1-\bk'_2)=  -\frac{a}{2} \ln \frac{\bk_1^2,
 (\bq_3-\bk_1-\bk_2)^2}{\bq_3^2 \bq_3^2}\nonumber\\
&&+
\frac{1}{2} \left(K^{(1)}(\bk_1,\bk_2;(\bk'_1,\bk'_2)
+K^{(1)}(\bk_2,\bq_3-\bk_1-\bk_2;\bk'_1,\bq_3-\bk'_1-\bk'_2)\right) 
\ea
as a sum of two infrared finite color singlet BFKL kernels $K^{(1)}$ plus an additional infrared finite term. It is this 3-gluon kernel in momentum space which defines the open string Hamiltonian consisting of three sites.

In the next step, (\ref{f-3-reggeon}) has to be cast into the conformal invariant $\nu,n$ representation, in particular
the production vertices $V^b$ and $V^c$. For this one needs the eigenfunctions of the 3-gluon states.
Details will be discussed in a forthcoming  paper.  

\subsection{Further terms of the order $(i\pi)^2$: $\Delta_{12}\Delta_{1234}$  and $\Delta_{12} \Delta_{34}\Delta_{1234}$}

Before we insert these results into our all order expressions (\ref{amp15reg}  - (\ref {amp12345reg}) we have to discuss the corrections of the order $(i\pi)^2$. For this we return to our discussion at the end of section 6.2 (and section 5.2). As we have said before, in the scattering amplitude the new contributions - the double cut and the 3-reggeon cut - come as real-valued terms, i.e.compared to the familiar leading order  
2-reggeon cut contributions they come with an extra factor $i\pi$. But in order to have a complete understanding of these terms we must consider also contributions that come from expanding the phases contained in the energy factors.  For the long cuts this has been discussed before (at the end of section 5.2, in particular in (\ref{W23R-final}) and  (\ref{W23L-final})): there are leading order contributions to the long cut amplitude, which cancel if we
disregard phases and consider only the sum $\widetilde{W}_{\omega_2\omega_3;L}+\widetilde{W}_{\omega_2\omega_3;R}$. However, if we expand  phases and compute terms with an extra $i \pi$, we need $\widetilde{W}_{\omega_2\omega_3;L}$ and $\widetilde{W}_{\omega_2\omega_3;R}$ separately. For this we need multiple energy discontinuities. To extend this discussion to the very long cut we  have to complete our investigations of double discontinuities and compute
also  $\Delta_{12}\Delta_{1234}$,  $\Delta_{34}\Delta_{1234}$, and even the triple discontinuity  $\Delta_{12} \Delta_{34}\Delta_{1234}$:
\begin{figure}[H]
\centering
\epsfig{file=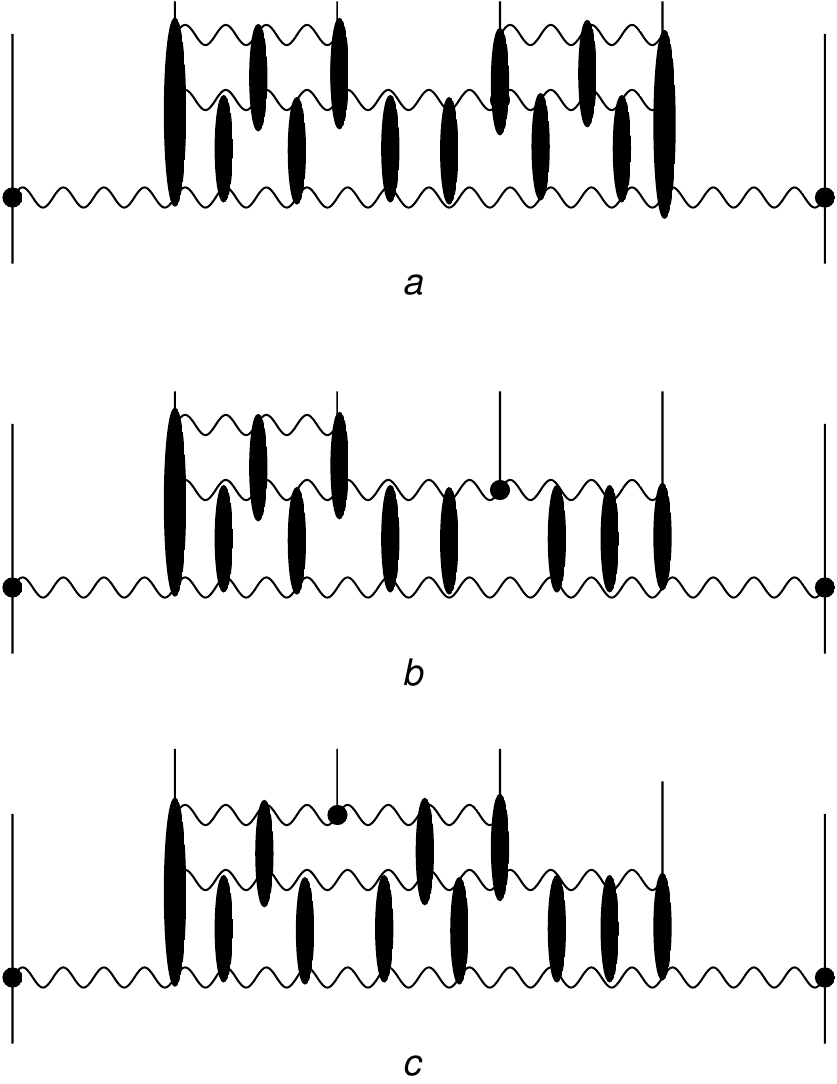,height=6cm,width=9cm}
\caption{Multiple discontinuities: (a)  $\Delta_{12} \Delta_{34}\Delta_{1234}$; (b) $\Delta_{12}\Delta_{1234}$;(c) $\Delta_{123}\Delta_{1234}$}.
\label{fig:multiple-discontinuities}
\end{figure}
We will be brief and list only the main results. Our main interest is devoted to  further contributions to the 
terms $\widetilde{W}_{RL}$, in particular from the very long cut and from the double cut. 
We begin with the triple discontinuity:
\ba
&& \Delta_{12} \Delta_{34}\Delta_{1234}= e^{-i\pi (\omega_1+\omega_5)}\Big[ \int \frac{d \omega'_2  d\omega'_3d \omega'_4}{(2 \pi i)^3}  s_{12}^{\omega'_2}  s_{23}^{\omega'_3}  s_{34}^{\omega'_4} 
e^{i\pi \omega'_3}  \Omega_{2'3'}\Omega_{4'3'} \widetilde{W}_{RL}-\nonumber\\
&&- \int \frac{d \omega'_2  d \omega'_4}{(2 \pi i)^2}  s_{12}^{\omega'_2}    s_{34}^{\omega'_4} \frac{\Omega_{2'3}\Omega_{4'3} }{\Omega_3}W_{24}\nonumber\\
&&
-\int   \frac{d \omega'_2}{2 \pi i}  s_{12}^{\omega'_2} \Omega_{2'3} W_{\omega_2} \frac{V_L(c) V_R(d)}{\Omega_4}  - \frac{V_L(a) V_R(b)}{\Omega_2} \int \frac{d \omega'_4}{2 \pi i}  s_{34}^{\omega'_4}
\Omega_{4'3} W_{\omega_4}\nonumber\\
&&-\frac{V_L(a) V_R(b) \Omega_3 V_L(c) V_R(d)}{\Omega_2 \Omega_4} \Big].
\ea
Inserting the subtractions we arrive at the regular expression:
\ba
\label{d12d34d1234-finite}
&& \Delta_{12} \Delta_{34}\Delta_{1234}=
e^{-i\pi (\omega_1+\omega_5)}\Big[ \int \frac{d \omega'_2  d\omega'_3d \omega'_4}{(2 \pi i)^3}  s_{12}^{\omega'_2}  s_{23}^{\omega'_3}  s_{34}^{\omega'_4} 
e^{i\pi \omega'_3}  \Omega_{2'3'}\Omega_{4'3'} \widetilde{W}_{RL}\nonumber\\
&&-\int   \frac{d \omega'_2}{2 \pi i}  s_{12}^{\omega'_2} \Omega_{2'3}  W_{\omega_2}^{reg}  e^{i\pi \omega_c} V_R(d)-  V_L(a) e^{i\pi \omega_b}  \int \frac{d \omega'_4}{2 \pi i}  s_{34}^{\omega'_4}\Omega_{4'3}  W_{\omega_4}^{reg}
 \nonumber\\
&&+ \int \frac{d \omega'_2  d \omega'_4}{(2 \pi i)^2}  s_{12}^{\omega'_2}    s_{34}^{\omega'_4}\left( \sin \pi (\omega'_2+\omega'_4-\omega_3) +i e^{i\pi \omega_3} \Omega_{2'} \Omega_{4‘} \right)\nonumber\\
&&\hspace{2cm} \cdot \left(W_{\omega_2}^{reg} - \frac{1}{2}\sin \pi (\omega_a-\omega_b)\right) \left(W_{\omega_4}^{reg} - \frac{1}{2}\sin \pi (\omega_d-\omega_c)\right)
\nonumber\\
&&-\frac{1}{2} \left( e^{i\pi \omega_a} V_R(b) V_R(c) V_R(d)+V_L(a) V_L(b) V_L(c) e^{i\pi \omega_d} \right) \Big].
\ea
Restricting ourselves to the leading order we find
\ba
\label{d12d34d1234-LL}
&&\Delta_{12} \Delta_{34}\Delta_{1234}=
\int \frac{d \omega'_2  d\omega'_3d \omega'_4}{(2 \pi i)^3}  s_{12}^{\omega'_2}  s_{23}^{\omega'_3}  s_{34}^{\omega'_4} \Omega_{2'3}\Omega_{4'3} \widetilde{W}_{RL}^{reg}\nonumber\\
&&
-\int   \frac{d \omega'_2}{2 \pi i}  s_{12}^{\omega'_2}\Omega_{2'3}  W_{\omega_2}^{reg} V_R(d)-V_L(a) \int   \frac{d \omega'_4}{2 \pi i}  s_{34}^{\omega'_4}\Omega_{4'3}     W_{\omega_4}^{reg}
 \nonumber\\
&&+\int \frac{d \omega'_2  d \omega'_4}{(2 \pi i)^2}  s_{12}^{\omega'_2}    s_{34}^{\omega'_4}\ \pi (\omega'_2+\omega'_4-\omega_3) 
 \left(W_{\omega_2}^{reg} - \frac{1}{2} \pi (\omega_a-\omega_b)\right) \left(W_{\omega_4}^{reg} - \frac{1}{2} \pi (\omega_d-\omega_c)\right)
\nonumber\\
&&-\frac{1}{2} \left(  V_R(b) V_R(c) V_R(d)+V_L(a) V_L(b) V_L(c)  \right) \,, 
\ea
where, for simplicity, we have not expanded  $\Omega_{23}$, and $V_L(a)$ etc. This triple discontinuity allows to determine $ \widetilde{W}_{RL}^{reg }$.

Next we consider the double discontinuity $\Delta_{12}\Delta_{1224}$:
\ba
&&\Delta_{12}\Delta_{1224}=e^{-i\pi (\omega_1+\omega_5)} \cdot
\nonumber\\
&&\cdot \Big[ -\int \frac{d \omega'_2  d\omega'_3d \omega'_4}{(2 \pi i)^3}  s_{12}^{\omega'_2}  s_{23}^{\omega'_3}  s_{34}^{\omega'_4}    \Omega_{2'3'}\left( 
e^{i\pi(2 \omega'_3 -\omega'_4)}\widetilde{W}_{RL}+e^{i\pi(2 \omega'_4 -\omega'_3)} \widetilde{W}_{RR}\right)
\nonumber\\
 &&+\int \frac{d \omega'_2  d \omega'_4}{(2 \pi i)^2}  s_{12}^{\omega'_2}    s_{34}^{\omega'_4} e^{i\pi( \omega_3 -\omega'_4)} \frac{\Omega_{2'3}}{\Omega_3}W_{\omega_2\omega_4}
+\int  \frac{d \omega'_2  d\omega'_3}{(2 \pi i)^2}  s_{12}^{\omega'_2}  s_{23}^{\omega'_3}  e^{i\pi( \omega_4 -\omega'_3)} \Omega_{2'3'}\widetilde{W}_{\omega_2\omega_3;R} e^{i\pi \omega_d}\nonumber\\
&&+\frac{V_L(a)V_R(b)}{\Omega_2}  \int  \frac{d \omega'_4}{2 \pi i}  s_{34}^{\omega'_4}
e^{i\pi( \omega_3 -\omega'_4)} W_{\omega_4}
- 2i \int   \frac{d \omega'_2}{2 \pi i}  s_{12}^{\omega'_2}W_{\omega_2} \Omega_{2'3} \frac{V_L(c) V_R(d)}{\Omega_4}\nonumber\\
&&+\frac{V_L(a) V_R(b)  V_R(d)}{\Omega_2\Omega_4}\left( e^{i\pi( \omega_3 +\omega_4)} \Omega_c -2i e^{i\pi \omega_c} \Omega_3 \Omega_4\right)
\Big]\,.
\ea
After inserting the subtractions this becomes:
\ba
\label{d12d1234-finite}
&&\Delta_{12}\Delta_{1224}=\nonumber\\
&&e^{-i\pi (\omega_1+\omega_5)} \Big[ 
-\int \frac{d \omega'_2  d\omega'_3d \omega'_4}{(2 \pi i)^3}  s_{12}^{\omega'_2}  s_{23}^{\omega'_3}  s_{34}^{\omega'_4}\Omega_{2'3'}\left(  e^{i\pi(2 \omega'_3 -\omega'_4)} \widetilde{W}_{RL}^{reg}+ e^{i\pi(2 \omega'_4 -\omega'_3)} \widetilde{W}_{RR}^{reg}\right) \nonumber\\
&&+\int  \frac{d \omega'_2  d\omega'_3}{(2 \pi i)^2}  s_{12}^{\omega'_2}  s_{23}^{\omega'_3} 
e^{i\pi( \omega_4 -\omega'_3)} \Omega_{2'3'}\widetilde{W}_{\omega_2\omega_3;R}^{reg} e^{i\pi \omega_d}\nonumber\\
&&\hspace{1cm}-2i\int   \frac{d \omega'_2}{2 \pi i}  s_{12}^{\omega'_2} \Omega_{23}W_{\omega_2}^{reg} e^{i\pi \omega_c} V_R(d)+ V_L(a) e^{i\pi \omega_b} \int  \frac{d \omega'_4}{2 \pi i}  s_{34}^{\omega'_4}
e^{i\pi(\omega_3-\omega'_4)}  W_{\omega_4}^{reg}\nonumber\\
&&\hspace{1cm}-  \int \frac{d \omega'_2  d \omega'_4}{(2 \pi i)^2}  s_{12}^{\omega'_2}    s_{34}^{\omega'_4}
(\cos \pi \omega'_2 e^{i\pi(\omega_3-\omega'_4)}-2i \Omega_{2'} \cos \pi \omega_3 e^{i\pi \omega'_4})\nonumber\\
&&\hspace{3cm} \cdot \left(W_{\omega_2}^{reg} - \frac{1}{2}\sin \pi (\omega_a-\omega_b)\right) \left(W_{\omega_4}^{reg} - \frac{1}{2}\sin \pi (\omega_d-\omega_c)\right)\nonumber\\
&&\hspace{1cm}-i\left( e^{i\pi \omega_a} V_R(b) V_R(c) V_R(d)+V_L(a) V_L(b) V_L(c) e^{i\pi \omega_d} \right) \Big].
\ea
To leading order this equals:
\ba
\label{d12d1234-LL}
&&\Delta_{12}\Delta_{1234}
=-  \int \frac{d \omega'_2  d\omega'_3d \omega'_4}{(2 \pi i)^3}  s_{12}^{\omega'_2}  s_{23}^{\omega'_3}  s_{34}^{\omega'_4} \Omega_{2'3'} \left( \widetilde{W}_{RL}^{reg}+\widetilde{W}_{RR}^{reg}\right) \nonumber\\ 
&&+ \int  \frac{d \omega'_2  d\omega'_3}{(2 \pi i)^2}  s_{12}^{\omega'_2}  s_{23}^{\omega'_3} \Omega_{2'3'}\widetilde{W}_{\omega_2\omega_3;R}^{reg}
+  V_L(a)  \int  \frac{d \omega'_4}{2 \pi i}  s_{34}^{\omega'_4}W_{\omega_4}^{reg}\nonumber\\
&&- \left(\int  \frac{d \omega'_2}{2 \pi i}  s_{12}^{\omega'_2} W_{\omega_2}^{reg} - \frac{1}{2} \pi (\omega_a-\omega_b)\right) \left( \int \frac{d \omega'_4}{2 \pi i}  s_{34}^{\omega'_4}W_{\omega_4}^{reg} - \frac{1}{2} \pi (\omega_d-\omega_c)\right)\,.
\ea
This double discontinuity allows to determine the sum of $ \widetilde{W}_{RL}^{reg }+ \widetilde{W}_{RR}^{reg }$. By symmetry arguments we can also find the double discontinuity $\Delta_{34}\Delta_{1224}$ which detemines the sum of $ \widetilde{W}_{RL}^{reg }$ and 
$ \widetilde{W}_{LL}^{reg }$.

Rather than going through the leading order calculations of the unitarity integral illustrated in Fig.\ref{fig:multiple-discontinuities}b
we only quote a few results. First we note that, because of the bootstrap property of the 
two gluon cut in the octet representation,  in both double discontinuities the three reggeon cuts collapse into two reggeon states (in contrast to the double discontinuity in Fig.\ref{fig:double-disc(1)} b). As a result, both double discontinuities contain the very long cut on which we will concentrate first. Correspondingly, also on the rhs of eqs (\ref{d12d34d1234-LL}) and (\ref{d12d1234-LL}) we ignore all terms other than those of the very long cut. We thus find for $ \widetilde{W}_{RL}^{reg }$:
\ba
\label{W_RL-full}
&& \int \frac{d \omega'_2  d\omega'_3d \omega'_4}{(2 \pi i)^3}  s_{12}^{\omega'_2}  s_{23}^{\omega'_3}  s_{34}^{\omega'_4}  \widetilde{W}_{RL}^{reg}= \\
&& \int \frac{d \omega'_2  d\omega'_3d \omega'_4}{(2 \pi i)^3}  s_{12}^{\omega'_2}  s_{23}^{\omega'_3}  s_{34}^{\omega'_4} 
\frac{ \pi^2 \omega'_2 \omega'_4 f_{\omega_2\omega_3\omega_4}^{int}
-\pi \omega'_2 f_{\omega_2\omega_3\omega_4}^{{c}} -\pi \omega'_4 f_{\omega_2\omega_3\omega_4}^{{b}} + f_{\omega_2\omega_3\omega_4}^{{bc}}}{\pi^2 (\omega'_2-\omega'_3)(\omega'_4-\omega'_3)} + ... \,,\nonumber
\ea
where the rhs of this equation is obtained from the unitarity integral of Fig.\ref{fig:multiple-discontinuities}a.
The use of (\ref{d12d1234-LL}) and of the corresponding equation for the double discontinuity $\Delta_{12}\Delta_{1234}$ lead to 
\ba
 &&\int \frac{d \omega'_2  d\omega'_3d \omega'_4}{(2 \pi i)^3}  s_{12}^{\omega'_2}  s_{23}^{\omega'_3}  s_{34}^{\omega'_4} \label{W_RR-full}
\widetilde{W}_{RR}^{reg}=\\
&&-  \int \frac{d \omega'_2  d\omega'_3d \omega'_4}{(2 \pi i)^3}  s_{12}^{\omega'_2}  s_{23}^{\omega'_3}  s_{34}^{\omega'_4} \frac{ \pi^2 \omega'_2 \omega'_3 f_{\omega_2\omega_3\omega_4}^{int}
-\pi \omega'_2 f_{\omega_2\omega_3\omega_4}^{{c}} -\pi \omega'_3 f_{\omega_2\omega_3\omega_4}^{{b}} + f_{\omega_2\omega_3\omega_4}^{{bc}}}{\pi^2 (\omega'_2-\omega'_3)(\omega'_4-\omega'_3)}+... \nonumber
\ea
and 
\ba
\label{W_LL-full}
&& \int \frac{d \omega'_2  d\omega'_3d \omega'_4}{(2 \pi i)^3}  s_{12}^{\omega'_2}  s_{23}^{\omega'_3}  s_{34}^{\omega'_4} \widetilde{W}_{LL}^{reg}= \\
&&-  \int \frac{d \omega'_2  d\omega'_3d \omega'_4}{(2 \pi i)^3}  s_{12}^{\omega'_2}  s_{23}^{\omega'_3}  s_{34}^{\omega'_4} \frac{ \pi^2 \omega'_4 \omega'_3 f_{\omega_2\omega_3\omega_4}
-\pi \omega'_4 f_{\omega_2\omega_3\omega_4}^{{b}} -\pi \omega'_3 f_{\omega_2\omega_3\omega_4}^{{c}} + f_{\omega_2\omega_3\omega_4}^{{bc}}}{\pi^2 (\omega'_2-\omega'_3)(\omega'_4-\omega'_3)}+... \,\,.\nonumber
\ea
Finally, from the single discontinuity  $\Delta_{1234}$ we have an expression for the sum
$\widetilde{W}_{LL}^{reg}+\widetilde{W}_{RL}^{reg}+\widetilde{W}_{LR}^{reg}+\widetilde{W}_{RR}^{reg}$.
From this we derive for $ \widetilde{W}_{LR}^{reg }$:
\ba
\label{W_LR-full}
&& \int \frac{d \omega'_2  d\omega'_3d \omega'_4}{(2 \pi i)^3}  s_{12}^{\omega'_2}  s_{23}^{\omega'_3}  s_{34}^{\omega'_4} \widetilde{W}_{LR}^{reg}=\\
&& \int \frac{d \omega'_2  d\omega'_3d \omega'_4}{(2 \pi i)^3}  s_{12}^{\omega'_2}  s_{23}^{\omega'_3}  s_{34}^{\omega'_4}  \frac{ \pi^2 {\omega'_3}^2  f_{\omega_2\omega_3\omega_4}^{int}
-\pi \omega'_3 f_{\omega_2\omega_3\omega_4}^{{c}} -\pi \omega'_3 f_{\omega_2\omega_3\omega_4}^{{b}} + f_{\omega_2\omega_3\omega_4}^{{bc}}}{\pi^2 (\omega'_2-\omega'_3)(\omega'_4-\omega'_3)}+... \,\,.\nonumber
\ea

As an example we illustrate the result for $  \int \frac{d \omega'_2  d\omega'_3d \omega'_4}{(2 \pi i)^3}  s_{12}^{\omega'_2}  s_{23}^{\omega'_3}  s_{34}^{\omega'_4} \widetilde{W}_{RR}^{reg}$:
 \begin{figure}[H]
\centering
\epsfig{file=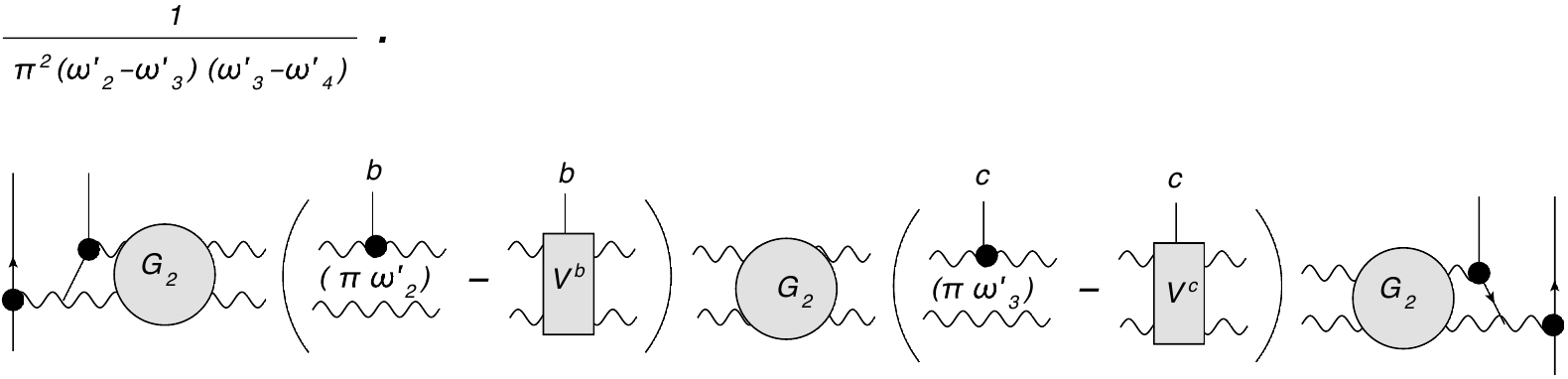,scale=0.9}
\caption{Illustration of $\widetilde{W}_{RR}^{reg}$: factorization of the production vertices}
\label{fig:WtildeRR}
\end{figure}
\noindent
We remind that the vertices denoted be $V^b$ and $V^c$ are identical to those found for the long cut (cf. Fig.\ref{fig:prod-vertex}). Comparing with Fig.\ref{fig:W23R} we recognize the factorization of the two production vertices for the particles $b$ and $c$.  

However, beyond this factorization of  the product of vertices  $b$ and $c$, there is another  new feature contained in the unitarity integral, to be more precise, in the centre of Fig.\ref{fig:WtildeRR}. Going back to Figs. \ref{fig:prod-vertex} and \ref{fig:W23R} and using the bootstrap property of the two reggeon state, one sees that Fig.\ref{fig:WtildeRR} contains a contribution in which the 2-reggeon Green's function between the produced particles $b$ and $c$ collapses into a single reggeon, i.e. 
Fig.\ref{fig:WtildeRR} contains the double cut product. To be more precise, one can show that, on the rhs of 
(\ref{W_RL-full}) -(\ref{W_LR-full}), the three last terms contribute as follows:
\ba
&&f_{\omega_2\omega_3\omega_4}^{bc} \to - \pi \omega'_3 W_{\omega_2}^{reg} W_{\omega_4}^{reg} \nonumber\\
&&f_{\omega_2\omega_3\omega_4}^{b} \to - W_{\omega_2}^{reg} W_{\omega_4}^{reg} \nonumber\\
&&f_{\omega_2\omega_3\omega_4}^{c} \to -W_{\omega_2}^{reg} W_{\omega_4}^{reg} .
\ea
This implies that the unitarity integal contains:
\ba
\label{WRL-doublecutpiece}
\text{rhs of (\ref{W_RL-full})}& \to&  \int \frac{d \omega'_2  d \omega'_4}{(2 \pi i)^2}  s_{12}^{\omega'_2}    s_{34}^{\omega'_4} \frac{\pi (\omega'_2+\omega'_4-\omega_3)}{\pi^2 (\omega'_2-\omega_3)(\omega'_4-\omega_3)}W_{\omega_2}^{reg} W_{\omega_4}^{reg}\\
\label{WRR-doublecutpiece}
\text{rhs of (\ref{W_RR-full})}& \to&   \int \frac{d \omega'_2  d \omega'_4}{(2 \pi i)^2}  s_{12}^{\omega'_2}   s_{34}^{\omega'_4}\frac{-\pi \omega'_2}{\pi^2 (\omega'_2-\omega_3)(\omega'_4-\omega_3)}W_{\omega_2}^{reg} W_{\omega_4}^{reg}\\
\label{WLL-doublecutpiece}
\text{rhs of (\ref{W_LL-full})}& \to&  \int \frac{d \omega'_2  d \omega'_4}{(2 \pi i)^2}  s_{12}^{\omega'_2}    s_{34}^{\omega'_4} \frac{-\pi \omega'_4}{\pi^2 (\omega'_2-\omega_3)(\omega'_4-\omega_3)}W_{\omega_2}^{reg} W_{\omega_4}^{reg}\\
\label{WLR-doublecutpiece}
\text{rhs of (\ref{W_LR-full})}& \to&  \int \frac{d \omega'_2  d \omega'_4}{(2 \pi i)^2}  s_{12}^{\omega'_2}    s_{34}^{\omega'_4} \frac{\pi \omega_3}{\pi^2 (\omega'_2-\omega_3)(\omega'_4-\omega_3)}W_{\omega_2}^{reg} W_{\omega_4}^{reg}:
\ea

This has to be compared with the remaining terms on the rhs of the discontinuity equations, eqs. (\ref{d12d34d1234-LL} and (\ref{d12d1234-LL}).
Proceeding in the same way as before  we find  that  $ \widetilde{W}_{RL}^{reg }$, in addition to the
long cut pieces in (\ref{W_RL-full}) which lead to (\ref{WRL-doublecutpiece}), must also contain
\ba
&&\int_{24} \widetilde{W}_{RL}^{reg}=\\&&... - \int_{24} \frac{ \pi (\omega'_2+\omega'_4-\omega_3)}{\pi^2( \omega_{2'}-\omega_3)(\omega_{4'}-\omega_3)} \left(W_{\omega_2}^{reg} - \frac{1}{2} \pi (\omega_a-\omega_b)\right) \left(W_{\omega_4}^{reg} - \frac{1}{2} \pi (\omega_d-\omega_c)\right)+...\,\,.\nonumber
\ea
Similarly:
\ba
&&\int_{24} \widetilde{W}_{RR}^{reg}=\\
&&...  \int_{24}  \frac{ \pi \omega'_2
}{\pi^2( \omega_{2'}-\omega_3)(\omega_{4'}-\omega_3)} \left(W_{\omega_2}^{reg} - \frac{1}{2} \pi (\omega_a-\omega_b)\right) \left(W_{\omega_4}^{reg} - \frac{1}{2} \pi (\omega_d-\omega_c)\right)+...\,\,.\nonumber
\ea\ba
&&\int_{24}  \widetilde{W}_{LL}^{reg}=\\
&&...   \int_{24} \frac{ \pi \omega'_4}{\pi^2( \omega_{2'}-\omega_3)(\omega_{4'}-\omega_3)} \left(W_{\omega_2}^{reg} - \frac{1}{2} \pi (\omega_a-\omega_b)\right) \left(W_{\omega_4}^{reg} - \frac{1}{2} \pi (\omega_d-\omega_c)\right)+...\,\,.\nonumber
\ea
\ba
&&\int_{24}  \widetilde{W}_{LR}^{reg}=\\
&&... - \int_{24}\frac{ \pi \omega_3)}{\pi^2( \omega_{2'}-\omega_3)(\omega_{4'}-\omega_3)} \left(W_{\omega_2}^{reg} - \frac{1}{2} \pi (\omega_a-\omega_b)\right) \left(W_{\omega_4}^{reg} - \frac{1}{2} \pi (\omega_d-\omega_c)\right)+...\,\,.\nonumber
\ea
where we have used the abbreviation
\be
\int_{24}=\int \frac{d \omega'_2  d \omega'_4}{(2 \pi i)^2}  s_{12}^{\omega'_2}    s_{34}^{\omega'_4}\,.
\ee
It is easy to see that, for each partial wave $\widetilde{W}_{RR}^{reg}$, the double cut products $W_{\omega_2}^{reg} W_{\omega_4}^{reg}$ cancel exactly. As a result, the partial waves $\widetilde{W}_{RR}^{reg}$ are free from double cut contributions..

As an important further check we mention that all these results are consistent with the double dicontinuities $\Delta_{123}\Delta_{1234}$ and  $\Delta_{234}\Delta_{1234}$.

The results of this (and the previous) section allow to find all terms of the order $i\pi $ and of the order $(i \pi)^2$ of the scattering amplitude. However, in our final formulae, for simplicity,  we will be complete only with all terms proportional to $i\pi$ whereas, among the real terms proportional to $(i \pi)^2$, we limit ourselves to the double cut and the 3-regggeon cut. All other 
terms of the order $(i \pi)^2$ are proportional to short or long cut contributions and will not be listed.

\section{The leading order amplitudes in different regions}

With these results we now return to section 6.2. In order to compare with the one loop results of the BDS formula we need  the BDS phases (\ref{interest})
which have been 
discussed in \cite{Bartels:2013jna}. There it has been derived  that for each kinematic region (characterized by a product of $\tau$ factors $\tau_i...\tau_k$) 
the BDS formula predicts a phase factor composed of two pieces:
\be 
\label{BDSphases}
e^{i( \pi \varphi_{i...k} +\delta_{i...k})}
\ee
where the phases $\delta_{i...k}$ are conformal invariant.
For the $2 \to 6$ amplitude a complete list of these phase factors is given in Appendix A. From this list we derive:
\ba
\delta_{15}-\delta_{25}&=&-\delta_{125}\nonumber\\
\delta_{15}-\delta_{45}&=&-\delta_{145}\nonumber\\
\delta_{15}-\delta_{13}-\delta_{35}&=&-\delta_{135}\nonumber\\
\delta_{15}-\delta_{25}-\delta_{13}+\delta_{35}&=&\delta_{1235}\nonumber\\
\delta_{15}-\delta_{14}-\delta_{35}+\delta_{13}&=&\delta_{1345}\nonumber\\
\delta_{15}-\delta_{14}-\delta_{25}+\delta_{24}&=&\delta_{1245}\nonumber\\
\delta_{15}-\delta_{14}-\delta_{35}+\delta_{13}+\delta_{13}+\delta_{35}&=&-\delta_{12345}.
\ea
Using these identities we find the following leading order expessions:
\ba
\label{amp15regLO}
\tau_1\tau_5:&& e^{-i\pi(\omega_2+\omega_3+\omega_4)} 
 \Big[ e^{i\pi (\omega_b+\omega_c)} \cos \pi(\omega_a-\omega_d)\nonumber\\
&&+2i  \Big[  f_{\omega_2\omega_3\omega_4}+\frac{ \delta_{15}}{2} \nonumber\\ 
&& - i \left( f_{\omega_2}+\frac{ \delta_{13}}{2} -\frac{ \pi (\omega_a-\omega_b)}{2}\right)
\left( f_{\omega_4} + \frac{ \delta_{35}}{2}-\frac{ \pi (\omega_d-\omega_c)}{2}  \right) \big] \Big]
 \\
\label{amp125regLO}
\tau_1\tau_2\tau_5:&& e^{-i\pi(\omega_3+\omega_4)} \Big[ -e^{i\pi (\omega_c+\omega_d)} \cos \pi(\omega_a-\omega_b) \nonumber\\
&&+2i\big[ f_{\omega_2\omega_3\omega_4}   - f_{\omega_3\omega_4}-\frac{ \delta_{125}}{2}
 \nonumber\\
&&-i \left( f_{\omega_2}+\frac{ \delta_{13}}{2} -\frac{ \pi (\omega_a-\omega_b)}{2}\right)
\left( f_{\omega_4} + \frac{ \delta_{35}}{2}-\frac{ \pi (\omega_d-\omega_c)}{2}  \right)  \big]\Big]
\\
\label{amp145regLO}
\tau_1\tau_4\tau_5:&& e^{-i\pi(\omega_2+\omega_3)} \Big[- e^{i\pi (\omega_a+\omega_b)}\cos \pi(\omega_d-\omega_c)\nonumber\\
&&
+2i\big[ f_{\omega_2\omega_3\omega_4}  - f_{\omega_2\omega_3}-\frac{ \delta_{145}}{2}
\nonumber\\
&&-i\left( f_{\omega_2}+\frac{ \delta_{13}}{2} -\frac{ \pi (\omega_a-\omega_b)}{2}\right)
\left( f_{\omega_4} + \frac{ \delta_{35}}{2}-\frac{ \pi (\omega_d-\omega_c)}{2}  \right) \big]\Big]
\ea
\ba
\label{amp135regLO}
\tau_1\tau_3\tau_5:&& e^{-i\pi(\omega_2+\omega_4)} \Big[- e^{i\pi (\omega_a+\omega_d)}\cos \pi(\omega_b-\omega_c)\nonumber\\
&&+2i\big[ f_{\omega_2\omega_3\omega_4} - f_{\omega_2} - f_{\omega_4}-\frac{ \delta_{135}}{2}
\nonumber\\
&&
- i\left( f_{\omega_2}+\frac{ \delta_{13}}{2} -\frac{ \pi (\omega_a-\omega_b)}{2}\right)
\left( f_{\omega_4} + \frac{ \delta_{35}}{2}-\frac{ \pi (\omega_d-\omega_c)}{2}  \right)\big] \Big]
\\
\label{amp1345regLO}
\tau_1\tau_3\tau_4\tau_5:&& e^{-i\pi\omega_2} \Big[ e^{i\pi (\omega_c-\omega_d)} \cos \pi(\omega_b-\omega_a)
\nonumber\\
&&
+2i \big[ f_{\omega_2\omega_3\omega_4} - f_{\omega_2\omega_3}+
 f_{\omega_2} - f_{\omega_4}+\frac{ \delta_{1345}}{2}
\nonumber\\
&&- i\left( f_{\omega_2}+\frac{ \delta_{13}}{2} -\frac{ \pi (\omega_a-\omega_b)}{2}\right)
\left( f_{\omega_4} + \frac{\delta_{35}}{2}-\frac{ \pi (\omega_d-\omega_c)}{2}  \right)
\big] \Big]
\\
\label{amp1235regLO}
\tau_1\tau_2\tau_3\tau_5:&& e^{-i\pi\omega_4} \Big[ e^{i\pi (\omega_b-\omega_a)}\cos \pi(\omega_c-\omega_d) \nonumber\\
&&+2i\big( f_{\omega_2\omega_3\omega_4}   - f_{\omega_3\omega_4}
 -f_{\omega_2} + f_{\omega_4}+\frac{ \delta_{1235}}{2}
\nonumber\\
&&- i \left( f_{\omega_2}+\frac{ \delta_{13}}{2} -\frac{ \pi (\omega_a-\omega_b)}{2}\right)
\left( f_{\omega_4} + \frac{ \delta_{35}}{2}-\frac{ \pi (\omega_d-\omega_c)}{2}  \right)
\Big]\Big]
\ea
\ba
\label{amp1245regLO}
\tau_1\tau_2\tau_4\tau_5:&& e^{-i\pi \omega_3} \Big[ e^{i\pi(\omega_b+\omega_c-\omega_a-\omega_d)}
\nonumber\\
&&
+2i\Big[ f_{\omega_2\omega_3\omega_4}
 - f_{\omega_2\omega_3}- f_{\omega_3\omega_4} + f_{\omega_3}+\frac{ \delta_{1245}}{2}
 \nonumber\\
&&
-i 
\left( f_{\omega_2}+\frac{ \delta_{13}}{2} -\frac{ \pi (\omega_a-\omega_b)}{2}\right)
\left( f_{\omega_4} + \frac{ \delta_{35}}{2}-\frac{ \pi (\omega_d-\omega_c)}{2}  \right)\nonumber\\ 
&&+2i \left( f_{\text{3 reggeon}}+{\text{two loop}} \right)  \Big] \Big]
\ea
\ba
\label{amp12345regLO}
\tau_1\tau_2\tau_3\tau_4\tau_5:&& -\cos \pi (\omega_b+\omega_d-\omega_a-\omega_c)\nonumber\\
&&+2i\Big[ f_{\omega_2\omega_3\omega_4}
 - f_{\omega_2\omega_3}- f_{\omega_3\omega_4} + f_{\omega_2}+ f_{\omega_3}+ f_{\omega_4}-\frac{ \delta_{12345}}{2} \nonumber\\
&&
-i\left( f_{\omega_2}+\frac{ \delta_{13}}{2} -\frac{ \pi (\omega_a-\omega_b)}{2}\right)
\left( f_{\omega_4} + \frac{ \delta_{35}}{2}-\frac{ \pi (\omega_d-\omega_c)}{2}  \right)\nonumber\\
&&-2i\left( f_{\text{3 reggeon}}+{\text{two loop}} \right)
\Big] \Big]\, .
\ea
One easily verifies that the one loop terms on the rhs coincide with the lowest order expansion of the phases (\ref{BDSphases}).

The expressions (\ref{amp15regLO}) - (\ref{amp12345regLO}) represent the main results of this paper.

\section{Summary and outlook}

In this paper we have studied in $N=4$ SYM the production process $2 \to 6$ in the multiregge limit in the planar approximation. 
As the main resultr we have calculated, in the leading logarithmic approximation, 
the novel Regge cut contributions: the  product of 
two short 2-reggeon cuts known from $2 \to 4$ process, and the Regge cut consisting of three reggeized gluons. The latter is of special interest for studying the integrable BFKL spin chain:
whereas in former studies only the shortest chain consisting of two sites had been found, the $2 fo \to 6$ process for the first time also exhibits the longer chain corresponding to three sites.

Technically speaking, the investigation described in this paper,\\
 (i) supports the analytic structure (Steinmann relations) on which the analysis of the scattering amplitude in the multiregge limit has been based. The amplitude as written as a sum of 
integrals  which are characterized by the maximal number of non-overlapping energy discontinuities\footnote{For $2 \to 2+n$ the number of terms is  given by the Catalan numbers 
$C_{n+1}$ with $C_n=1,1,2,5,14,42,...$ for $n=0,1.2,...$ and  $C_{n+1}=\sum_{i=0}^{i=n} C_i C_{n-i}$.}. For the calculation of the novel Regge cuts we had to make use of  double (and even triple) energy discontinuities: this provides an even more stringent  test of the ansatz. \\ 
(ii) confirms that the singularities which are a feature of the planar approximation, can be cancelled by adding subtraction terms to the angular momentum integrals of the Regge cut contributions. For the $2\to6$ amplitude 
the calculation of these subtractions has been a challenging task, much more demanding than for the simpler $2 \to4$ and $2 \to5$ processes. In order to generalize to higher order processes we definitely need a deeper understanding of the structure of these subtraction terms.

A  very important task will be the comparison of our results with those obtained with other methods  \cite{Bargheer:2015djt,DelDuca:2018hrv,DelDuca:2018raq,Caron-Huot:2019vjl,Bargheer:2019lic,DelDuca:2019tur}.
Although some of our final results are restricted to the leading order (leading log approximation), the obtained analytic structure of the scattering amplitides is expected to be valid to all orders. We believe that this information 
will be valuable also for higher loop resuits computed by novel methods.

\begin{figure}[H]
\centering
\epsfig{file=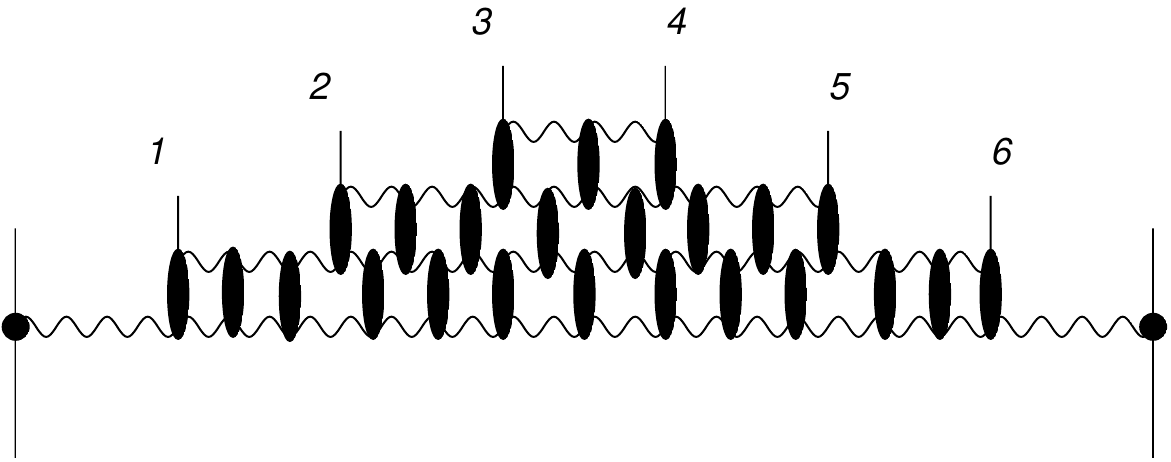,scale=0.7}
\caption{Regge cut composed of 4 reggeized gluons}.
\label{fig:4-reggeoncut}
\end{figure}

\begin{figure}[H]
\centering
\epsfig{file=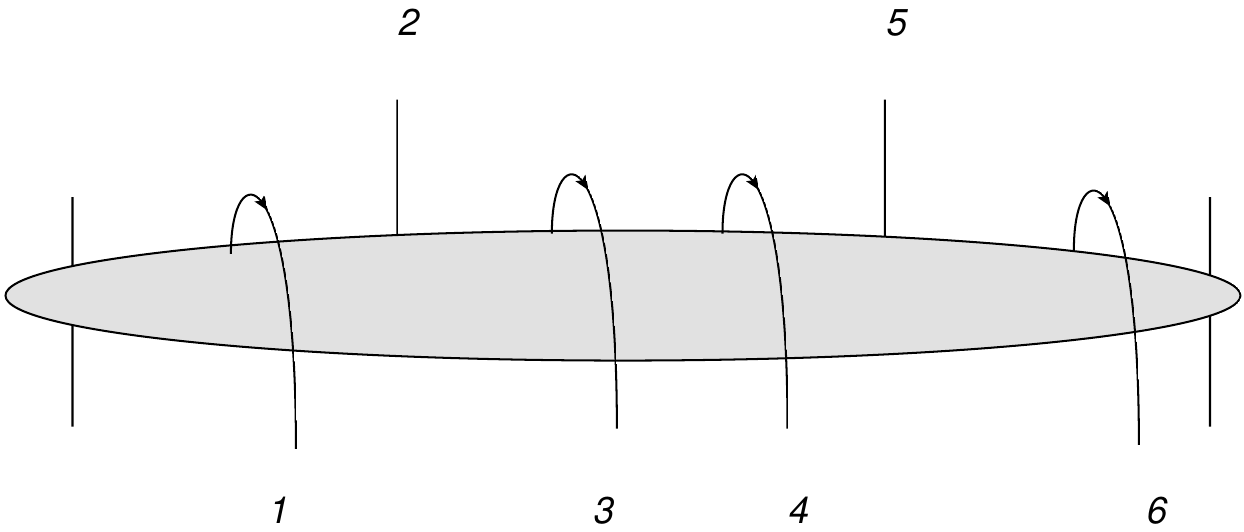,scale=0.7}
\caption{Kinematic region where the four-reggeon cut appears}.
\label{fig:4-reggeoncut-kinematics}
\end{figure}

Based upon the results obained in this (and previous) papers, a few general statemants can be made about higher order amplitudes. First, for the appearance of Regge cuts with  $n$ reggeized gluons we need the process $2 \to 2n$ (Fig.\ref{fig:4-reggeoncut}). 
This contribution is expected to show up in the kinematic region where the energy of the produced particles have alternating signs (Fig.\ref{fig:4-reggeoncut-kinematics}). 
\noindent
For a Regge cut composed of n reggeized gluons the leading order BFKL kernel will have the form:
\ba
&&K^{n;planar)}(\bk_1,...,\bk_n;\bk'_1,...,\bk'_n)= -\frac{a}{2} \ln \frac{\bk_1^2 \bk_n^2}{\bq^2 \bq^2} \nonumber\\
&&+K^{(1)}(\bk_1,\bk_2;\bk'_1,\bk'_2) + ...+K^{(1)}(\bk_{n-1},\bk_n;\bk'_{n-1},\bk'_n)
\ea
with $\bq=\bk_1+...+\bk_n$. As indicated in Fig.\ref{fig:double-disc(1)}b, beyond leading order there will be interactions between more than two reggeized gluons \cite{Bartels:2012sw}.

\begin{figure}[H]
\centering
\epsfig{file=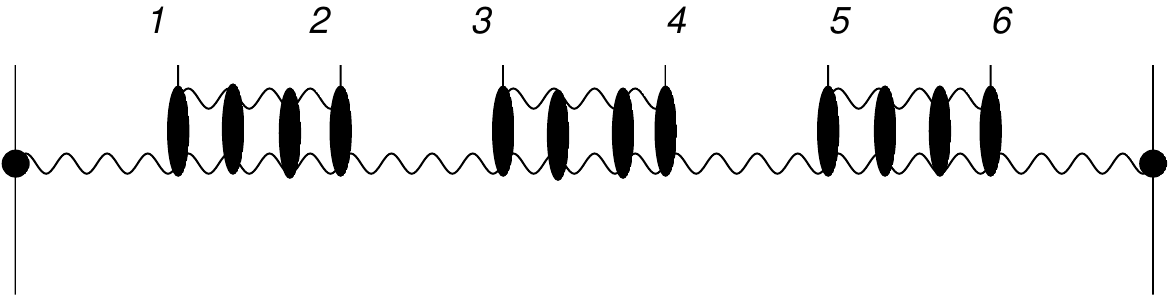,scale=0.7}
\caption{Factorization of the short two-reggeon cut}.
\label{fig:Factorization}
\end{figure}

Second, starting with $2 \to 6$ we expect to have multiple products of Regge cuts. For example, for $2 \to 8$ we expect the triple product of the shortest cut (Fig.\ref{fig:Factorization})
It  will be interesting to see
whether these products are consistent with exponentiation. \\  \\
{\bf Acknowledgments:}\\ I wish to express my deep gratitude to Lev Nikolaevich Lipatov who initiated this paper and provided substantial contributions. I also thank A.Kormilitzin for his valuable help. For valuable discussions I want to thank T.Bargheer, G.Papathanasiou, and A.Sabio Vera.

%%%%%%%%%%%%%%%%%%%%%%%%%%%%
\newpage

\appendix
\section{BDS phases of the $2 \to 6$ amplitude}

As explained in \cite{Bartels:2013jna} for the $2 \to 5$ amplitude, there are phase factors
related to Regge pole terms and to the $Li_2$ functions contained in the BDS amplitude (\ref{interest}). Their derivation has been described in detail in \cite{Bartels:2013jna}, and in the following we present a list for those different kinematic regions of the $2 \to 6$ amplitude which contain Regge cuts. We list the phases $e^{i \pi  \varphi_{\tau_i,..,\tau_j}}$ and $ \delta_{\tau_i...\tau_j} $ using the notation of \cite{Bartels:2013jna}:
\ba
\tau_1\tau_3:&e^{-i\pi(\omega_2+\omega_4+\omega_5)} e^{i\pi\omega_c}e^{i\pi \omega_d} e^{i\delta_{13}},&\,\,\delta_{13} = \pi \frac{\gamma_K}{4} \ln \frac{|q_1||q_3||k_a||k_b|}{|k_a+k_b|^2|q_2|^2}\nonumber\\
\tau_1\tau_2\tau_3:&e^{-i\pi(\omega_4+\omega_5)} e^{i\pi\omega_c}e^{i\pi \omega_d} e^{i\delta_{123}},&\,\,\delta_{123}=-\delta_{13}\nonumber\\ 
\tau_2\tau_4:&e^{-i\pi(\omega_1+\omega_3+\omega_5)} e^{i\pi\omega_a}e^{i\pi \omega_d} e^{i\delta_{24}},&\,\,\delta_{24} = \pi \frac{\gamma_K}{4} \ln \frac{|q_2||q_4||k_b||k_c|}{|k_b+k_c|^2|q_3|^2}\nonumber\\
\tau_2\tau_3\tau_4:&e^{-i\pi(\omega_1+\omega_5)}e^{i\pi\omega_c}e^{i\pi \omega_d} e^{i\delta_{234}},&\,\,\delta_{234}=-\delta_{24}\nonumber\\
\tau_3\tau_5:&e^{-i\pi(\omega_2+\omega_4)}e^{i\pi\omega_a}e^{i\pi \omega_b} e^{i\delta_{35}},&\,\,\delta_{35} = \pi \frac{\gamma_K}{4} \ln \frac{|q_3||q_5||k_c||k_d|}{|k_c+k_d|^2|q_4|^2}\nonumber\\
\tau_3\tau_4\tau_5:&e^{-i\pi(\omega_1+\omega_2)}e^{i\pi\omega_a}e^{i\pi \omega_b} e^{i\delta_{345}},&\,\,\delta_{345}=-\delta_{35}\nonumber\\
\tau_1\tau_4:&e^{-i\pi(\omega_2+\omega_3+\omega_5)} e^{i\pi \omega_b}e^{i\pi \omega_d} e^{i\delta_{14}},&\,\,\delta_{14}= \pi \frac{\gamma_K}{4} \ln \frac{|q_1||q_4||k_a||k_c|}{|k_a+k_b+k_c|^2|q_2||q_3|}\nonumber\\
\tau_1\tau_2\tau_4:&e^{-i\pi(\omega_3+\omega_5)} e^{i\pi \omega_c}e^{i\pi \omega_d} e^{i\delta_{124}},&\,\,
\delta_{124}= \pi \frac{\gamma_K}{4} \ln \frac{|k_a+k_b+k_c|^2|q_2|^2 |k_b|}{|k_b+k_c|^2|q_1||q_3||k_a|}\nonumber\\
\tau_1\tau_3\tau_4:&e^{-i\pi(\omega_2+\omega_5)} e^{i\pi \omega_a}e^{i\pi \omega_d} e^{i\delta_{134}},&\,\,
\delta_{134}= \pi \frac{\gamma_K}{4} \ln \frac{|k_a+k_b+k_c|^2|q_3|^2 |k_b|}{|k_a+k_b|^2|q_2||q_4||k_c|}\nonumber\\
\tau_1\tau_2\tau_3\tau_4:&e^{-i\pi\omega_5} e^{-i\pi \omega_b}e^{i\pi \omega_d} e^{i\delta_{1234}},&\,\,\delta_{1234}= \pi \frac{\gamma_K}{4} \ln \frac{|k_a+k_b|^2|k_b+k_c|^2 |q_1||q_4|}{|k_a+k_b+k_c|^2|k_a||k_c||q_2||q_3|}\nonumber\\
\tau_1\tau_5:&e^{-i\pi(\omega_2+\omega_3+\omega_4)} e^{i\pi(\omega_b+\omega_c)} e^{i\delta_{15}},&\,\,\delta_{15} = \pi \frac{\gamma_K}{4} \ln \frac{|q_1||q_5||k_a||k_d|}{|k_a+k_b+k_c+k_d|^2|q_2||q_4|}\nonumber\\
\tau_1\tau_2\tau_5:&e^{-i\pi(\omega_3+\omega_4)}e^{i\pi(\omega_c+\omega_d)} e^{i\delta_{125}},&\,\,\delta_{125}= \pi \frac{\gamma_K}{4} \ln \frac{|k_a+k_b+k_c+k_d|^2 |q_2|^2
|k_b|}{|k_b+k_c+k_d|^2|q_1| |q_3|  |k_a|}
\nonumber\\
\tau_1\tau_4\tau_5:&e^{-i\pi(\omega_2+\omega_3)}e^{i\pi(\omega_a+\omega_b)} e^{i\delta_{145}},&\,\,\delta_{145}= \pi \frac{\gamma_K}{4} \ln \frac{|k_a+k_b+k_c+k_d|^2 |q_4|^2|k_c|}{|k_a+k_b+k_c|^2 |q_3||q_5||k_d|}\nonumber\\
\tau_1\tau_3\tau_5:&e^{-i(\pi\omega_2+\omega_4)} e^{i\pi(\omega_a+\omega_d)} e^{i\delta_{135}},&\,\,\delta_{135} = \pi \frac{\gamma_K}{4} \ln \frac{|k_a+k_b+k_c+k_d|^2|q_3|^2 |k_b||k_c|
}{|k_a+k_b|^2|k_c+k_d|^2|q_2||q_4|}\nonumber\\
\tau_1\tau_3\tau_4\tau_5:&e^{-i\pi\omega_2} e^{i\pi(\omega_c-\omega_d)} e^{i\delta_{1345}},&\nonumber\\
&&\hspace{-2cm}\delta_{1345} = \pi \frac{\gamma_K}{4} 
\ln \frac{|k_a+k_b+k_c|^2|k_c+k_d|^2 |k_a| |k_b| |q_1| | q_3|}
{|k_a+k_b+k_c+k_d|^2|k_a+k_b|^2 |k_c|^2 |q_2|^2}\nonumber\\
\tau_1\tau_2\tau_3\tau_5:&e^{-i\pi\omega_4} e^{i\pi(\omega_b-\omega_a)} e^{i\delta_{1235}},\nonumber\\
&&\hspace{-2cm}\delta_{1235} = \pi \frac{\gamma_K}{4} 
\ln \frac{|k_b+k_c+k_d|^2|k_a+k_b|^2 |k_c| |k_d| |q_5| | q_3|}
{|k_a+k_b+k_c+k_d|^2|k_c+k_c|^2 |k_b|^2 |q_4|^2}\nonumber
\ea
\ba
\tau_1\tau_2\tau_4\tau_5:&e^{-i\pi\omega_3} e^{i\pi(-\omega_a+\omega_b+\omega_c-\omega_d)} e^{i\delta_{1245}},&\nonumber\\
&&\hspace{-6cm}\delta_{1245} = \pi \frac{\gamma_K}{4} \ln \frac{|k_a+k_b+k_c|^2|k_b+k_c+k_d|^2}{|k_a+k_b+k_c+k_d|^2|k_b+k_c|^2}\nonumber\\
\tau_1\tau_2\tau_3\tau_4\tau_5:&  e^{i\delta_{12345}},&\nonumber\\
&&\hspace{-6cm}\delta_{12345} = \pi \frac{\gamma_K}{4} \ln 
\frac{|k_a+k_b+k_c+k_d|^2|q_2|^2|q_4|^2}{ |q_1||q_3|^2 |q_5||k_a||k_b|k_c||k_d|}
\frac{|k_a+k_b|^2|k_b+k_c|^2|k_c+k_d|^2}
{|k_a+k_b+k_c|^2|k_b+k_c+k_d|^2}\nonumber\\
\ea
%\newpage
%%%%%%%%%%%%%%%%%%%%%%%%%%%%%%%%%%%%%%%%%%%%%%%%%%%%%%%%%%%%%%%%%%% 
\section{Trigonometric factors for the $2\to 6$ amplitude}
\subsection{Regge pole contributions}
In this appendix we list the trigonometric prefactors for the Regge pole contributions. The Regge poles contribute to all
42 partial waves. We  present the different groups. 
\subsubsection{Five singlets: LLLL,RRLL,RRRR,RRRL,RLLL}
\begin{eqnarray}
F_{LLLL}^{Pole}\;&=&\;\frac{V_L(a)}{\Omega_{21}}\frac{V_L(b)}{\Omega_{32}}\frac{V_L(c)}{\Omega_{43}}\frac{V_L(d)}{\Omega_{54}}\nonumber\\
F_{RRLL}^{Pole}\;&=&\;\frac{V_R(a)}{\Omega_{12}}\frac{V_R(b)}{\Omega_{23}}\frac{V_L(c)}{\Omega_{43}}\frac{V_L(d)}{\Omega_{54}}\nonumber\\
F_{RRRR}^{Pole}\;&=&\;\frac{V_R(a)}{\Omega_{12}}\frac{V_R(b)}{\Omega_{23}}\frac{V_R(c)}{\Omega_{34}}\frac{V_R(d)}{\Omega_{45}}\nonumber\\
F_{RRRL}^{Pole}\;&=&\;\frac{V_R(a)}{\Omega_{12}}\frac{V_R(b)}{\Omega_{23}}\frac{V_R(c)}{\Omega_{34}}\frac{V_L(d)}{\Omega_{54}}\nonumber\\
F_{RLLL}^{Pole}\;&=&\;\frac{V_R(a)}{\Omega_{12}}\frac{V_L(b)}{\Omega_{32}}\frac{V_L(c)}{\Omega_{43}}\frac{V_L(d)}{\Omega_{54}}
\end{eqnarray}

\subsubsection{Three doublets: LRLL, RLRL, RRLR}

\begin{eqnarray}
F_{LRLL(1)}^{Pole}\;&=&\;\frac{\Omega_{3}}{\Omega_{2}}\frac{\Omega_{21}}{\Omega_{31}}\;\;\;\frac{V_L(a)}{\Omega_{21}}\frac{V_R(b)}{\Omega_{23}}\frac{V_L(c)}{\Omega_{43}}\frac{V_L(d)}{\Omega_{54}}\nonumber\\
F_{LRLL(2)}^{Pole}\;&=&\;\frac{\Omega_{1}}{\Omega_{2}}\frac{\Omega_{23}}{\Omega_{13}}\;\;\;\frac{V_L(a)}{\Omega_{21}}\frac{V_R(b)}{\Omega_{23}}\frac{V_L(c)}{\Omega_{43}}\frac{V_L(d)}{\Omega_{54}}
\end{eqnarray}
and
\begin{eqnarray}
F_{RLRL(1)}^{Pole}\;&=&\;\frac{\Omega_{4}}{\Omega_{3}}\frac{\Omega_{32}}{\Omega_{42}}\;\;\;\frac{V_R(a)}{\Omega_{12}}\frac{V_L(b)}{\Omega_{32}}\frac{V_R(c)}{\Omega_{34}}\frac{V_L(d)}{\Omega_{54}}\nonumber\\
F_{RLRL(2)}^{Pole}\;&=&\;\frac{\Omega_{2}}{\Omega_{3}}\frac{\Omega_{34}}{\Omega_{24}}\;\;\;\frac{V_R(a)}{\Omega_{12}}\frac{V_L(b)}{\Omega_{32}}\frac{V_R(c)}{\Omega_{34}}\frac{V_L(d)}{\Omega_{54}}
\end{eqnarray}
and
\begin{eqnarray}
F_{RRLR(1)}^{Pole}\;&=&\;\frac{\Omega_{5}}{\Omega_{4}}\frac{\Omega_{43}}{\Omega_{53}}\;\;\;\frac{V_R(a)}{\Omega_{12}}\frac{V_R(b)}{\Omega_{23}}\frac{V_L(c)}{\Omega_{43}}\frac{V_R(d)}{\Omega_{45}}\nonumber\\
F_{RRLR(2)}^{Pole}\;&=&\;\frac{\Omega_{3}}{\Omega_{4}}\frac{\Omega_{45}}{\Omega_{35}}\;\;\;\frac{V_R(a)}{\Omega_{12}}\frac{V_R(b)}{\Omega_{23}}\frac{V_L(c)}{\Omega_{43}}\frac{V_R(d)}{\Omega_{45}}
\end{eqnarray}

\subsubsection{Four triplets: LLRL, LRRL, RLLR, RLRR}

\begin{eqnarray}
F_{LLRL(1)}^{Pole}\;&=&\;\frac{\Omega_{4}}{\Omega_{3}}\frac{\Omega_{32}}{\Omega_{42}}\;\;\;\;\;\;\;\;\;\frac{V_L(a)}{\Omega_{21}}\frac{V_L(b)}{\Omega_{32}}\frac{V_R(c)}{\Omega_{34}}\frac{V_L(d)}{\Omega_{54}}\nonumber\\
F_{LLRL(2)}^{Pole}\;&=&\;\frac{\Omega_{4}}{\Omega_{3}}\frac{\Omega_{21}}{\Omega_{41}}\frac{\Omega_{34}}{\Omega_{24}}\;\;\;\frac{V_L(a)}{\Omega_{21}}\frac{V_L(b)}{\Omega_{32}}\frac{V_R(c)}{\Omega_{34}}\frac{V_L(d)}{\Omega_{54}}\nonumber\\
F_{LLRL(3)}^{Pole}\;&=&\;\frac{\Omega_{1}}{\Omega_{3}}\frac{\Omega_{34}}{\Omega_{14}}\;\;\;\;\;\;\;\;\;\frac{V_L(a)}{\Omega_{21}}\frac{V_L(b)}{\Omega_{32}}\frac{V_R(c)}{\Omega_{34}}\frac{V_L(d)}{\Omega_{54}}
\end{eqnarray}

\begin{eqnarray}
F_{LRRL(1)}^{Pole}\;&=&\;\frac{\Omega_{1}}{\Omega_{2}}\frac{\Omega_{23}}{\Omega_{13}}\;\;\;\;\;\;\;\;\;\frac{V_L(a)}{\Omega_{21}}\frac{V_R(b)}{\Omega_{23}}\frac{V_R(c)}{\Omega_{34}}\frac{V_L(d)}{\Omega_{54}}\nonumber\\
F_{LRRL(2)}^{Pole}\;&=&\;\frac{\Omega_{4}}{\Omega_{2}}\frac{\Omega_{21}}{\Omega_{41}}\;\;\;\;\;\;\;\;\;\frac{V_L(a)}{\Omega_{21}}\frac{V_R(b)}{\Omega_{23}}\frac{V_R(c)}{\Omega_{34}}\frac{V_L(d)}{\Omega_{54}}\nonumber\\
F_{LRRL(3)}^{Pole}\;&=&\;\frac{\Omega_{1}}{\Omega_{2}}\frac{\Omega_{34}}{\Omega_{14}}\frac{\Omega_{21}}{\Omega_{31}}
\;\;\;\frac{V_L(a)}{\Omega_{21}}\frac{V_R(b)}{\Omega_{23}}\frac{V_R(c)}{\Omega_{34}}\frac{V_L(d)}{\Omega_{54}}\
\end{eqnarray}

\begin{eqnarray}
F_{RLLR(1)}^{Pole}\;&=&\;\frac{\Omega_{5}}{\Omega_{4}}\frac{\Omega_{43}}{\Omega_{53}}\;\;\;\;\;\;\;\;\;\frac{V_R(a)}{\Omega_{12}}\frac{V_L(b)}{\Omega_{32}}\frac{V_L(c)}{\Omega_{43}}\frac{V_R(d)}{\Omega_{45}}\nonumber\\
F_{RLLR(2)}^{Pole}\;&=&\;\frac{\Omega_{2}}{\Omega_{4}}\frac{\Omega_{45}}{\Omega_{25}}\;\;\;\;\;\;\;\;\;\frac{V_R(a)}{\Omega_{12}}\frac{V_L(b)}{\Omega_{32}}\frac{V_L(c)}{\Omega_{43}}\frac{V_R(d)}{\Omega_{45}}\nonumber\\
F_{RLLR(3)}^{Pole}\;&=&\;\frac{\Omega_{5}}{\Omega_{4}}\frac{\Omega_{32}}{\Omega_{52}}\frac{\Omega_{45}}{\Omega_{35}}\;\;\;\frac{V_R(a)}{\Omega_{12}}\frac{V_L(b)}{\Omega_{32}}\frac{V_L(c)}{\Omega_{43}}\frac{V_R(d)}{\Omega_{45}}
\end{eqnarray}

\begin{eqnarray}
F_{RLRR(1)}^{Pole}\;&=&\;\frac{\Omega_{2}}{\Omega_{3}}\frac{\Omega_{34}}{\Omega_{24}}\;\;\;\;\;\;\;\;\;\frac{V_R(a)}{\Omega_{12}}\frac{V_L(b)}{\Omega_{32}}\frac{V_R(c)}{\Omega_{34}}\frac{V_R(d)}{\Omega_{45}}\nonumber\\
F_{RLRR(2)}^{Pole}\;&=&\;\frac{\Omega_{2}}{\Omega_{3}}\frac{\Omega_{45}}{\Omega_{25}}\frac{\Omega_{32}}{\Omega_{42}}\;\;\;
\frac{V_R(a)}{\Omega_{12}}\frac{V_L(b)}{\Omega_{32}}\frac{V_R(c)}{\Omega_{34}}\frac{V_R(d)}{\Omega_{45}}\nonumber\\
F_{RLRR(3)}^{Pole}\;&=&\;\frac{\Omega_{5}}{\Omega_{3}}\frac{\Omega_{32}}{\Omega_{52}}\;\;\;\;\;\;\;\;\;\frac{V_R(a)}{\Omega_{12}}\frac{V_L(b)}{\Omega_{32}}\frac{V_R(c)}{\Omega_{34}}\frac{V_R(d)}{\Omega_{45}}
\end{eqnarray}

\subsubsection{Two quartets: LLLR, LRRR}

\begin{eqnarray}
F_{LLLR(1)}^{Pole}\;&=&\;\frac{\Omega_{5}}{\Omega_{4}}\frac{\Omega_{43}}{\Omega_{53}}\;\;\;\;\;\;\;\;\;\frac{V_L(a)}{\Omega_{21}}\frac{V_L(b)}{\Omega_{32}}\frac{V_L(c)}{\Omega_{43}}\frac{V_R(d)}{\Omega_{45}}\nonumber\\
F_{LLLR(2)}^{Pole}\;&=&\;\frac{\Omega_{5}}{\Omega_{4}}\frac{\Omega_{32}}{\Omega_{52}}\frac{\Omega_{45}}{\Omega_{35}}\;\;\;\frac{V_L(a)}{\Omega_{21}}\frac{V_L(b)}{\Omega_{32}}\frac{V_L(c)}{\Omega_{43}}\frac{V_R(d)}{\Omega_{45}}\nonumber\\
F_{LLLR(3)}^{Pole}\;&=&\;\frac{\Omega_{5}}{\Omega_{4}}\frac{\Omega_{21}}{\Omega_{51}}\frac{\Omega_{45}}{\Omega_{25}}\;\;\;\frac{V_L(a)}{\Omega_{21}}\frac{V_L(b)}{\Omega_{32}}\frac{V_L(c)}{\Omega_{43}}\frac{V_R(d)}{\Omega_{45}}\nonumber\\
F_{LLLR(4)}^{Pole}\;&=&\;\frac{\Omega_{1}}{\Omega_{4}}\frac{\Omega_{45}}{\Omega_{15}}\;\;\;\;\;\;\;\;\;\frac{V_L(a)}{\Omega_{21}}\frac{V_L(b)}{\Omega_{32}}\frac{V_L(c)}{\Omega_{43}}\frac{V_R(d)}{\Omega_{45}}
\end{eqnarray}

\begin{eqnarray}
F_{LRRR(1)}^{Pole}\;&=&\;\frac{\Omega_{1}}{\Omega_{2}}\frac{\Omega_{23}}{\Omega_{13}}\;\;\;\;\;\;\;\;\;\frac{V_L(a)}{\Omega_{21}}\frac{V_R(b)}{\Omega_{23}}\frac{V_R(c)}{\Omega_{34}}\frac{V_R(d)}{\Omega_{45}}\nonumber\\
F_{LRRR(2)}^{Pole}\;&=&\;\frac{\Omega_{1}}{\Omega_{2}}\frac{\Omega_{34}}{\Omega_{14}}\frac{\Omega_{21}}{\Omega_{31}}\;\;\;\frac{V_L(a)}{\Omega_{21}}\frac{V_R(b)}{\Omega_{23}}\frac{V_R(c)}{\Omega_{34}}\frac{V_R(d)}{\Omega_{45}}\nonumber\\
F_{LRRR(3)}^{Pole}\;&=&\;\frac{\Omega_{1}}{\Omega_{2}}\frac{\Omega_{45}}{\Omega_{15}}\frac{\Omega_{21}}{\Omega_{41}}\;\;\;\frac{V_L(a)}{\Omega_{21}}\frac{V_R(b)}{\Omega_{23}}\frac{V_R(c)}{\Omega_{34}}\frac{V_R(d)}{\Omega_{45}}\nonumber\\
F_{LRRR(4)}^{Pole}\;&=&\;\frac{\Omega_{5}}{\Omega_{2}}\frac{\Omega_{21}}{\Omega_{51}}\;\;\;\;\;\;\;\;\;\frac{V_L(a)}{\Omega_{21}}\frac{V_R(b)}{\Omega_{23}}\frac{V_R(c)}{\Omega_{34}}\frac{V_R(d)}{\Omega_{45}}
\end{eqnarray}

\subsubsection{A quintet: LRLR}

\begin{eqnarray}
F_{LRLR(1)}^{Pole}\;&=&\;\frac{\Omega_{3}}{\Omega_{2}}\frac{\Omega_{5}}{\Omega_{4}}\frac{\Omega_{21}}{\Omega_{31}}\frac{\Omega_{43}}{\Omega_{53}}\;\;\;\frac{V_L(a)}{\Omega_{21}}\frac{V_R(b)}{\Omega_{23}}\frac{V_L(c)}{\Omega_{43}}\frac{V_R(d)}{\Omega_{45}}\nonumber\\
F_{LRLR(2)}^{Pole}\;&=&\;\frac{\Omega_{1}}{\Omega_{2}}\frac{\Omega_{5}}{\Omega_{4}}\frac{\Omega_{23}}{\Omega_{13}}\frac{\Omega_{43}}{\Omega_{53}}\;\;\;\frac{V_L(a)}{\Omega_{21}}\frac{V_R(b)}{\Omega_{23}}\frac{V_L(c)}{\Omega_{43}}\frac{V_R(d)}{\Omega_{45}}\nonumber\\
F_{LRLR(3)}^{Pole}\;&=&\;\frac{\Omega_{1}}{\Omega_{2}}\frac{\Omega_{3}}{\Omega_{4}}\frac{\Omega_{23}}{\Omega_{13}}\frac{\Omega_{45}}{\Omega_{35}}\;\;\;\frac{V_L(a)}{\Omega_{21}}\frac{V_R(b)}{\Omega_{23}}\frac{V_L(c)}{\Omega_{43}}\frac{V_R(d)}{\Omega_{45}}\nonumber\\
F_{LRLR(4)}^{Pole}\;&=&\;\frac{\Omega_{3}}{\Omega_{2}}\frac{\Omega_{5}}{\Omega_{4}}\frac{\Omega_{21}}{\Omega_{51}}\frac{\Omega_{45}}{\Omega_{35}}\;\;\;\frac{V_L(a)}{\Omega_{21}}\frac{V_R(b)}{\Omega_{23}}\frac{V_L(c)}{\Omega_{43}}\frac{V_R(d)}{\Omega_{45}}\nonumber\\
F_{LRLR(5)}^{Pole}\;&=&\;\frac{\Omega_{1}}{\Omega_{2}}\frac{\Omega_{3}}{\Omega_{4}}\frac{\Omega_{21}}{\Omega_{31}}\frac{\Omega_{45}}{\Omega_{15}}\;\;\;\frac{V_L(a)}{\Omega_{21}}\frac{V_R(b)}{\Omega_{23}}\frac{V_L(c)}{\Omega_{43}}\frac{V_R(d)}{\Omega_{45}}
\end{eqnarray}

\subsubsection{A sextet: LLRR}

\begin{eqnarray}
\label{LLRLpole}
F_{LLRR(1)}^{Pole}\;&=&\;\frac{\Omega_{5}}{\Omega_{3}}\frac{\Omega_{32}}{\Omega_{52}}\;\;\;\;\;\;\;\;\;\;\;\;\;\;\;\;\frac{V_L(a)}{\Omega_{21}}\frac{V_L(b)}{\Omega_{32}}\frac{V_R(c)}{\Omega_{34}}\frac{V_R(d)}{\Omega_{45}}\nonumber\\
F_{LLRR(2)}^{Pole}\;&=&\;\frac{\Omega_{1}}{\Omega_{3}}\frac{\Omega_{34}}{\Omega_{14}}\;\;\;\;\;\;\;\;\;\;\;\;\;\;\;\;\frac{V_L(a)}{\Omega_{21}}\frac{V_L(b)}{\Omega_{32}}\frac{V_R(c)}{\Omega_{34}}\frac{V_R(d)}{\Omega_{45}}\nonumber\\
F_{LLRR(3)}^{Pole}\;&=&\;\frac{\Omega_{5}}{\Omega_{3}}\frac{\Omega_{21}}{\Omega_{51}}\frac{\Omega_{32}}{\Omega_{42}}\frac{\Omega_{45}}{\Omega_{25}}\;\;\;\frac{V_L(a)}{\Omega_{21}}\frac{V_L(b)}{\Omega_{32}}\frac{V_R(c)}{\Omega_{34}}\frac{V_R(d)}{\Omega_{45}}\nonumber\\
F_{LLRR(4)}^{Pole}\;&=&\;\frac{\Omega_{1}}{\Omega_{3}}\frac{\Omega_{21}}{\Omega_{41}}\frac{\Omega_{34}}{\Omega_{24}}\frac{\Omega_{45}}{\Omega_{15}}\;\;\;\frac{V_L(a)}{\Omega_{21}}\frac{V_L(b)}{\Omega_{32}}\frac{V_R(c)}{\Omega_{34}}\frac{V_R(d)}{\Omega_{45}}\nonumber\\
F_{LLRR(5)}^{Pole}\;&=&\;\frac{\Omega_{5}}{\Omega_{3}}\frac{\Omega_{21}}{\Omega_{51}}\frac{\Omega_{34}}{\Omega_{24}}\;\;\;\;\;\;\;\;\;\;\frac{V_L(a)}{\Omega_{21}}\frac{V_L(b)}{\Omega_{32}}\frac{V_R(c)}{\Omega_{34}}\frac{V_R(d)}{\Omega_{45}}\nonumber\\
F_{LLRR(6)}^{Pole}\;&=&\;\frac{\Omega_{1}}{\Omega_{3}}\frac{\Omega_{32}}{\Omega_{42}}\frac{\Omega_{45}}{\Omega_{15}}\;\;\;\;\;\;\;\;\;\;\frac{V_L(a)}{\Omega_{21}}\frac{V_L(b)}{\Omega_{32}}\frac{V_R(c)}{\Omega_{34}}\frac{V_R(d)}{\Omega_{45}}
\end{eqnarray}

\subsection{Regge cut contributions}
In the following we list the trigonometric factors of Regge cut contributions in the partial waves. 
Let us begin with a general remark. When making an ansatz for a Regge cut contribution,
 we initially should allow for independent contributions in each partial wave. As an example (see eq.(C.1) below), the contribution of the Regge cut in $\omega_2$ inside $F_{LRLL(1)}$ could be different from the one inside $F_{LRLL(2)}$, i.e. our ansatz should allow for 
$W_{\omega_2;(1)}\ne W_{\omega_2;(2)}$. However, from the requirement that in the region of positive energies this cut contribution must cancel in the sum of the two partial waves we conclude that the two Regge cut functions must be equal: $W_{\omega_2;(1)}=W_{\omega_2;(2)}=W_{\omega_2}$. In the following we will make repeated use of this argument to simplify our discussion.  For simplicity we use un-primed  $\omega_i$  variables; when inserting the Regge cut terms into the integral (\ref{struct-LLLL}) we have to switch to 
the primed variables $\omega'_i$.

\subsubsection{Short Regge cut in the $t_2$ channel}
The short Regge cut in the $t_2$ channel contributes to the doublet $LRLL(i)\, (i=1,2)$, to the triplet
$LRRL(i)\,(i=1,2,3)$, to the quartet $LRRR(i)\,(i=1,...,4)$, and to the quintet $LRLR(i)\,(i=1,...,5)$. The results are:
\ba
\label{short2-1}
F_{LRLL(1)}^{\text{short cut};\omega_2}&=&\frac{W_{\omega_2}}{\Omega_{31}} \frac{V_L(c)}{\Omega_{43}} \frac{V_L(d)}{\Omega_{54}} \nonumber\\
F_{LRLL(2)}^{\text{short cut};\omega_2}&=&\frac{W_{\omega_2}}{\Omega_{13}} \frac{V_L(c)}{\Omega_{43}} \frac{V_L(d)}{\Omega_{54}}\\ \nonumber\\
\label{short2-2}
F_{LRRL(1)}^{\text{short cut};\omega_2}&=&\hspace{1.5cm} \frac{W_{\omega_2}}{\Omega_{13}} \frac{V_R(c)}{\Omega_{34}} \frac{V_L(d)}{\Omega_{54}}\nonumber\\
F_{LRRL(2)}^{\text{short cut};\omega_2}&=& \frac{\Omega_4}{\Omega_3} \frac{\Omega_{31}}{\Omega_{41}}  \;\;\;   \frac{W_{\omega_2}}{\Omega_{31}} \frac{V_R(c)}{\Omega_{34}} \frac{V_L(d)}{\Omega_{54}} \nonumber\\
F_{LRRL(3)}^{\text{short cut};\omega_2}&=& \frac{\Omega_1}{\Omega_3} \frac{\Omega_{34}}{\Omega_{14}}  \;\;\;  \frac{W_{\omega_2}}{\Omega_{31}} \frac{V_R(c)}{\Omega_{34}} \frac{V_L(d)}{\Omega_{54}} \\ \nonumber\\
\label{short2-3}
F_{LRRR(1)}^{\text{short cut};\omega_2}&=&\hspace{2.1cm} \frac{W_{\omega_2}}{\Omega_{13}}\, \frac{V_R(c)}{\Omega_{34}}\, \frac{V_R(d)}{\Omega_{45}} \nonumber\\
F_{LRRR(2)}^{\text{short cut};\omega_2}&=& \frac{\Omega_1}{\Omega_3} \frac{\Omega_{34}}{\Omega_{14}} \hspace{0.9cm}\frac{W_{\omega_2}}{\Omega_{31}} \frac{V_R(c)}{\Omega_{34}}\, \frac{V_R(d)}{\Omega_{45}} \nonumber\\
F_{LRRR(3)}^{\text{short cut};\omega_2}&=&\frac{\Omega_1}{\Omega_3} \frac{\Omega_{13}\Omega_{45}}{\Omega_{15}\Omega_{14}}\;\;\;\frac{W_{\omega_2}}{\Omega_{31}} \frac{V_R(c)}{\Omega_{34}} \frac{V_R(d)}{\Omega_{45}}\nonumber\\
F_{LRRR(4)}^{\text{short cut};\omega_2}&=&\frac{\Omega_5}{\Omega_3} \frac{\Omega_{31}}{\Omega_{51}}\hspace{0.9cm}\frac{W_{\omega_2}}{\Omega_{31}} \frac{V_R(c)}{\Omega_{34}} \frac{V_R(d)}{\Omega_{45}}\\ \nonumber \\ 
\label{short2-4}
F_{LRLR(1)}^{\text{short cut};\omega_2}&=& \frac{\Omega_5}{\Omega_4} \frac{\Omega_{34}}{\Omega_{35}} \hspace{1cm} \frac{W_{\omega_2}}{\Omega_{31}} \frac{V_L(c)}{\Omega_{43}} \frac{V_R(d)}{\Omega_{45}}\nonumber\\
F_{LRLR(2)}^{\text{short cut};\omega_2}&=& \frac{\Omega_5}{\Omega_4} \frac{\Omega_{34}}{\Omega_{35}} \hspace{1cm}\frac{W_{\omega_2}}{\Omega_{13}}\frac{V_L(c)}{\Omega_{43}} \frac{V_R(d)}{\Omega_{45}}\nonumber\\
F_{LRLR(3)}^{\text{short cut};\omega_2}&=&  \frac{\Omega_3}{\Omega_4} \frac{\Omega_{45}}{\Omega_{35}} \hspace{1cm}\frac{W_{\omega_2}}{\Omega_{13}}\frac{V_L(c)}{\Omega_{43}} \frac{V_R(d)}{\Omega_{45}}\nonumber\\
F_{LRLR(4)}^{\text{short cut};\omega_2}&=&  \frac{\Omega_5}{\Omega_4} \frac{\Omega_{45}\Omega_{13}}{\Omega_{51}\Omega_{35}}\;\;\;\; \frac{W_{\omega_2}}{\Omega_{13}}\frac{V_L(c)}{\Omega_{43}} \frac{V_R(d)}{\Omega_{45}} \nonumber\\
F_{LRLR(5)}^{\text{short cut};\omega_2}&=& \frac{\Omega_1}{\Omega_4} \frac{\Omega_{45}}{\Omega_{15}} \hspace{1cm}\frac{W_{\omega_2}}{\Omega_{31}}\frac{V_L(c)}{\Omega_{43}} \frac{V_R(d)}{\Omega_{45}}
\ea

\subsubsection{Short Regge cut in the $t_3$ channel}

The short Regge cut in the $t_3$ channel appears in the doublet $RLRL(i),\,(i=1,2)$, 
the two triplets $LLRL(i),\,(i=1,2,3)$ and  $RLRR(i),\,(i=1,2,3)$, and in the sixtet $LLRR(i),\,(i=1,...,6)$. 
One finds:
\ba
\label{short3-1}
F_{RLRL(1)}^{\text{short cut};\omega_3}\;&=& \frac{V_R(a)}{\Omega_{12}}\, \frac{W_{\omega_3}}{\Omega_{42}}\, \frac{V_L(d)} {\Omega_{54}} \nonumber\\
F_{RLRL(2)}^{\text{short cut};\omega_3}\;&=& \frac{V_R(a)}{\Omega_{12}} \,\frac{W_{\omega_3}}{\Omega_{24}}\, \frac{V_L(d)} {\Omega_{54}} \\ \nonumber \\ 
\label{short3-2}
F_{LLRL(1)}^{\text{short cut};\omega_3}\;&=& \hspace{1.5cm}\frac{V_L(a)}{\Omega_{21}} \,\frac{W_{\omega_3}}{\Omega_{42}} \,\frac{V_L(d)} {\Omega_{54}} \nonumber\\
F_{LLRL(2)}^{\text{short cut};\omega_3}\;&=& \frac{\Omega_4}{\Omega_2} \frac{\Omega_{21}}{\Omega_{41}}\,\hspace{0.2cm}\frac{V_L(a)}{\Omega_{21}} \,
\frac{W_{\omega_3}}{\Omega_{24}}\, \frac{V_L(d)} {\Omega_{54}} \nonumber\\
F_{LLRL(3)}^{\text{short cut};\omega_3}\;&=& \frac{\Omega_1}{\Omega_2} \frac{\Omega_{24}}{\Omega_{14}}\hspace{0.3cm}\frac{V_L(a)}{\Omega_{21}} \,\frac{W_{\omega_3}}{\Omega_{24}} \,\frac{V_L(d)} {\Omega_{54}}\\ \nonumber \\
\label{short3-3}
F_{RLRR(1)}^{\text{short cut};\omega_3}\;&=&\hspace{1.4cm} \frac{V_R(a)}{\Omega_{12}} \,\frac{W_{\omega_3}}{\Omega_{24}} \,\frac{V_R(d)} {\Omega_{45}} \nonumber\\
F_{RLRR(2)}^{\text{short cut};\omega_3}\;&=& \frac{\Omega_2}{\Omega_4} \frac{\Omega_{45}}{\Omega_{25}}\;\;\frac{V_R(a)}{\Omega_{12}} \,
\frac{W_{\omega_3}}{\Omega_{42}}\, \frac{V_R(d)} {\Omega_{45}} \nonumber\\
F_{RLRR(3)}^{\text{short cut};\omega_3}\;&=& \frac{\Omega_5}{\Omega_4} \frac{\Omega_{42}}{\Omega_{52}}\;\;\frac{V_R(a)}{\Omega_{12}} \,\frac{W_{\omega_3}}{\Omega_{42}} \,\frac{V_R(d)} {\Omega_{45}}\\\nonumber  \\
\label{short3-4}
F_{LLRR(1)}^{\text{short cut};\omega_3}\;&=&\frac{\Omega_5}{\Omega_4} \frac{\Omega_{24}}{\Omega_{25}}\hspace{0.8cm} \frac{V_L(a)}{\Omega_{21}} \,
\frac{W_{\omega_3}}{\Omega_{42}}\, \frac{V_R(d)} {\Omega_{45}} \nonumber\\
F_{LLRR(2)}^{\text{short cut};\omega_3}\;&=& \frac{\Omega_1}{\Omega_2} \frac{\Omega_{24}}{\Omega_{14}}\hspace{0.8cm}\frac{V_L(a)}{\Omega_{21}} \,
\frac{W_{\omega_3}}{\Omega_{24}}\, \frac{V_R(d)} {\Omega_{45}} \nonumber\\
F_{LLRR(3)}^{\text{short cut};\omega_3}\;&=& \frac{\Omega_5}{\Omega_4} \frac{\Omega_{21}\Omega_{45}}{\Omega_{51}\Omega_{25}}\;\;\frac{V_L(a)}{\Omega_{21}} \,
\frac{W_{\omega_3}}{\Omega_{42}}\, \frac{V_R(d)} {\Omega_{45}} \nonumber\\
F_{LLRR(4)}^{\text{short cut};\omega_3}\;&=&\frac{\Omega_1}{\Omega_2} \frac{\Omega_{21}\Omega_{45}}{\Omega_{41}\Omega_{15}}\;\; \frac{V_L(a)}{\Omega_{21}} \,
\frac{W_{\omega_3}}{\Omega_{24}}\, \frac{V_R(d)} {\Omega_{45}} \nonumber\\
F_{LLRR(5)}^{\text{short cut};\omega_3}\;&=& \frac{\Omega_5}{\Omega_2} \frac{\Omega_{21}}{\Omega_{51}}\hspace{0.8cm}\frac{V_L(a)}{\Omega_{21}} \,
\frac{W_{\omega_3}}{\Omega_{24}}\, \frac{V_R(d)} {\Omega_{45}} \nonumber\\
F_{LLRR(6)}^{\text{short cut};\omega_3}\;&=& \frac{\Omega_1}{\Omega_4} \frac{\Omega_{45}}{\Omega_{15}}\hspace{0.8cm}\frac{V_L(a)}{\Omega_{21}} \,
\frac{W_{\omega_3}}{\Omega_{42}}\, \frac{V_R(d)} {\Omega_{45}} \\ \nonumber
\ea

\subsubsection{Long Regge cut in the $t_2$ and $t_3$ channels}

This cut appears in the triplets $LLRL(i)\,(i=2,3)$ and  $LRRL(i)\,(i=2,3)$, in the quartet $LRRR\,(i=2,3,4)$, and 
in the sextet $LLRR(i)\,(i=2,4,5)$. The form of the partial waves is the following:
\ba
\label{long23-1}
F_{LLRL(2)}^{\text{long cut};\omega_2,\omega_3}&=&\frac{W_{\omega_2\omega_3;L}}{\Omega_{32} \Omega_{41}}\frac{V_L(d)}{\Omega_{54}}\nonumber\\
F_{LLRL(3)}^{\text{long cut};\omega_2,\omega_3}&=&\frac{W_{\omega_2\omega_3;L}}{\Omega_{32} \Omega_{14}}\frac{V_L(d)}{\Omega_{54}} 
\ea
\ba
\label{long23-2}
F_{LRRL(2)}^{\text{long cut};\omega_2,\omega_3}&=&\frac{W_{\omega_2\omega_3;R}}{\Omega_{23} \Omega_{41}}\frac{V_L(d)}{\Omega_{54}}\nonumber\\
F_{LRRL(3)}^{\text{long cut};\omega_2,\omega_3}&=&\frac{W_{\omega_2\omega_3;R}}{\Omega_{23} \Omega_{14}}\frac{V_L(d)}{\Omega_{54}} 
\ea
\ba
\label{long23-3}
F_{LRRR(2)}^{\text{long cut};\omega_2,\omega_3}&=&\hspace{1.4cm}\frac{W_{\omega_2\omega_3;R}}{\Omega_{23} \Omega_{14}}\frac{V_R(d)}{\Omega_{45}}\nonumber\\
F_{LRRR(3)}^{\text{long cut};\omega_2,\omega_3}&=&\frac{\Omega_1}{\Omega_4}\frac{\Omega_{45}}{\Omega_{15}}\;\;
\frac{W_{\omega_2\omega_3;R}}{\Omega_{23} \Omega_{41}}\frac{V_R(d)}{\Omega_{45}}\nonumber\\
F_{LRRR(4)}^{\text{long cut};\omega_2,\omega_3}&=&\frac{\Omega_5}{\Omega_4}\frac{\Omega_{41}}{\Omega_{51}}\;\;\frac{W_{\omega_2\omega_3;R}}{\Omega_{23} \Omega_{41}}\frac{V_R(d)}{\Omega_{45}}
\ea
\ba
\label{long23-4}
F_{LLRR(2)}^{\text{long cut};\omega_2,\omega_3}&=&\hspace{1.4cm}\frac{W_{\omega_2\omega_3;L}}{\Omega_{32} \Omega_{14}}\frac{V_R(d)}{\Omega_{45}}\nonumber\\
F_{LLRR(4)}^{\text{long cut};\omega_2,\omega_3}&=&\frac{\Omega_1}{\Omega_4}\frac{\Omega_{45}}{\Omega_{15}}\;\;\frac{W_{\omega_2\omega_3;L}}{\Omega_{32} \Omega_{41}}\frac{V_R(d)}{\Omega_{45}}\nonumber\\
F_{LLRR(5)}^{\text{long cut};\omega_2,\omega_3}&=&\frac{\Omega_5}{\Omega_4}\frac{\Omega_{41}}{\Omega_{51}}\;\;\frac{W_{\omega_2\omega_3;L}}{\Omega_{32} \Omega_{41}}\frac{V_R(d)}{\Omega_{45}}
\ea

\subsubsection{Regge cuts in the $t_2$ and in the $t_4$  channels} 
This contribution appears in the quintet: $LRLR(i),\, (i=1,...,5)$.
\begin{eqnarray}
F_{LRLR(1)}^{\text{short cuts in }\omega_2, \omega_4}\;&=& \hspace{1.5cm}\frac{W_{\omega_2, \omega_4}}{\Omega_{31}\Omega_{53}} \nonumber\\
F_{LRLR(2)}^{\text{short cuts in }\omega_2, \omega_4}\;&=& \hspace{1.5cm}\frac{W_{\omega_2, \omega_4}}{\Omega_{13}\Omega_{53}}\nonumber\\
F_{LRLR(3)}^{\text{short cuts in }\omega_2, \omega_4}\;&=& \hspace{1.5cm}\frac{W_{\omega_2, \omega_4}}{\Omega_{13}\Omega_{35}}\nonumber\\
F_{LRLR(4)}^{\text{short cuts in }\omega_2, \omega_4}\;&=& 
\frac{\Omega_5}{\Omega_3} \frac{\Omega_{13}}{\Omega_{15}}\;\;
\frac{W_{\omega_2, \omega_4}}{\Omega_{31}\Omega_{35}}\nonumber\\
F_{LRLR(5)}^{\text{short cuts in }\omega_2, \omega_4}\;&=& 
\frac{\Omega_1}{\Omega_3} \frac{\Omega_{35}}{\Omega_{15}} \;\;\frac{W_{\omega_2, \omega_4}}{\Omega_{31}\Omega_{35}}.
\end{eqnarray}

\subsubsection{Very long Regge cut in the  $t_2$, $t_3$, and $t_4$ channels}
This cut contributes to partial waves of the two quartets: $LLLR(i)\, (i=3,4)$,  $LRRR(i)\, (i=3,4)$, of the quintet:
$LRLR(i)\, (i=4,5)$, and of the sixtet $LLRR(i)\, (i=3,4,5,6)$. They have the following form:
\begin{eqnarray}
F_{LLLR(3)}^{\text{very long cut}}\;&=&\;\frac{W_{\text{very long cut};LL}}{\Omega_{32}\Omega_{43}\Omega_{51}} \nonumber\\
F_{LLLR(4)}^{\text{very long cut}}\;&=&\; \frac{W_{\text{very long cut};LL}}{\Omega_{32}\Omega_{43}\Omega_{15}}
\label{verylong1}
\end{eqnarray}
\begin{eqnarray}
F_{LRRR(3)}^{\text{very long cut}}\;&=&\; \frac{W_{\text{very long cut};RR}}{\Omega_{23}\Omega_{34}\Omega_{15}}\nonumber\\
F_{LRRR(4)}^{\text{very long cut}}\;&=&\; \frac{W_{\text{very long cut};RR}}{\Omega_{23}\Omega_{34}\Omega_{51}}
\label{verylong2}
\end{eqnarray}
\begin{eqnarray}
F_{LRLR(4)}^{\text{very long cut}}\;&=&\; \frac{W_{\text{very long cut};RL}}{\Omega_{23}\Omega_{43}\Omega_{51}}\nonumber\\
F_{LRLR(5)}^{\text{very long cut}}\;&=&\; \frac{W_{\text{very long cut};RL}}{\Omega_{23}\Omega_{43}\Omega_{15}}
\label{verylong3}
\end{eqnarray}
\begin{eqnarray}
\label{verylong4}
F_{LLRR(3)}^{\text{very long cut}}\;&=&\; \frac{\Omega_4}{\Omega_3} \frac{\Omega_{23}}{\Omega_{42}}\;\; \frac{W_{\text{very long cut};LR}}{\Omega_{23}\Omega_{34}\Omega_{51}}\nonumber\\
F_{LLRR(4)}^{\text{very long cut}}\;&=&\;  \frac{\Omega_2}{\Omega_3}\frac{\Omega_{34}}{\Omega_{24}} \;\;\frac{W_{\text{very long cut};LR}}{\Omega_{32}\Omega_{34}\Omega_{15}}\nonumber\\
F_{LLRR(5)}^{\text{very long cut}}\;&=&\; \frac{\Omega_2}{\Omega_3}\frac{\Omega_{34}}{\Omega_{24}} \;\;\frac{W_{\text{very long cut};LR}}{\Omega_{32}\Omega_{34}\Omega_{51}} \nonumber\\
F_{LLRR(6)}^{\text{very long cut}}\;&=&\;  \frac{\Omega_4}{\Omega_3} \frac{\Omega_{23}}{\Omega_{42}}\;\; \frac{W_{\text{very long cut};LR}}{\Omega_{23}\Omega_{34}\Omega_{15}}\\ \nonumber
\end{eqnarray}

\subsubsection{The Regge cut consisting of 3 reggeized gluons}
This cut contributes to four partial waves of the sixtet, $LLRR(i)\,(i=3,...,6)$. They have the following form:  
\begin{eqnarray}
\label{3-reggeon-cut}
F_{LLRR(3)}^{\text{3-reggeon cut}}\;&=&\; \frac{W_{\text{3-reggeon cut}}}{\Omega_{42}\Omega_{51}}\nonumber\\
F_{LLRR(4)}^{\text{3-reggeon cut}}\;&=&\;\frac{W_{\text{3-reggeon cut}}}{\Omega_{24}\Omega_{15}} \nonumber\\
F_{LLRR(5)}^{\text{3-reggeon c ut}}\;&=&\; \frac{W_{\text{3-reggeon cut}}}{\Omega_{24}\Omega_{51}}\nonumber\\
F_{LLRR(6)}^{\text{3-reggeon cut}}\;&=&\; \frac{W_{\text{3-reggeon cut}}}{\Omega_{42}\Omega_{15}}
\end{eqnarray}

%%%%%%%%%%%%%%%%%%%%%%%%%%%%%%%%%%%%%%%%%%%%%%%%%%%%%%%%%%%%%%%%%%%
 
\section{Regge pole contributions in different kinematic regions}

%In the following we list the Regge pole contributions, $P_{red;2\rightarrow6}\;=\;$, in different kinematic regions  
\begin{table}[H]
\begin{center}
\begin{tabular}{r l}
\vspace{0.3cm}
$=\;e^{i\pi\left(\omega_a+\omega_b+\omega_c+\omega_d\right)}e^{-i\pi\left(\omega_1+\omega_2+\omega_3+\omega_4+\omega_5\right)}$ & (free term)  \\ 
\vspace{0.3cm}
$-e^{i\pi\left(\omega_a+\omega_b+\omega_c+\omega_d\right)}e^{-i\pi\left(\omega_2+\omega_3+\omega_4+\omega_5\right)}$ & $\tau_1$ \\ 
\vspace{0.3cm}
$-e^{i\pi\left(\omega_a+\omega_b+\omega_c+\omega_d\right)}e^{-i\pi\left(\omega_1+\omega_3+\omega_4+\omega_5\right)}$ & $\tau_2$\\ 
\vspace{0.3cm}
$-e^{i\pi\left(\omega_a+\omega_b+\omega_c+\omega_d\right)}e^{-i\pi\left(\omega_1+\omega_2+\omega_4+\omega_5\right)}$ & $\tau_3$\\
\vspace{0.3cm}
$-e^{i\pi\left(\omega_a+\omega_b+\omega_c+\omega_d\right)}e^{-i\pi\left(\omega_1+\omega_2+\omega_3+\omega_5\right)}$ & $\tau_4$\\
\vspace{0.3cm}
$-e^{i\pi\left(\omega_a+\omega_b+\omega_c+\omega_d\right)}e^{-i\pi\left(\omega_1+\omega_2+\omega_3+\omega_4\right)}$ & $\tau_5$\\
\vspace{0.3cm}
$e^{i\pi\left(-\omega_a+\omega_b+\omega_c+\omega_d\right)}e^{-i\pi\left(\omega_3+\omega_4+\omega_5\right)}$ & $\tau_1\tau_2$\\ 
\vspace{0.3cm}
$e^{i\pi\left(\omega_a-\omega_b +\omega_c+\omega_d\right)}e^{-i\pi\left(\omega_1+\omega_4+\omega_5\right)}$ & $\tau_2\tau_3$\\
\vspace{0.3cm}
$e^{i\pi\left(\omega_a+\omega_b-\omega_c+\omega_d\right)}e^{-i\pi\left(\omega_1+\omega_2+\omega_5\right)}$ & $\tau_3\tau_4$\\
\vspace{0.3cm}
$e^{i\pi\left(\omega_a+\omega_b+\omega_c-\omega_d\right)}e^{-i\pi\left(\omega_1+\omega_2+\omega_3\right)}$ & $\tau_4\tau_5$\\
\vspace{0.3cm}
$e^{-i\pi\left(\omega_2+\omega_4+\omega_5\right)}e^{i\pi\left(\omega_c+\omega_d\right)}\left[e^{i\pi\left(\omega_a+\omega_b\right)} - 2ie^{i\pi\omega_2}\frac{\sin(\pi\omega_a)\sin(\pi\omega_b)}{\sin(\pi\omega_2)} \right]$ & $\tau_1\tau_3$\\
\vspace{0.3cm}
$e^{-i\pi\omega_5}e^{i\pi\omega_d}\left[e^{i\pi\left(\omega_a+\omega_b+\omega_c\right)}e^{-i\pi\left(\omega_2+\omega_3\right)} - 2i\frac{\sin(\pi\omega_a)\sin(\pi\omega_b)\sin(\pi\omega_c)}{\sin(\pi\omega_2)\sin(\pi\omega_3)}\right]$ & $\tau_1\tau_4$\\
\vspace{0.3cm}
$\left[e^{i\pi\left(\omega_a+\omega_b+\omega_c+\omega_d\right)}e^{-i\pi\left(\omega_2+\omega_3+\omega_4\right)} - 2i\frac{\sin(\pi\omega_a)\sin(\pi\omega_b)\sin(\pi\omega_c)\sin(\pi\omega_d)}{\sin(\pi\omega_2)\sin(\pi\omega_3)\sin(\pi\omega_4)}\right]$ & $\tau_1\tau_5$\\
\vspace{0.3cm}
$e^{-i\pi\left(\omega_1+\omega_3+\omega_5\right)}e^{i\pi\left(\omega_a+\omega_d\right)}\left[e^{i\pi\left(\omega_b+\omega_c\right)} - 2ie^{i\pi\omega_3}\frac{\sin(\pi\omega_b)\sin(\pi\omega_c)}{\sin(\pi\omega_3)} \right]$ & $\tau_2\tau_4$\\
\vspace{0.3cm}
$e^{-i\pi\omega_1}e^{i\pi\omega_a}\left[e^{i\pi\left(\omega_b+\omega_c+\omega_d\right)}e^{-i\pi\left(\omega_3+\omega_4\right)} - 2i\frac{\sin(\pi\omega_b)\sin(\pi\omega_c)\sin(\pi\omega_d)}{\sin(\pi\omega_3)\sin(\pi\omega_4)} \right]$ & $\tau_2\tau_5$\\
\vspace{0.3cm}
$e^{i\pi\left(\omega_a+\omega_b\right)}e^{-i\pi\left(\omega_1+\omega_2+\omega_4\right)}\left[e^{i\pi\left(\omega_c+\omega_d\right)} - 2ie^{i\pi\omega_4}\frac{\sin(\pi\omega_c)\sin(\pi\omega_d)}{\sin(\pi\omega_4)}\right]$ & $\tau_3\tau_5$\\
\vspace{0.3cm}
$-e^{-i\pi\left(\omega_4+\omega_5\right)}e^{i\pi\left(\omega_c+\omega_d\right)}\left[e^{-i\pi\left(\omega_a+\omega_b\right)} + 2ie^{-i\pi\omega_2}\frac{\sin(\pi\omega_a)\sin(\pi\omega_b)}{\sin(\pi\omega_2)} \right]$ & $\tau_1\tau_2\tau_3$\\
\vspace{0.3cm}
$-e^{-i\pi\omega_5}e^{i\pi\omega_d}\left[e^{i\pi\left(-\omega_a+\omega_b+\omega_c\right)}e^{-i\pi\omega_3} - 2i\frac{\sin(\pi\omega_2-\pi\omega_a)\sin(\pi\omega_b)\sin(\pi\omega_c)}{\sin(\pi\omega_2)\sin(\pi\omega_3)} \right]$ & $\tau_1\tau_2\tau_4$\\
\vspace{0.3cm}
$-\left[e^{i\pi\left(-\omega_a+\omega_b+\omega_c+\omega_d\right)}e^{-i\pi\left(\omega_3+\omega_4\right)} - 2i\frac{\sin(\pi\omega_2-\pi\omega_a)\sin(\pi\omega_b)\sin(\pi\omega_c)\sin(\pi\omega_d)}{\sin(\pi\omega_2)\sin(\pi\omega_3)\sin(\pi\omega_4)} \right]$ & $\tau_1\tau_2\tau_5$\\
\vspace{0.3cm}
$-\left[e^{i\pi\left(\omega_a+\omega_b+\omega_c+\omega_d\right)}e^{-i\pi\left(\omega_2+\omega_4\right)} - 2ie^{-i\pi\omega_2}e^{i\pi\left(\omega_a+\omega_b\right)}\frac{\sin(\pi\omega_c)\sin(\pi\omega_d)}{\sin(\pi\omega_4)}\right.-$ & \\
\vspace{0.3cm}
$ - 2ie^{-i\pi\omega_4}e^{i\pi\left(\omega_c+\omega_d\right)}\frac{\sin(\pi\omega_a)\sin(\pi\omega_b)}{\sin(\pi\omega_2)} + 2i\frac{\sin(\pi\omega_a)\sin(\pi\omega_b)e^{-i\pi \omega_3} \sin(\pi\omega_c)\sin(\pi\omega_d)}{\sin(\pi\omega_2)  \sin(\pi\omega_3)  \sin(\pi\omega_4)} + $ & \\
\vspace{0.3cm}
$\left. + (2i)^2 \frac{\sin(\pi\omega_a)\sin(\pi\omega_b)\sin(\pi\omega_c)\sin(\pi\omega_d)}{\sin(\pi\omega_2)\sin(\pi\omega_4)}\right]$ & $\tau_1\tau_3\tau_5$\\
\end{tabular} 
\end{center}
\end{table}

\begin{table}[H]
\begin{center}
\begin{tabular}{r l}
\vspace{0.3cm}
$-\left[e^{i\pi\left(\omega_a+\omega_b+\omega_c-\omega_d\right)}e^{-i\pi\left(\omega_2+\omega_3\right)} - 2i\frac{\sin(\pi\omega_a)\sin(\pi\omega_b)\sin(\pi\omega_c)\sin(\pi \omega_4-\pi\omega_d)}{\sin(\pi\omega_2)\sin(\pi\omega_3)\sin(\pi\omega_4)} \right]$ & $\tau_1\tau_4\tau_5$\\
\vspace{0.3cm}
$-e^{-i\pi\left(\omega_2\right)}e^{i\pi\left(\omega_a\right)}\left[e^{i\pi\left(-\omega_b+\omega_c-\omega_d\right)}e^{-i\pi\left(\omega_4\right)} - 2i\frac{\sin(\pi\omega_3-\pi\omega_b)\sin(\pi\omega_c)\sin(\pi\omega_4-\pi\omega_d)}{\sin(\pi\omega_3)\sin(\pi\omega_4)} \right]$ & $\tau_2\tau_3\tau_5$\\
\vspace{0.3cm}
$-e^{-i\pi\left(\omega_4\right)}e^{i\pi\left(\omega_d\right)}\left[e^{i\pi\left(\omega_a+\omega_b-\omega_c\right)}e^{-i\pi\left(\omega_2\right)} - 2i\frac{\sin(\pi\omega_3-\pi\omega_c)\sin(\pi\omega_a)\sin(\omega_4-\pi\omega_b)}{\sin(\pi\omega_2)\sin(\pi\omega_3)} \right]$ & $\tau_1\tau_3\tau_4$\\
\vspace{0.3cm}
$-e^{-i\pi\left(\omega_1\right)}e^{i\pi\left(\omega_a\right)}\left[e^{i\pi\left(\omega_b+\omega_c-\omega_d\right)}e^{-i\pi\left(\omega_3\right)} - 2i\frac{\sin(\pi\omega_4-\pi\omega_a)\sin(\pi\omega_b)\sin(\pi\omega_c)}{\sin(\pi\omega_3)\sin(\pi\omega_4)} \right]$ & $\tau_2\tau_4\tau_5$\\
\vspace{0.3cm}
$-e^{-i\pi\left(\omega_1+\omega_5\right)}e^{i\pi\left(\omega_a+\omega_d\right)}\left[e^{-i\pi\left(\omega_b+\omega_c\right)} + 2ie^{-i\pi\left(\omega_3\right)}\frac{\sin(\pi\omega_b)\sin(\pi\omega_c)}{\sin(\pi\omega_3)} \right]$ & $\tau_2\tau_3\tau_4$\\
\vspace{0.3cm}
$-e^{-i\pi\left(\omega_1+\omega_2\right)}e^{i\pi\left(\omega_a+\omega_b\right)}\left[e^{-i\pi\left(\omega_c+\omega_d\right)} + 2ie^{-i\pi\left(\omega_4\right)}\frac{\sin(\pi\omega_c)\sin(\pi\omega_d)}{\sin(\pi\omega_4)} \right]$ & $\tau_3\tau_4\tau_5$\\
\vspace{0.3cm}
$-e^{-i\pi\left(\omega_5\right)}e^{i\pi\left(\omega_d\right)}\left[e^{i\pi\left(-\omega_a+\omega_b-\omega_c\right)} - 2i\frac{\sin(\pi\omega_2-\pi\omega_a)\sin(\pi\omega_b)\sin(\pi\omega_3-\pi\omega_c)}{\sin(\pi\omega_2)\sin(\pi\omega_3)} \right]$ & $\tau_1\tau_2\tau_3\tau_4$\\
\vspace{0.3cm}
$-e^{-i\pi\left(\omega_1\right)}e^{i\pi\left(\omega_a\right)}\left[e^{i\pi\left(-\omega_b+\omega_c-\omega_d\right)} - 2i\frac{\sin(\pi\omega_3-\pi\omega_b)\sin(\pi\omega_c)\sin(\pi\omega_4-\pi\omega_d)}{\sin(\pi\omega_3)\sin(\pi\omega_4)} \right]$ & $\tau_2\tau_3\tau_4\tau_5$\\
\vspace{0.3cm}
$\left[e^{i\pi\left(-\omega_a-\omega_b+\omega_c+\omega_d\right)}e^{-i\pi\left(\omega_4\right)} - 2i\frac{\cos(\pi\omega_b)e^{-i\pi\left(\omega_a\right)}\sin(\pi\omega_c)\sin(\pi\omega_d)}{\sin(\pi\omega_4)}  \right.$ & \\
\vspace{0.3cm}
$\left.  + 2i\frac{\sin(\pi\omega_a)\sin(\pi\omega_b)e^{-i\pi\left(\omega_2+\omega_4\right)}e^{i\pi\left(\omega_c+\omega_d\right)}}{\sin(\pi\omega_2)}\right.$&\\
\vspace{0.3cm}
$\left.
 - (2i)i e^{-i\pi\left(\omega_2\right)}\frac{\sin(\pi\omega_a)\sin(\pi\omega_b)\sin(\pi\omega_c)\sin(\pi\omega_d)}{\sin(\pi\omega_2)\sin(\pi\omega_4)}\right.$ & \\
\vspace{0.3cm}
$\left.
+ 2i\frac{\sin(\pi\omega_2-\pi\omega_a)\sin(\pi\omega_b)\cos(\pi\omega_3)\sin(\pi\omega_c)\sin(\pi\omega_d)}{\sin(\pi\omega_2)\sin(\pi\omega_3)\sin(\pi\omega_4)}
\right]$ & $\tau_1\tau_2\tau_3\tau_5$\\
\vspace{0.3cm}
$\left[e^{i\pi\left(\omega_a+\omega_b-\omega_c-\omega_d\right)}e^{-i\pi\left(\omega_2\right)} - 2i\frac{\cos(\pi\omega_c)e^{-i\pi\left(\omega_d\right)}\sin(\pi\omega_b)\sin(\pi\omega_a)}{\sin(\pi\omega_2)}  \right.$ & \\
\vspace{0.3cm}
$\left. + 2i\frac{\sin(\pi\omega_c)\sin(\pi\omega_d)e^{-i\pi\left(\omega_2+\omega_4\right)}e^{i\pi\left(\omega_a+\omega_b\right)}}{\sin(\pi\omega_4)} \right.$ & \\
\vspace{0.3cm}
$\left. -(2i) ie^{-i\pi\left(\omega_4\right)}\frac{\sin(\pi\omega_a)\sin(\pi\omega_b)\sin(\pi\omega_c)\sin(\pi\omega_d)}{\sin(\pi\omega_2)\sin(\pi\omega_4)}\right.$ & \\
\vspace{0.3cm}
$\left.
+ (2i) \frac{\sin(\pi\omega_a)\sin(\pi\omega_b)\cos(\pi\omega_3)\sin(\pi\omega_c)
\sin(\pi \omega_4-\pi\omega_d)}{\sin(\pi\omega_2)\sin(\pi\omega_3)\sin(\pi\omega_4)}
\right]$ & $\tau_1\tau_3\tau_4\tau_5$\\
\vspace{0.3cm}
$\left[e^{i\pi\left(-\omega_a+\omega_b+\omega_c-\omega_d\right)}e^{-i\pi\omega_3} - 2i\frac{\sin(\pi\omega_2-\pi\omega_a)\sin(\pi\omega_b)\sin(\pi\omega_c)\sin(\pi\omega_4-\pi\omega_d)}{\sin(\pi\omega_2)\sin(\pi\omega_3)\sin(\pi\omega_4)} \right]$ & $\tau_1\tau_2\tau_4\tau_5$\\
\vspace{0.3cm}
$-\left[e^{-i\pi\left(\omega_a+\omega_b+\omega_c+\omega_d\right)}
+2i\frac{\sin(\pi\omega_2-\pi\omega_a)\sin(\pi\omega_b)\cos(\pi\omega_3)\sin(\pi\omega_c)
\sin(\pi\omega_4-\pi\omega_d)}{\sin(\pi\omega_2)\sin(\pi\omega_3)\sin(\pi\omega_4)} \right.$ & \\
\vspace{0.3cm}
$\left.  
+2ie^{-i\pi \omega_2}\frac{\sin(\pi\omega_a)\sin(\pi\omega_b)\cos(\pi\omega_c)e^{-i\pi\omega_d}}{\sin(\pi\omega_2)}  \right.$ & \\
\vspace{0.3cm}
$\left.  
+ 2i\frac{e^{-i\pi\omega_4} e^{-i\pi\omega_a}\cos(\pi\omega_b)\sin(\pi\omega_c)\sin(\pi\omega_d)}{\sin(\pi\omega_4)} \right.$ & \\
\vspace{0.3cm}
$\left.
+ (2i)i e^{-i\pi\left(\omega_2+\omega_4\right)}\frac{\sin(\pi\omega_a)\sin(\pi\omega_b)\sin(\pi\omega_c)\sin(\pi\omega_d)}{\sin(\pi\omega_2)\sin(\pi\omega_4)}  
 \right.$ & \\
\vspace{0.3cm}
$\left.  
+2e^{-i\pi (\omega_a+\omega_d)} \sin(\pi \omega_b) \sin (\pi \omega_d)
\right]$ & $\tau_1\tau_2\tau_3\tau_4\tau_5$\\
\end{tabular} 
\end{center}
\end{table}
\newpage

\newpage

\section{Singular pieces of the long cut, the very long cut, the double cut, and the 3-reggeon cut}

In this section we adopt a slightly different way of our discussion. Namely, we return to the starting point (\ref{struct-LLLL}) which we write in the form (\ref{struct-LLLL-mod})
%\ba
%T_{LLLL} &=& s \int ...\int \frac{d \omega'_1d \omega'_2 d \omega'_3d \omega'_4 d\omega'_5} {(2\pi i)^5} -s_{45}^{\omega'_{54}}s_{35}^{\omega'_{43}}s_{25}^{\omega'_{32}}s_{15}^{\omega'_{21}}s^{\omega'_1}
%\nonumber\\
%&&\cdot e^{-i\pi(\omega'_5 -\omega'_4) }
% F_{LLL}(t_1,t_2,t_3,t_4,t_5;\omega'_1,\omega'_2,\omega'_3,\omega'_4,\omega'_5) ,
%\label{struct-LLLL-mod}
%\ea
and focus on the last line: phase $\times$ $\omega'$-dependent partial wave. This implies that, for $t_i$-channels without a Regge cut, we have a pole factor $1/(\omega'_i - \omega_i)$ which, for simplicity, we will not write. 
The $t_1$ and $t_5$ channels are always free from Regge cuts, therefore we do not need to introduce 
$\omega'_1$ and $\omega'_5$. 
At the end of the discussion, when we will carry out the $\omega'_i$ integrations: for  $t_i$-channels with a  Regge pole only we have $\omega'_i \to \omega_i$, whereas for  $t_i$-channels with a Regge cut we perform the shift
$\omega'_i= \omega''_i + \omega_i$. This provides that, for each $t_i$ channel, we have the Regge pole factor
$s_{i-1 i}^{\omega_i}$ which is part of the BDS part. The same procedure will also be applied for energy discontinuities. 

\subsection{The long cut}

This long cut has already been studied for the $2 \to 5$ amplitude, and we can simply recapitulate the discussion  
\cite{Bartels:2013jna,Bartels:2014jya} and apply to the long cut in the $t_2$ and $t_3$ channels. The equations for the long $\omega_2 \omega_3$-Regge cut are:
\ba
\label{tau14}
\tau_1\tau_4:&\hspace{0.5cm}&2i e^{-i\pi(\omega'_2+\omega'_3+\omega_5)} e^{i \pi \omega_d}\Big[
e^{i\pi \omega'_2} \widetilde{W}_{\omega_2\omega_3;L}+  e^{i\pi \omega'_3}  \widetilde{W}_{\omega_2\omega_3;R}\Big]\\
\label{tau124}
\tau_1\tau_2\tau_4:&\hspace{0.5cm}&2i e^{-i\pi\omega'_3+\omega_5)}  e^{i \pi \omega_d}\Big[
e^{i\pi \omega'_2}  \widetilde{W}_{\omega_2\omega_3;L}+  e^{i\pi \omega'_3}  \widetilde{W}_{\omega_2\omega_3;R}-e^{i\pi \omega_a}  W_{\omega_3}\Big]\\
\label{tau134}
\tau_1\tau_3\tau_4:&\hspace{0.5cm}&2i e^{-i\pi\omega'_2+\omega_5)}  e^{i \pi \omega_d}\Big[
e^{i\pi \omega'_2}  \widetilde{W}_{\omega_2\omega_3;L}+  e^{i\pi \omega'_3} \widetilde{W}_{\omega_2\omega_3;R}- W_{\omega_2}e^{i\pi \omega_c}\Big]\\
\label{tau1234}
\tau_1\tau_2\tau_3\tau_4:&\hspace{0.5cm}&2i e^{-i\pi \omega_5} e^{i \pi \omega_d} \Big[
e^{-i\pi \omega'_2} \widetilde{W}_{\omega_2\omega_3;L}+  e^{-i\pi \omega'_3}  \widetilde{W}_{\omega_2\omega_3;R}- e^{-i\pi \omega_a} W_{\omega_3}-  W_{\omega_2}e^{-i\pi \omega_c}\Big]\,. \nonumber\\
\ea
When combining with Regge pole terms it was found in \cite{Bartels:2013jna} that we should start from the last kinematic region, 
$\tau_1\tau_2\tau_3\tau_4$,
in which the most singular piece of the Regge pole contribution has the simple form 
\ba
-2i \frac{V_L(a)\Omega_b V_R(c)}{\Omega_{2'} \Omega_{3'}}\,.
\ea
Together with the singular pieces in  (\ref{tau1234}) the sum of all singular terms which have to be compensated by the $\delta \widetilde{W}$ becomes
\ba
\label{lc-singular}
e^{-i\pi \omega'_2} \delta \widetilde{W}_{\omega_2\omega_3;L}+  e^{-i\pi \omega'_3} \delta \widetilde{W}_{\omega_2\omega_3;R}&=&
 \Big[ \frac{V_L(a)\Omega_b V_R(c)}{\Omega_{2'} \Omega_{3'}}+ e^{-i\pi \omega_a} \delta W_{\omega_3}+ \delta W_{\omega_2}e^{-i\pi \omega_c} \Big]\nonumber\\
&=& e^{-i\pi \omega_a} \delta f_{23}^a  +\delta f_{23}^c e^{-i\pi \omega_c}
\ea
with 
\ba
 \delta f_{23}^a &=& - \frac{\Omega_c}{\Omega_{ac}} \frac{V_L(a) \Omega_b V_R(c)}{\Omega_{2'} \Omega_{3'}}+ \delta W_{\omega_3}\\
 \delta f_{23}^c &=&  \frac{\Omega_a}{\Omega_{ac}} \frac{V_L(a) \Omega_b V_R(c)}{\Omega_{2'} \Omega_{3'}}+ \delta W_{\omega_2}\,.
\ea

Since all partial waves $\widetilde{W}_{\omega_2\omega_3;L}$ etc. are real-valued, we are also searching for we real-valued corrections 
$ \delta \widetilde{W}_{\omega_2\omega_3;L}$; together with the complex conjugate of (\ref{lc-singular}) 
\be
\label{lc-singular-cc}
e^{i\pi \omega'_2} \delta \widetilde{W}_{\omega_2\omega_3;L}+  e^{i\pi \omega'_3} \delta \widetilde{W}_{\omega_2\omega_3;R}= e^{i\pi \omega_a} \delta f_{23}^a  +\delta f_{23}^c e^{i\pi \omega_c}
\ee
we thus have two equations which allow to find $ \delta \widetilde{W}_{\omega_2\omega_3;L}$ and  $ \delta \widetilde{W}_{\omega_2\omega_3;R}$:
\ba
\label{transfinv}
\Omega_{3'2'}  \delta \widetilde{W}_{\omega_2\omega_3;L}&=& \Omega_{3'a}  \delta f_{23}^a + \Omega_{3'c}  \delta f_{23}^c \nonumber\\
\Omega_{2'3'}  \delta \widetilde{W}_{\omega_2\omega_3;R}&=& \Omega_{2'a}  \delta f_{23}^a + \Omega_{2'c}  \delta f_{23}^c\, ,
\ea
or,  in more detail:
\ba
\Omega_{2'3'} \delta\widetilde{W}_{\omega_2\omega_3;L}&=&\frac{\Omega_a\Omega_b\Omega_c}{\Omega_{3'}}-\frac{1}{2}\Big[\cos\pi(\omega'_3-\omega_b) \cos\pi(\omega_a-\omega_c)-\cos\pi(\omega'_3-\omega_a-\omega_b-\omega_c)
\Big]\nonumber\\ 
\Omega_{3'2'} \delta\widetilde{W}_{\omega_2\omega_3;R}&=&\frac{\Omega_a\Omega_b\Omega_c}{\Omega_{2'}} -\frac{1}{2}\Big[\cos\pi(\omega'_2-\omega_b)
 \cos\pi(\omega_a-\omega_c)-\cos\pi(\omega'_2-\omega_a-\omega_b-\omega_c)\Big]  \,.\nonumber\\ 
\ea

Here all subtraction terms are free from Regge cut pieces, i.e.proportional to Regge pole factors. Therefore the
$\omega'$-integrations simply imply $\omega'_i \to \omega_i$, $(i=2,3,4)$.
Similar subtractions apply to the long $\omega_3\omega_4$- cut.The complete list of finite results is given  in section 4.2.

\subsection{The very long cut, the double cut, and the 3-reggeon cut}

\subsubsection{The very long cut for the regions $\tau_1\tau_5$, $\tau_1\tau_2\tau_5$, $\tau_1\tau_4\tau_5$, $\tau_1\tau_3\tau_5$}

We begin with the very long cut in the region  $\tau_1\tau_5$.
Looking at the singularities of the Regge pole terms in Appendix B, we recognize that the region $\tau_1 \tau_2 \tau_4 \tau_5$  may play the same role as the region  $\tau_1 \tau_2 \tau_3 \tau_4$ above has been  playing for the long cut: in this region the Regge pole singularity is of the form
\be 
\label{vl-sing-1}
- 2 i e^{i\pi \omega'_3} \frac{V_L(a) \Omega_b \Omega_c V_R(d)}{\Omega_{2'} \Omega_{3'} \Omega_{4'}} \,.
\ee
Beginning with (\ref{amp15}) and disregarding, for the moment, the double cut $W_{\omega_2\omega_4}$ we make the following ansatz for $\delta VLC^{++}$:
\ba
\label{deltapp-prel}
&&e^{i \pi \omega_3}\frac{V_L(a) \Omega_b \Omega_c V_R(d)}{\Omega_{2'} \Omega_{3'} \Omega_{4'}} \\
&&+\Big[ \delta f_{23}^a e^{i\pi \omega_a}+\delta f_{23}^c e^{i\pi \omega_c}\Big] e^{i\pi \omega_d}
 +e^{i\pi \omega_a}  \Big[ \delta f_{34}^b e^{i\pi \omega_b}+\delta f_{34}^d e^{i\pi \omega_d}\Big]\
- e^{i\pi \omega_a} \delta W_{\omega_3}e^{i\pi \omega_d} \,. \nonumber\\
\ea
When combining this with the Regge pole  of the region $\tau_1\tau_5$ we obtain
the conformal invariant Regge pole written in (\ref{amp15reg}).

However, there is still the double cut piece, $W_{\omega_2 \omega_4}$, which introduces additonal singularities. So the singular terms we have to cancel are
\be
\label{vl-sing-2}
-2i \left(  e^{i\pi \omega'_3} \frac{V_L(a) \Omega_b \Omega_c V_R(d)}{\Omega_{2'} \Omega_{3'} \Omega_{4'}} 
 + e^{i\pi \omega'_3}  \frac{W_{\omega_2 \omega_4}}{\Omega_{3‘}}\right)\,,
\ee
where $W_{\omega_2 \omega_4}$ depends upon $\omega'_2$ and $\omega'_4$, As shown in section 6.2,   $W_{\omega_2 \omega_4}$ factorizes:
\ba 
W_{\omega_2 \omega_4} & =& W_{\omega_2}  W_{\omega_4}\nonumber\\
&=& \left(W_{\omega_2}^{reg} + \delta W_{\omega_2}\right) \left(  W_{\omega_4}^{reg} + \delta W_{\omega_4}\right) \nonumber\\
&=&  W_{\omega_2}^{reg} W_{\omega_4}^{reg} + \delta W_{\omega_2 \omega_4}
\ea
with
\be
 \delta W_{\omega_2 \omega_4}= W_{\omega_2}^{reg}  \delta W_{\omega_4} + 
\delta W_{\omega_2} W_{\omega_4}^{reg} +  \delta W_{\omega_2}\delta W_{\omega_4}\,.
\ee
We write the brackets in (\ref{vl-sing-2}) as
\be
\label{vl-sing-1+2}
S_{r1}+i\left(\frac{V_L(a) \Omega_b}{\Omega_{2'}} \frac{\Omega_c V_R(d)}{\Omega_{4'}} + W_{\omega_2 \omega_4}\right),
\ee
where the real part is
\be
S_{r1}=\frac{\cos \pi \omega'_3}{\Omega_{3'}} \left(  \frac{V_L(a) \Omega_b \Omega_c V_R(d)}{\Omega_{2'}\Omega_{4'}}+
W_{\omega_2 \omega_4}\right)\,
\ee
and the imaginary part can be written as 
\ba
&&\left(\frac{V_L(a) \Omega_b}{\Omega_{2'}} \frac{\Omega_c V_R(d)}{\Omega_{4'}} + W_{\omega_2 \omega_4}\right) \nonumber \\
&&=-\left(S_2+S_4\right) + \left(  W_{\omega_2}^{reg} -\frac{1}{2} \sin \pi (\omega_a-\omega_b)\right) 
\left( W_{\omega_4}^{reg} -\frac{1}{2} \sin \pi (\omega_d-\omega_c)\right) 
\ea
with 
\be
S_2=  W_{\omega_2} \frac{\Omega_c V_R(d)}{\Omega_{4'}},\,\,S_4= \frac{V_L(a) \Omega_b}{\Omega_{2'}}  W_{\omega_4}\, .
\ee
We thus need to cancel the singular terms $-S_{r1} +i (S_2+S_4)$, and arrive at the subtraction $\delta VLC^{++}$, instead of (\ref{deltapp-prel}):
\ba 
\label{deltapp}
\delta VLC^{++}&=& S_{r1} - i(S_2+S_4\\
&& \Big[ \delta f_{23}^a e^{i\pi \omega_a}+\delta f_{23}^c e^{i\pi \omega_c}\Big] e^{i\pi \omega_d}
 +e^{i\pi \omega_a}  \Big[ \delta f_{34}^b e^{i\pi \omega_b}+\delta f_{34}^d e^{i\pi \omega_d}\Big]- 
e^{i\pi \omega_a} \delta W_{\omega_3}  e^{i\pi \omega_d} .\nonumber
\ea
Here we observe that, in contrast to all previous subtractions, now $\delta VLC^{++}$ being a subtraction to the 
three dimensional integral $VLC^{++}$ contains also integrals: in $S_{r1}$ and in $S_2$, $S_4$.
For later purposes it is sometimes convenient to write $S_{r1}$ as 
\ba 
\label{Sr1-identity}
S_{r1}=\frac{\cos \pi \omega'_3}{\Omega_{3'}} \left( -(S_2+S_4) +\left(  W_{\omega_2}^{reg} -\frac{1}{2} \sin \pi (\omega_a-\omega_b)\right) 
\left( W_{\omega_4}^{reg} -\frac{1}{2} \sin \pi (\omega_d-\omega_c)\right)\right).\nonumber\\
\ea

Performing the $\omega'$-integrals (including the shifts for $\omega'_2$ and $\omega'_4$) and inserting  into the square brackets of (\ref{amp15}) (and adding  the Regge pole term), all singular terms cancel and we find the final expression
 \ba
\tau_1\tau_5:&& e^{-i\pi(\omega_2+\omega_3+\omega_4)} 
 \Big[ e^{i\pi (\omega_b+\omega_c)} \cos \pi(\omega_a-\omega_d)\nonumber\\
&&+2i \Big[ \int VLC^{++;reg}   \nonumber\\ 
&& - i \left( \int W_{\omega_2}^{reg} -\frac{1}{2} \sin \pi (\omega_a-\omega_b)\right)
\left(\int W_{\omega_4}^{reg} -\frac{1}{2} \sin \pi (\omega_d-\omega_c)  \right) \big] \Big] \,.
\ea 
The same applies to the regions $\tau_1\tau_2\tau_5$, $\tau_1\tau_4\tau_5$, and  $\tau_1\tau_3\tau_5$. The finite results are listed in (\ref{amp15reg}) - (\ref{amp135reg}),

\subsubsection{The very long cut for the regions $\tau_1\tau_2\tau_3\tau_5$, $\tau_1\tau_3\tau_4\tau_5$, }

For the regions (\ref{amp1345}) and (\ref{amp1235}) we have to modify the ansatz.
Addition of the last term of  (\ref{amp1345}) leads, in (\ref{deltapp}), to the change $-i(S_2+S_4) \to -i(S_2-S_4))$:
\ba 
\delta VLC^{+-}&=& \Big[ \delta f_{23}^a e^{i\pi \omega_a}+\delta f_{23}^c e^{i\pi \omega_c}\Big] e^{-i\pi \omega_d}
 +e^{i\pi \omega_a}  \Big[ \delta f_{34}^b e^{-i\pi \omega_b}+\delta f_{34}^d e^{-i\pi \omega_d}\Big] \nonumber\\
&&- e^{i\pi \omega_a} \delta W_{\omega_3}
e^{-i\pi \omega_d}
+ S_{r1} - i(S_2-S_4)\,.
\ea
and the cancellation  of singularities proceeds in the same way as before. The same arguments appliy to the  region  $\tau_1\tau_2\tau_3\tau_5$
in (\ref{amp1235}):
\ba
\delta VLC^{-+}&=& \Big[ \delta f_{23}^a e^{-i\pi \omega_a}+\delta f_{23}^c e^{-i\pi \omega_c}\Big] e^{i\pi \omega_d}
 +e^{-i\pi \omega_a}  \Big[ \delta f_{34}^b e^{i\pi \omega_b}+\delta f_{34}^d e^{i\pi \omega_d}\Big] \nonumber\\
&&- e^{-i\pi \omega_a} \delta W_{\omega_3}
e^{i\pi \omega_d} + S_{r1} - i(-S_2+S_4)\,.
\ea

The three equations for $\delta VLC^{++}$,  $\delta VLC^{+-}$ and $\delta VLC^{-+}$, together with 
 $\delta VLC^{--}$ (which is the complex conjugate of $\delta VLC^{++}$) can be used to find the expressions for
$\delta \widetilde{W}_{LL}$ etc:
\ba
&&\Omega_{2'3'}\Omega_{3'4'} \delta \widetilde{W}_{LL}= \Big[\Omega_{3'a}\delta f _{23}^a + \Omega_{3'c} \delta f _{23}^c \Big] \Omega_{4d}+
 \Omega_{3'a}\Big[\Omega_{4'b}\delta f _{34}^b + \Omega_{4'd} \delta f _{34}^d \Big] \\
&&\hspace{4cm}-\Omega_{'a} \delta W_{\omega_3} \Omega_{4'd}+\Omega_{3'} \Omega_{4'} S_{r1} +\cos \pi \omega'_3 \Omega_{4'} S_2 +\cos \pi \omega'_4 \Omega_{3'} S_4\nonumber \\
&&\Omega_{2'3'}\Omega_{3'4'} \delta \widetilde{W}_{RL}= -\Big[\Omega_{ 2'a}\delta f _{23}^a + \Omega_{2'c}\delta f _{23}^c \Big] 
\Omega_{4'd}
- \Omega_{2'a}\Big[\Omega_{4'b}\delta f _{34}^b + \Omega_{4'd} \delta f _{34}^d \Big] \\
&&\hspace{4cm}+\Omega_{2'a} \delta W_{\omega_3} \Omega_{4'd}-\Omega_{2'} \Omega_{4'} S_{r1} - \cos \pi \omega'_2 \Omega_{4'} S_2 -\cos \pi \omega'_4 \Omega_{2'} S_4 \nonumber\\
&&\Omega_{2'3'}\Omega_{3'4'} \delta \widetilde{W}_{LR}=- \Big[\Omega_{3'a} \delta f _{23}^a + \Omega_{3'c} \delta f _{23}^c \Big] \Omega_{3'd}
- \Omega_{3'a}\Big[\Omega_{3'b}\delta f _{34}^b + \Omega_{3'd}\delta f _{34}^d \Big]\\
&&\hspace{4cm} +\Omega_{3'a} \delta W_{\omega_3} \Omega_{3'd}-\Omega_{3'}^2 S_{r1} - \cos \pi \omega'_3 \Omega_{3'}( S_2+S_4) \nonumber\\
&& \Omega_{2'3'}\Omega_{3'4'} \delta \widetilde{W}_{RR}= \Big[\Omega_{2'a} \delta f _{23}^a + \Omega_{2'c}\delta f _{23}^c \Big] \Omega_{3'd}
+ \Omega_{2'a}\Big[\Omega_{3'b} \delta f _{34}^b + \Omega_{3'd}\delta f _{34}^d \Big] \\
&&\hspace{4cm}-\Omega_{2'a} \delta W_{\omega_3} \Omega_{3'd}+\Omega_{2'} \Omega_{3'} S_{r1} +\cos \pi \omega'_2 \Omega_{3'} S_2 + \cos \pi \omega'_3 \Omega_{2'} S_4 \,.\nonumber
\ea
Making use of the relations:
\ba
\Omega_{3'2'}  \delta \widetilde{W}_{\omega_2\omega_3;L}&=& \Omega_{3'a}  \delta f_{23}^a + \Omega_{3'c}  \delta f_{23}^c \nonumber\\
\Omega_{2'3'}  \delta \widetilde{W}_{\omega_2\omega_3;R}&=& \Omega_{2'a}  \delta f_{23}^a + \Omega_{2'c}  \delta f_{23}^c
\ea
and 
\ba
\Omega_{4'3'}  \delta \widetilde{W}_{\omega_3\omega_4;L}&=& \Omega_{4'b}  \delta f_{34}^b + \Omega_{4'd}  \delta f_{23}^d \nonumber\\
\Omega_{3'4'}  \delta \widetilde{W}_{\omega_3\omega_4;R}&=& \Omega_{3'b}  \delta f_{34}^b + \Omega_{3'd}  \delta f_{34}^d
\ea
we can also write:
\ba
\label{sub-WLL}
&&\Omega_{2'3'}\Omega_{3'4'} \delta \widetilde{W}_{LL}= \Omega_{3'2'} \delta \widetilde{W}_{\omega_2\omega_3;L} \Omega_{4d} +
\Omega_{3'a} \Omega_{4'3'}  \delta \widetilde{W}_{\omega_3\omega_4;L} -\Omega_{3'a} \delta W_{\omega_3} \Omega_{4'd}\nonumber\\
&&\hspace{3cm}+\Omega_{3'} \Omega_{4'} S_{r1} +\cos \pi \omega'_3 \Omega_{4'} S_2 +\cos \pi \omega'_4 \Omega_{3'} S_4 \\
%%%%%%%%%%%%%%%%
\label{sub-WRL}
&&\Omega_{2'3'}\Omega_{3'4'} \delta \widetilde{W}_{RL}= -\Omega_{2'3'}  \delta \widetilde{W}_{\omega_2\omega_3;R}\Omega_{4'd}
-\Omega_{2'a}\Omega_{4'3'}  \delta \widetilde{W}_{\omega_3\omega_4;L}
+\Omega_{2'a} \delta W_{\omega_3} \Omega_{4'd}\nonumber\\
&&\hspace{3cm}-\Omega_{2'} \Omega_{4'} S_{r1} - \cos \pi \omega'_2 \Omega_{4'} S_2 -\cos \pi \omega'_4 \Omega_{2'} S_4\\
%%%%%%%%%
\label{sub-WLR}
&& \Omega_{2'3'}\Omega_{3'4'} \delta \widetilde{W}_{LR}=- 
\Omega_{3'2'}  \delta \widetilde{W}_{\omega_2\omega_3;L} \Omega_{3'd}
- \Omega_{3'a}\Omega_{3'4'}  \delta \widetilde{W}_{\omega_3\omega_4;R} +\Omega_{3'a} \delta W_{\omega_3} \Omega_{3'd}\nonumber\\
&&\hspace{3cm}-\Omega_{3'}^2 S_{r1} - \cos \pi \omega'_3 \Omega_{3'} ( S_2+ S_4)  \\
%%%%%%%%%%%%%
\label{sub-WRR}
&& \Omega_{2'3'}\Omega_{3'4'} \delta \widetilde{W}_{RR}=\Omega_{2'3'}  \delta \widetilde{W}_{\omega_2\omega_3;R} \Omega_{3'd}
+ \Omega_{2'a}\Omega_{3'4'}  \delta \widetilde{W}_{\omega_3\omega_4;R}-\Omega_{2'a} \delta W_{\omega_3} \Omega_{3'd} \nonumber\\
&&\hspace{3cm}+\Omega_{2''} \Omega_{3'} S_{r1} +\cos \pi \omega'_2 \Omega_{3'} S_2 + \cos \pi \omega'_3 \Omega_{2'} S_4 \,.
\ea
We should stress that the subtractions of the partial waves $ \widetilde{W}_{LL}$ etc. are all real-valued, as it should be.

\subsubsection{The very long cut for the regions $\tau_1\tau_2\tau_4\tau_5$, $\tau_1\tau_2\tau_3\tau_4\tau_5$, }

For the remaining kinematic regions $\tau_1\tau_2\tau_4\tau_5$ and $\tau_1\tau_2\tau_3\tau_4\tau_5$ we first need to analyse the first two lines of  (\ref{amp1245})  and  (\ref{amp12345}), then find the subtraction of the three reggeon cut contribution, $W_{\text{3-reggeon cut}}$.
We beginn with the the first two lines of (\ref{amp1245}), disregarding for the moment the double cut term $W_{\omega_2\omega_4}$. First we notice that the simple ansatz
\ba
\label{delta--}
&&e^{i \pi \omega'_3}\frac{V_L(a) \Omega_b \Omega_c V_R(d)}{\Omega_{2'} \Omega_{3'} \Omega_{4'}}\\
&&
+ \Big[e^{i \pi \omega_a} \delta f_{23}^a + e^{i \pi \omega_c} \delta f_{23}^c \Big] e^{-i\pi \omega_d}+e^{-i\pi \omega_a}\Big[ e^{i \pi \omega_b} \delta f_{34}^b + e^{i \pi \omega_d} \delta f_{34}^d \Big] - e^{i \pi \omega_a} \delta W_{\omega_3}e^{i \pi \omega_d}\nonumber
\ea
removes the singular terms. Here the first term compensates the most singular piece of the Regge pole; it has the same form as in the region $\tau_1\tau_5$.
On the other hand, our subtraction must be consistent with the complex complex conjugate of the Regge cut subtractions of $VLC^{++}$ in (\ref{deltapp-prel}). We therefore  write 
\ba
&&e^{i \pi \omega'_3}\frac{V_L(a) \Omega_b \Omega_c V_R(d)}{\Omega_{2'} \Omega_{3'} \Omega_{4'}}\nonumber\\
&&+ \Big[e^{i \pi \omega_a} \delta f_{23}^a + e^{i \pi \omega_c} \delta f_{23}^c \Big] e^{-i\pi \omega_d}+e^{-i\pi \omega_a}\Big[ e^{i \pi \omega_b} \delta f_{34}^b + e^{i \pi \omega_d} \delta f_{34}^d \Big] - e^{i \pi \omega_a} \delta W_{\omega_3}e^{i \pi \omega_d}\nonumber\\
&&\hspace{1cm}+\Big[(e^{2 i \pi \omega'_3} e^{-i \pi \omega_a}- e^{i \pi \omega_a}) \delta f_{23}^a + (e^{2 i \pi \omega'_3} e^{-i \pi \omega_c}- e^{i \pi \omega_c}) \delta f_{23}^c \Big] e^{-i\pi \omega_d}\nonumber\\
&&\hspace{1cm}+e^{-i\pi \omega_a}\Big[(e^{2 i \pi \omega'_3} e^{-i \pi \omega_b}- e^{i \pi \omega_b}) \delta f_{34}^b + (e^{2 i \pi \omega'_3} e^{-i \pi \omega_d}- e^{i \pi \omega_d}) \delta f_{34}^d \Big] \nonumber\\
&&\hspace{1cm}-\delta W_{\omega_3} (e^{2 i \pi \omega'_3} e^{-i\pi (\omega_a+\omega_d})-e^{i\pi (\omega_a+\omega_d})) \,.
\ea
The last three lines can be written in the more compact form
\ba
\label{delta--sing}
2i \Big[ S_{r2} + i \left( \Omega_{3'd} \Omega_{3'2'}   \delta \widetilde{W}_{\omega_2\omega_3;L}+ \Omega_{3'a} \Omega_{3'4'}   \delta \widetilde{W}_{\omega_3\omega_4;R}
-\Omega_{3'} \sin \pi (\omega'_3 -\omega_a-\omega_d) \right)\Big]
\ea
with 
\ba
&&S_{r2}=
\cos \pi (\omega'_3 - \omega_d)
\Omega_{3'2'}  \delta \widetilde{W}_{\omega_2\omega_3;L}+ \Omega_{3'4'}  \delta \widetilde{W}_{\omega_3\omega_4;R} \cos \pi (\omega'_3 - \omega_a) \nonumber\\
&&   + \cos \pi \omega'_3  \sin \pi (\omega_a +\omega_d - \omega'_3) 
 \delta W_{\omega_3} \Big] \,.
\ea

Next we include the double cut. Repeating our discussion given in the context of the region $\tau_1\tau_5$, we replace the first term in (\ref{delta--}) by $S_{r1}-i(S_2+S_4)$, and the first 2 lines inside the square brackets become:
\ba
\label{delta--next}
&&e^{i\pi  (\omega'_3-\omega'_2)}\widetilde{W}_{LL}^{reg}+\widetilde{W}_{RL}^{reg}+e^{i\pi  (2 \omega'_3-\omega'_2-\omega'_4)}\widetilde{W}_{LR}^{reg}+e^{i\pi  (\omega'_3-\omega'_4)}\widetilde{W}_{RR}^{reg}\\
&&-e^{i\pi \omega_a} LC(34)^{reg}-e^{i\pi \omega_d} LC(23)^{reg}+ e^{i\pi (\omega_a+\omega_d)} W_{\omega_3}^{reg}\nonumber\\
&&-i \left( W_{\omega_2}^{reg}- \frac{1}{2} \sin \pi (\omega_a-\omega_b)\right)
\left( W_{\omega_4}^{reg} -\frac{1}{2} \sin \pi (\omega_d-\omega_c)  \right)\nonumber\\
&&+2i \Big[ S_{r2}+i \left( \Omega_{3'd} \Omega_{3'2'}   \delta \widetilde{W}_{\omega_2\omega_3;L}+ \Omega_{3'a} \Omega_{3'4'}   \delta \widetilde{W}_{\omega_3\omega_4;R}
-\Omega_{3'} \sin \pi (\omega'_3 -\omega_a-\omega_d) \right)\Big] \,.\nonumber
\ea
 
Next we turn to the two last lines of (\ref{amp1245}) and collect the singular pieces which need to be compensated.
First, in the third line we decompose the second and third terms:
\ba 
\label{Sr2}
&&-\frac{e^{i\pi \omega'_3}}{\Omega_{3'}}\left( \Omega_{2'3'} \Omega_{4'3'} \widetilde{W}_{LR}+ \Omega_a\Omega_d W_{\omega_3} \right) 
\nonumber\\
 &&=-\frac{\cos \pi \omega'_3}{\Omega_{3'}} \left( \Omega_{2'3'} \Omega_{4'3'}\widetilde{W}_{LR} + \Omega_a\Omega_d W_{\omega_3} \right)
 - i \left( \Omega_{2'3'} \Omega_{4'3'} \delta \widetilde{W}_{LR}+\Omega_a\Omega_d \delta W_{\omega_3} \right)
\nonumber\\
&& - i \left( \Omega_{2'3'} \Omega_{4'3'}  \widetilde{W}_{LR}^{reg} + \Omega_a\Omega_d W_{\omega_3}^{reg} \right)\,.
\ea 
The real part together  with the last line of  (\ref{amp1245}) suggests to define 
\be
S_{r3}=
  \frac{\cos \pi \omega'_3}{\Omega_{3'}} \left(\Omega_{2'3'} \Omega_{4'3'} \widetilde{W}_{LR}+\Omega_a\Omega_d W_{\omega_3}\right)
+ \left(  \widetilde{W}_{\omega_2\omega_3;L}\frac{ \Omega_{3'2'}\Omega_d}{\Omega_{3'}}
+ \frac{\Omega_a \Omega_{3'4'}}{\Omega_{3'}} \widetilde{W}_{\omega_3\omega_4;R} \right) \,.
\ee
In  (\ref{amp1245}) the sum of these terms then becomes:
\ba
 &&2i \Big[ W_{\text{3 reggeon cut}}-S_{r3} -i \left( \Omega_{2'3'} \Omega_{4'3'} \delta \widetilde{W}_{LR}+\Omega_a\Omega_d \delta W_{\omega_3} \right)\nonumber\\
&&\hspace{2cm}- i \left( \Omega_{2'3'} \Omega_{4'3'}  \widetilde{W}_{LR}^{reg} + \Omega_a\Omega_d W_{\omega_3}^{reg} \right)\Big] \,.
\ea
Here the first line contains the singular terms which have to be compensated.  Using our result for $ \delta \widetilde{W}_{LR}$ the sum of the singular terms  can be written as:
\ba
\label{sub1}
&&2 i \Big[W_{\text{3 reggeon cut}}-S_{r3} -i \left( \Omega_{2'3'} \Omega_{4'3'} \delta \widetilde{W}_{LR}+\Omega_a\Omega_d \delta W_{\omega_3} \right)\Big]
=\nonumber\\
&&2i \Big[W_{\text{3 reggeon cut}} -S_{r3}-i\left( \Omega_{3'2'} \Omega_{3'd} \delta \widetilde{W}_{23;L} + \Omega_{3'4'} \Omega_{3'a} \delta \widetilde{W}_{34;R}\right. \nonumber\\ 
&&\left.+\Omega_{3'}^2 S_{r1}+\cos \pi \omega'_3 \Omega_{3'} (S_2+S_4)\right.,\nonumber\\
&&\hspace{1cm}\left.+\Omega_{3'} \sin \pi (\omega_a+\omega_d+\omega'_3)\delta W_{\omega_3}\right)\Big]\,.
\ea

Now we combine  (\ref{sub1}) with the last line of (\ref{delta--next}). We notice that the second term in the last line of  (\ref{delta--next}) cancels part of the second part of 
(\ref{sub1}). Using the identity 
\be
\Omega_{3'}  S_{r1}+\cos \pi \omega'_3 (S_2+S_4) = \cos \pi \omega'_3 \left( W_{\omega_2}^{reg}- \frac{1}{2} \sin \pi (\omega_a-\omega_b)\right)
\left( W_{\omega_4}^{reg} -\frac{1}{2} \sin \pi (\omega_d-\omega_c)  \right)
\ee
we are left with:
\ba
&& (\ref{sub1})+\text{last line of (\ref{delta--next})}=+2 i \Big[ W_{\text{3 reggeon cut}} + S_{r2}- S_{r3} \nonumber\\
&&-i  \Omega_{3'} \cos \pi \omega'_3 \left( W_{\omega_2}^{reg}- \frac{1}{2} \sin \pi (\omega_a-\omega_b)\right)
\left( W_{\omega_4}^{reg} -\frac{1}{2} \sin \pi (\omega_d-\omega_c)  \right)\nonumber\\
&& -2 i \Omega_{3'}^2 \cos \pi (\omega_a + \omega_d) \delta W_{\omega_3} \Big] \,.
\ea
Defining the subtraction
\be
\label{delta-3reggeon}
\delta W_{\text{3 reggeon cut}}=S_{r3}-S_{r2}
\ee
we finally arrive at: 
\ba
&& (\ref{sub1})+\text{last line of (\ref{delta--next})}= \nonumber\\
&&+2 i \Big[ W_{\text{3 reggeon cut}}^{reg}-i  \Omega_{3'} \cos \pi \omega'_3 \left( W_{\omega_2}^{reg}- \frac{1}{2} \sin \pi (\omega_a-\omega_b)\right)
\left( W_{\omega_4}^{reg} -\frac{1}{2} \sin \pi (\omega_d-\omega_c)  \right)\nonumber\\
&& -2 i \Omega_{3'}^2 \cos \pi (\omega_a + \omega_d) \delta W_{\omega_3} \Big] \,.
\ea

Finally the region $\tau_1\tau_2\tau_3\tau_4\tau_5$. Starting from (\ref{amp12345}) we proceed  in the same way. Addressing the first  two lines (ignoring for the moment 
$W_{\omega_2 \omega_4}$) we observe that the ansatz
\ba
\label{delta++}
&&e^{i \pi \omega'_3}\frac{V_L(a) \Omega_b \Omega_c V_R(d)}{\Omega_{2'} \Omega_{3'} \Omega_{4'}}\\
&&
+ \Big[e^{-i \pi \omega_a} \delta f_{23}^a + e^{-i \pi \omega_c} \delta f_{23}^c \Big] e^{i\pi \omega_d}+e^{i\pi \omega_a}\Big[ e^{-i \pi \omega_b} \delta f_{34}^b + e^{-i \pi \omega_d} \delta f_{34}^d \Big] - e^{i \pi \omega_a} \delta W_{\omega_3}e^{i \pi \omega_d}\nonumber
\ea
after combination with the Regge singular pole term removes the singular terms. Here the second line is the complex conjugate of the second line of (\ref{delta--}). The analogue of
(\ref{delta--sing}), after some algebra, has the form
\ba
\label{delta++sing}
- 2i \Big[ S_{r2} - i \left( \Omega_{3'd} \Omega_{3'2'}   \delta \widetilde{W}_{\omega_2\omega_3;L}+ \Omega_{3'a} \Omega_{3'4'}   \delta \widetilde{W}_{\omega_3\omega_4;R}
-\Omega_{3'} \sin \pi (\omega'_3 -\omega_a-\omega_d) \right)\Big] \,.\nonumber\\
\ea
Including next the double cut term  $W_{\omega_2 \omega_4}$, the singularities add up to 
\be 
-S_{r1} + i(S_2+S_4)- i \left( W_{\omega_2}^{reg}- \frac{1}{2} \sin \pi (\omega_a-\omega_b)\right)
\left( W_{\omega_4}^{reg} -\frac{1}{2} \sin \pi (\omega_d-\omega_c)  \right) \,.
\ee
Adding the last two lines of (\ref{amp12345}) the first part is modifiied:
\be 
-S_{r1} + i(S_2+S_4) \to   -S_{r1}-i(S_2+S_4)\,.
\ee
In (\ref{delta++}) we therefore replace the first term by $S_{r1}+i(S_2+S_4)$.
In analogy with (\ref{delta--next}) we then have
\ba
\label{delta++next}
&&e^{i\pi  (\omega'_2-\omega'_3)}\widetilde{W}_{LL}^{reg}+\widetilde{W}_{RL}^{reg}+e^{-i\pi  (2 \omega'_3-\omega'_2-\omega'_4)}\widetilde{W}_{LR}^{reg}+e^{i\pi  (\omega'_4-\omega'_3)}\widetilde{W}_{RR}^{reg}\\
&&-e^{-i\pi \omega_a} {LC(34)^{reg}}^*-e^{-i\pi \omega_d} {LC(23)^{reg}}^*+ e^{-i\pi (\omega_a+\omega_d)} W_{\omega_3}^{reg}\nonumber\\
&&-i \left( W_{\omega_2}^{reg}- \frac{1}{2} \sin \pi (\omega_a-\omega_b)\right)
\left( W_{\omega_4}^{reg} -\frac{1}{2} \sin \pi (\omega_d-\omega_c)  \right)\nonumber\\
&&-2i \Big[ S_{r2}-i \left( \Omega_{3'd} \Omega_{3'2'}   \delta \widetilde{W}_{\omega_2\omega_3;L}+ \Omega_{3'a} \Omega_{3'4'}   \delta \widetilde{W}_{\omega_3\omega_4;R}
-\Omega_{3'} \sin \pi (\omega'_3 -\omega_a-\omega_d) \right)\Big] \,.\nonumber
\ea
Turning to the 3rd and 4th line of (\ref{amp12345}) we find, in analogy with (\ref{sub1}),
\ba
&&-2 i \Big[W_{\text{3 reggeon cut}}-S_{r3} +i \left( \Omega_{2'3'} \Omega_{4'3'} \delta \widetilde{W}_{LR}+\Omega_a\Omega_d \delta W_{\omega_3} \right)\Big]
=\nonumber\\
&&-2i \Big[W_{\text{3 reggeon cut}} -S_{r3}+i\left( \Omega_{3'2'} \Omega_{3'd} \delta \widetilde{W}_{23;L} + \Omega_{3'4'} \Omega_{3'a} \delta \widetilde{W}_{34;R}\right. \nonumber\\
&&\left.+\Omega_{3'}^2 S_{r1}+\cos \pi \omega'_3 \Omega_{3'} (S_2+S_4)\right.,\nonumber\\
&&\hspace{1cm}\left.+\Omega_{3'} \sin \pi (\omega_a+\omega_d+\omega'_3)\delta W_{\omega_3}\right)\Big] \,.
\ea
Combining with the last line of (\ref{delta++next}) we finally have:
\ba
&&-2 i \Big[ W_{\text{3 reggeon cut}}^{reg}+i  \Omega_{3'} \cos \pi \omega'_3 \left( W_{\omega_2}^{reg}- \frac{1}{2} \sin \pi (\omega_a-\omega_b)\right)
\left( W_{\omega_4}^{reg} -\frac{1}{2} \sin \pi (\omega_d-\omega_c)  \right)\nonumber\\
&& +2 i \Omega_{3'}^2 \cos \pi (\omega_a + \omega_d) \delta W_{\omega_3} \Big] \,.
\ea

Inserting, after performing the $\omega'_i$-integrations, the results of this appendix into 
(\ref{amp15}) - (\ref{amp12345}), we arrive at the results listed in
eqs.(\ref{amp15reg}) - (\ref{amp12345reg}).

\subsubsection{The discontinuity $\Delta_{1234}$}

We complete this section by inserting the subtractions into the discontinuities  $\Delta_{1234}$ and  $\Delta_{23}\Delta_{1234}$ and demonstrating the cancellation of the singularities.. We return to the rhs of (\ref{disc-1234}) and undo the $\omega'$-integrations. We then 
insert the subtractions derived in the previous subsections.
To show the cancellation of all singular terms requires some algebra, and we only scetch the main steps.
For the first three lines (without the double cut term, $W_{\omega_2 \omega_4}$)  we have the subtractions listed in (\ref{sub-WLL}) -  (\ref{sub-WRR}).
We combine the last term of the third  line, $W_{\omega_2 \omega_4}$, with the first term of the product in the next line and 
with  the two terms in the 7th  line. Using our previous notation we arrive at:
\ba
\label{W24pieces}
&&- e^{i\pi(\omega'_3-\omega'_2-\omega'_4)} \left( e^{i\pi \omega_3} \frac{W_{\omega_2\omega_4}}{\Omega_{3'}}
+ e^{i\pi \omega'_3} \frac{V_L(a) \Omega_b \Omega_c V_R(d)}{\Omega_{2'} \Omega_{3'} \Omega_{4'}}
+ 2i (S_2+S_4)\right)\nonumber\\
=&&- e^{i\pi(\omega'_3-\omega'_2-\omega'_4)} \left( S_{r1} + i(S_2+S_4)  \right.\nonumber\\ 
&&\hspace{2cm} \left. +i \left( W_{\omega_2}^{reg} -\frac{1}{2} \sin \pi (\omega_a-\omega_b)\right) 
\left( W_{\omega_4}^{reg} -\frac{1}{2} \sin \pi (\omega_d-\omega_c)\right) \right) \,.
\ea
In a similar way we rewrite the curley bracket in the 5th  line and obtain:
\ba
\label{3reggeonpieces}
&&2i e^{i\pi(\omega_2+\omega_4-\omega_3)} \left( -W_{\text{3 reggeon}}^{reg} \right.\nonumber\\
&&\left.
+S_{r2}+\Big[ \cos \pi \omega'_3 \sin \pi(\omega'_2+\omega'_4-\omega'_3)- i \Omega_{2'} \Omega_{4'} \Big] \widetilde{W}_{LR} - i \Omega_a \Omega_d W_{\omega_3}\right) \,.
\ea

Our expression for the discontinuity $\Delta _{1234}$ thus takes the somewhat modified form:
\ba
\label{disc-1234mod}
&&\Delta_{1234}^{int} =\nonumber\\
&& e^{-i\pi(\omega_1+\omega_5)} 
\cdot \Big[ e^{i\pi(2\omega'_2-\omega'_4)} \widetilde{W}_{LL} + e^{i\pi(3\omega'_3-\omega'_2-\omega'_4)}  \widetilde{W}_{RL}+
 e^{i\pi(2\omega'_4-\omega'_2)}  \widetilde{W}_{RR}  \nonumber\\
&&\hspace{2cm}+\left( e^{-i\pi \omega'_3} +2i e^{i\pi(\omega'_2+\omega'_4-\omega'_3)} (cos \pi \omega'_3
 \sin \pi(\omega'_2+\omega'_4-\omega'_3)- i \Omega_{2'} \Omega_{4'}
\right) \widetilde{W}_{LR} \nonumber\\
&&\hspace{1cm}-e^{i\pi (\omega'_4-\omega'_2-\omega'_3)}  LC(23)  e^{i\pi \omega_d} -
e^{i\pi (\omega'_2-\omega'_3-\omega'_4)}  e^{i\pi\omega_a} LC(34)
\nonumber\\
&&\hspace{2cm}+e^{i\pi (\omega'_2+\omega'_4 - \omega'_3)}  e^{-i\pi \omega_a} W_{\omega_3} e^{-i\pi \omega_d}
\nonumber\\
&&\hspace{1cm} - e^{i\pi (\omega'_3-\omega'_2 - \omega'_4)} \left( S_{r1}+ i(S_2+S_4) \right.\nonumber\\ 
&&\left.\hspace{2cm}+i  \left( W_{\omega_2}^{reg} -\frac{1}{2} \sin \pi (\omega_a-\omega_b)\right) 
\left( W_{\omega_4}^{reg} -\frac{1}{2} \sin \pi (\omega_d-\omega_c)\right) \right)\nonumber\\
&&\hspace{1cm}+2i e^{i\pi (\omega'_2+\omega'_4-\omega'_3)}\left( - W_{\text{3 reggeon}}^{reg}+S_{r2} - i \Omega_a \Omega_d W_{\omega_3}\right)
\nonumber\\
&&\hspace{1cm}+\frac{1}{\Omega_{3'}} \left( 2i  e^{i\pi (\omega'_3-\omega'_2)}  \frac{V_L(a)  \Omega_b}{\Omega_{2'}} V_R(c)V_R(d) 
+2i e^{i\pi (\omega'_3-\omega'_4)} V_L(a) V_L(b)\frac{ \Omega_cV_R(d)}{\Omega_{4'}}\right. \nonumber \\
&&\left.\hspace{2cm} + 4 V_L(a) V_L(b)V_R(c)V_R(d) \right)
\nonumber\\
&&\hspace{1cm} +2i e^{i \pi (\omega'_3-\omega'_2-\omega'_4)} \left( 
W_{\omega_2} e^{i\pi(\omega'_4-\omega_c)}V_R(d) +e^{i\pi(\omega'_2-\omega_b)}V_L(a) W_{\omega_4}\right) \Big] \,. \nonumber\\
\ea

Now we are ready to verify the cancellation of the different singularities. All subtractions $\delta \widetilde{W}_{LL}$ etc. have 
singulartities due to the  denominatoirs  $1/\Omega_{2'3'}\Omega_{3'4'}$. After some algebra one finds that in the 
sum of the first two lines of (\ref{disc-1234mod}) these  singulartities cancel completely.

Next we collect the terms proportional to $S_{r1}$, $S_2$ and $S_4$; they contain short cut contributions, $W_{\omega_2}$ or $W_{\omega_4}$, multiplied by a singular pole term.  They are contained in the
first two lines of  (\ref{disc-1234mod}) (in the $\delta \widetilde{W}_{LL}$ etc.),  in the 5th line, and in $S_{r2}$. With the identity (\ref{Sr1-identity})
%\be
%S_{r1}= \frac{\cos \pi \omega_3}{\Omega_3} \Big[-(S_2+S_4) + \left( W_{\omega_2}^{reg} -\frac{1}{2} \sin \pi (\omega_a-\omega_b)\right) 
%\left( W_{\omega_4}^{reg} -\frac{1}{2} \sin \pi (\omega_d-\omega_c)\right) \Big]
%\ee
the sum of terms proportional to  $S_{r1}$, $S_2$ and $S_4$ becomes
\ba
&&-2i\cos \pi \omega'_3 \cdot\nonumber\\
&&\cdot\left( e^{-i\pi (\omega'_2+\omega'_4} + 4 \Omega_{2'} \Omega_{4'}\right) \left( W_{\omega_2}^{reg} -\frac{1}{2} \sin \pi (\omega_a-\omega_b)\right) 
\left( W_{\omega_4}^{reg} -\frac{1}{2} \sin \pi (\omega_d-\omega_c)\right) \nonumber\\
&&\hspace{5cm}- 4  \left(  e^{i\pi (\omega'_3 - \omega'_2)}  \Omega_{4'} S_2 +  e^{i\pi (\omega'_3 - \omega'_4)} \Omega_{2'} S_4 \right) \,.
\ea

Next we collect terms proportional to $\frac{1}{\Omega_{2'}  \Omega_{3'}}$. They are contained in $\delta f_{23}^a$ and $\delta f_{23}^c$, in $S_{r2}$ in the 7th line, and in the 8th line of 
of (\ref{disc-1234mod}). For the sum of all 
$\delta f_{23}^a$ and $\delta f_{23}^c$ the coefficients are :
\be
c23a=-2i V_R(d) e^{i\pi (\omega_a-\omega'_2-\omega'_3)} \Big[ e^{2 i \pi \omega_a} + 4 e^{i\pi (\omega'_2+\omega'_3)} \Omega_{2'a} \Omega_{3'a}\Big]
\ee
and 
\be
c23c=-2i V_R(d) e^{i\pi (\omega_c-\omega'_2-\omega'_3)} \Big[ e^{2 i \pi \omega_c} + 4 e^{i\pi (\omega'_2+\omega'_3)} \Omega_{2'c} \Omega_{3'c}\Big] \,.
\ee
Making use of our results for $\delta f_{23}^a$ and $\delta f_{23}^c$ and adding  the last line of (\ref{disc-1234mod}) we find 
for the sum:
\ba
&&c23a\cdot \delta f_{23}^a +c23c \cdot \delta f_{23}^c + 2i e^{i \pi (\omega'_3-\omega'_2})\frac{  V_L(a) \Omega_b V_R(c) V_R(d)}{\Omega_{2'} \Omega_{3'}}\\
&& = (2i)^3 V_L(a) \Omega_b V_R(c) V_R(d) +( 2i)^2  e^{-i\pi \omega'_2} \frac{  V_L(a) \Omega_b V_R(c) V_R(d)}{\Omega_{2'}}\nonumber\\ 
&&-2ie^{i \pi (\omega_a-\omega'_2-\omega'_3)} \Big[ e^{2 i \pi \omega_a} + 4  e^{i\pi (\omega'_2+\omega'_3)}\Omega_{2'a} \Omega_{3'a}\Big] \delta W_{\omega_3}\nonumber\\
&&\hspace{1cm}-2ie^{i \pi (\omega_c-\omega'_2-\omega'_3}) \Big[ e^{2 i \pi \omega_c} + 4 e^{i\pi (\omega'_2+\omega'_3)} \Omega_{2'c} \Omega_{3'c}\Big] \delta W_{\omega_2} \,.\nonumber\\
\ea
Similar results hold for  $\delta f_{34}^b$ and $\delta f_{34}^d$. As expected, all double poles cancel.

Finally we collect terms proportional to simple poles contained in $ \delta W_{\omega_2}$,  $ \delta W_{\omega_3}$, and $ \delta W_{\omega_4}$. For the sum of terms containing
$ \delta W_{\omega_2}$ we obtain:
\be
(2i)^3  \Omega_{2'} \delta W_{\omega_2} V_R(c) V_R(d)+  (2i)^2 \frac{1}{2} \sin \pi (\omega_b - \omega_a)  V_R(c) V_R(d) \,.
\ee
For the sum of terms proportional to $ \delta W_{\omega_3}$ we find:
\ba
&&-4i \Omega_{3'}  V_L(a)  \delta W_{\omega_3} V_R(d)\nonumber\\
&& + 2 V_L(a) \left( 2 \Omega_{3'} \cos \pi \omega_b \cos \pi \omega_c - \cos \pi \omega'_3 \sin \pi (\omega_b+\omega_c) \right) V_R(d) \,.
\ea
This completes our proof that, after inserting all the subtractions and performing the $\omega'$-integrations, we end up with a finite expression for  the single discointinuity $\Delta_{1234}$, 
as written in (\ref{disc-1234-finite}).

\subsubsection{The double discontinuity $\Delta_{23} \Delta_{1234}$}

We start from (\ref{d-23-1234}). As the first step, using (\ref{delta-3reggeon}), we subsitute 
\ba
W_{\text{3 reggeon-cut}} 
= W_{\text{3 reggeon-cut }}^{reg}+ S_{r3}-S_{r2}
\ea
and combine $S_{r3}$ with the folllowing three lines. We obtain:
\ba
\label{d-23-1234-mod1}
&&(\Delta_{23} \Delta_{1234})^{int}=e^{-i\pi(\omega_1+\omega_5)} \nonumber\\
&&\cdot \Big[\left(- e^{i\pi(\omega'_2+\omega'_4-\omega'_3)} + 2 \cos \pi (\omega'_2+\omega'_4)\right)
(W_{\text{3 reggeon}}^{reg} -S_{r2}) \nonumber\\
&&+i \Omega_{2'3'}  \Omega_{4'3'} \left( e^{i\pi (\omega'_2+\omega'_4)}+2e^{2 i\pi  \omega'_3 \cos \pi(\omega'_2-\omega'_4)} \right) \widetilde{W}_{LR}\nonumber\\
&&-e^{i\pi \omega_d} e^{i\pi (\omega'_4-\omega'_2)} \Omega_{2'3'} \widetilde{W}_{23;L}  - e^{i\pi \omega_a} e^{i\pi (\omega'_2-\omega'_4)} \Omega_{4'3'} \widetilde{W}_{34;R}
\nonumber\\
&&+ \left(e^{-i\pi(\omega'_2+\omega'_4)}\sin \pi (\omega_a+\omega_d-\omega'_3)+2i (e^{i\pi(\omega'_3-\omega_a-\omega_d)} \Omega_{2'3'}\Omega_{4'3'} +\frac{1}{2} \Omega_a \Omega_d) \right.\nonumber\\
&&\hspace{2cm}\left. +4i e^{i\pi \omega'_3} \Omega_{2'3'} \Omega_{4'3'} \Omega_a\Omega_d
\right) W_{\omega_3} \nonumber\\ 
&&-2i \frac{V_L(a) V_L(b)V_R(c)V_R(d)}{\Omega_{3'}} 
 \Big] \,. \nonumber\\
\ea 
As a result, all singular terms $\sim 1/\Omega_{3'}$ have been cancelled, except for the Regge pole term in the last line. 
Next we use the decomposition of the Regge cut amplitudes $\widetilde{W}_{LR} = \widetilde{W}_{LR}^{reg} + \delta  \widetilde{W}_{LR}$ etc 
%\ba
%\widetilde{W}_{LR} &\rightarrow& \widetilde{W}_{LR}^{reg} + \delta  \widetilde{W}_{LR}\nonumber\\
%\widetilde{W}_{23;L} &\rightarrow& \widetilde{W}_{23;L}^{reg} + \delta  \widetilde{W}_{23;L}\nonumber\\
%\widetilde{W}_{34;R} &\rightarrow& \widetilde{W}_{34;R}^{reg} + \delta  \widetilde{W}_{34;R}\nonumber\\
%W_{\omega_3} &\rightarrow& W_{\omega_3}^{reg}+   \delta W_{\omega_3}\,.
%\ea 
and obtain for the regular pieces  (disregarding the overall phase factor $e^{-i\pi(\omega_1+\omega_5)}$)
\ba
\label{d23d1234-reg}
&&\left(- e^{i\pi(\omega'_2+\omega'_4-\omega'_3)} + 2 \cos \pi (\omega'_2+\omega'_4)\right) W_{\text{3 reggeon}}^{reg}\nonumber\\
&&+i \Omega_{2'3'}  \Omega_{4'3'} \left( e^{i\pi (\omega'_2+\omega'_4)}+2e^{2 i\pi  \omega'_3 \cos \pi(\omega'_2-\omega'_4)} \right) \widetilde{W}_{LR}^{reg}\nonumber\\
&&-e^{i\pi \omega_d} e^{i\pi (\omega'_4-\omega'_2)} \Omega_{2'3'} \widetilde{W}_{23;L}^{reg}  - e^{i\pi \omega_a} e^{i\pi (\omega'_2-\omega'_4)} \Omega_{4'3'} \widetilde{W}_{34;R}^{reg}\nonumber\\
&&+ \left(e^{-i\pi(\omega'_2+\omega'_4)}\sin \pi (\omega_a+\omega_d-\omega'_3)+2i (e^{i\pi(\omega'_3-\omega_a-\omega_d)} \Omega_{2'3'}\Omega_{4'3'} +\frac{1}{2} \Omega_a \Omega_d) \right.\nonumber\\
&&\hspace{2cm}\left. +4i e^{i\pi \omega'_3} \Omega_{2'3'} \Omega_{4'3'} \Omega_a\Omega_d
\right) W_{\omega_3}^{reg}\,.
\ea
Collecting the remaining singular terms, we 
use 
\ba
S_{r2}&=&
\cos \pi (\omega'_3 - \omega_d)
\Omega_{3'2'}  \delta \widetilde{W}_{\omega_2\omega_3;L}+ \Omega_{3'4'}  \delta \widetilde{W}_{\omega_3\omega_4;R} \cos \pi (\omega'_3 - \omega_a)\nonumber\\
&& + \Omega_{3'} \cos \pi \omega'_3  \sin \pi (\omega_a +\omega_d - \omega'_3)  \delta W_{\omega_3} 
\ea
and
\ba
 \Omega_{2'3'} \Omega_{4'3‘} \delta \widetilde{W}_{LR}&=&\Omega_{3'2'} \Omega_{3'd} \delta \widetilde{W}_{23;L} +\Omega_{3'4'} \Omega_{3'a} \delta \widetilde{W}_{34;R}
 - \Omega_{3'a}  \Omega_{3'd} \delta W_{\omega_3}\\
&&+ \Omega_{3'} \cos \pi \omega'_3  \left( W_{\omega_2}^{reg} -\frac{1}{2} \sin \pi (\omega_a-\omega_b)\right) 
\left( W_{\omega_4}^{reg} -\frac{1}{2} \sin \pi (\omega_d-\omega_c)\right)\,.\nonumber
\ea
Here the last line belongs to the regular pieces, i.e. has to be part of (\ref{d23d1234-reg}).
For the sum of all remaining terms we first collect all singular terms (proportional to $ \delta \widetilde{W}_{23;L}$,  $ \delta \widetilde{W}_{34;R}$,  $ \delta W_{\omega_3}$),
together with the the last line:
\ba
\label{sumsing}
&&-2i e^{i\pi \omega'_2} V_R(d) \Omega_{23}  \delta \widetilde{W}_{23;L}-2i e^{i\pi \omega'_4} V_L(a)  \Omega_{4'3'}  \delta \widetilde{W}_{34;R}\nonumber\\
&&+\left( -e^{i\pi \omega'_3} \sin \pi (\omega'_2+\omega'_4-\omega_a-\omega_d)+e^{-i\pi \omega'_3} (e^{i\pi (\omega'_4+\omega_d)} V_L(a) + e^{i\pi (\omega'_2+\omega_a)} V_R(d) 
\right)\delta W_{\omega_3}\nonumber\\
&&\hspace{1cm}-2i \frac{V_L(a) V_L(b)V_R(c)V_R(d)}{\Omega_{3'}} \,.
\ea
Next we insert:
\ba
\Omega_{3'2*} \delta \widetilde{W}_{23;L}&=& \frac{1}{2} \sin \pi (\omega_b-\omega_a) V_R(c) +\Omega_{3'a} \delta W_{\omega_3}  \\
\Omega_{3'4'} \delta \widetilde{W}_{34;R}&=& \frac{1}{2} \sin \pi (\omega_c-\omega_d)V_L(b)+\Omega_{3'd} \delta W_{\omega_3} \\
\ea
and obtain
\ba
&&2i e^{i\pi \omega'_3}V_L(a) V_R(d) \delta W_{\omega_3} --2i \frac{V_L(a) V_L(b)V_R(c)V_R(d)}{\Omega_{3'}} \nonumber\\
&&+i \left( e^{i\pi \omega'_2} \sin \pi (\omega_b-\omega_a) V_R(c) V_R(d)+  V_L(a) V_L(b) e^{i\pi \omega'_4} \sin \pi (\omega_c-\omega_d) \right) \,.
\ea
Finally  we use
\be
\delta W_{\omega_3}= -\frac{1}{2} \sin \pi (\omega_b+\omega_c) + \cos \pi \omega'_3 \frac{\Omega_b \Omega_c}{\Omega_{3'}}
\ee
and find complete cancellation of all singular terms:
\ba
\text{eq(\ref{sumsing})}&=& -iV_L(a) \left( e^{i\pi \omega_b} V_R(c) +V_L(b)  e^{i\pi \omega_c} \right) V_R(d)\\
&&+i \left( e^{i\pi \omega'_2} \sin \pi (\omega_b-\omega_a) V_R(c) V_R(d)+  V_L(a) V_L(b) e^{i\pi \omega'_4} \sin \pi (\omega_c-\omega_d) \right) \,.\nonumber
\ea
The final sum, after perfoming the  $\omega'_i$-integrations,  is given in  (\ref{d-23-1234-final}).

\end{document}